\documentclass[twocolumn,trackchanges]{aastex62}

\newcommand\kms{km s$^{-1}$}
\newcommand\masyr{mas yr$^{-1}$}
\newcommand\teff{$T_{eff}$}
\newcommand\logg{$\log g$}
\newcommand\vsini{$v\sin i$}
\newcommand\chisq{$\chi^2$}
\newcommand{\fbol}{\ensuremath{F_{\rm bol}}}
\newcommand{\lbol}{\ensuremath{L_{\rm bol}}}

\begin{document}
\title{The APOGEE-2 Survey of the Orion Star Forming Complex II: Six-dimensional structure}

\author[0000-0002-5365-1267]{Marina Kounkel}
\affil{Department of Physics and Astronomy, Western Washington University, 516 High St, Bellingham, WA 98225}
\author[0000-0001-6914-7797]{Kevin Covey}
\affil{Department of Physics and Astronomy, Western Washington University, 516 High St, Bellingham, WA 98225}
\author[0000-0002-2011-4924]{Genaro Su\'arez}
\affil{Instituto de Astronom\'{i}a, Universidad Nacional Aut\'{o}noma de M\'{e}xico, Unidad Acad\'{e}mica en Ensenada, Ensenada 22860, Mexico}
\author[0000-0001-8600-4798]{Carlos Rom\'{a}n-Z\'{u}\~{n}iga}
\affil{Instituto de Astronom\'{i}a, Universidad Nacional Aut\'{o}noma de M\'{e}xico, Unidad Acad\'{e}mica en Ensenada, Ensenada 22860, Mexico}
\author[0000-0001-9797-5661]{Jesus Hernandez}
\affil{Instituto de Astronom\'{i}a, Universidad Nacional Aut\'{o}noma de M\'{e}xico, Unidad Acad\'{e}mica en Ensenada, Ensenada 22860, Mexico}
\author[0000-0002-3481-9052]{Keivan Stassun}
\affil{Department of Physics and Astronomy, Vanderbilt University, VU Station 1807, Nashville, TN 37235, USA}
\author[0000-0002-7916-1493]{Karl O Jaehnig}
\affil{Department of Physics and Astronomy, Vanderbilt University, VU Station 1807, Nashville, TN 37235, USA}
\author[0000-0002-5077-6734]{Eric D. Feigelson}
\affil{Department of Astronomy and Astrophysics, 525 Davey Laboratory, Pennsylvania State University, University Park, PA 16802, USA}
\author[0000-0002-5855-401X]{Karla Pe\~{n}a Ram\'{i}rez}
\affil{Unidad de Astronom\'{i}a, Universidad de Antofagasta, Avenida Angamos 601, Antofagasta 1270300, Chile}
\author[0000-0002-1379-4204]{Alexandre Roman-Lopes}
\affil{Departamento de F\'{i}sica, Facultad de Ciencias, Universidad de La Serena, Cisternas 1200, La Serena, Chile}
\author{Nicola Da Rio}
\affil{Department of Astronomy, University of Virginia, Charlottesville, VA 22904, USA}
\author[0000-0003-1479-3059]{Guy S Stringfellow}
\affil{Center for Astrophysics and Space Astronomy, Department of Astrophysical and Planetary Sciences, University of Colorado, 389 UCB, Boulder, CO 80309-0389, USA}
\author[0000-0001-6072-9344]{J. Serena Kim}
\affil{Steward Observatory, University of Arizona, 933 North Cherry Avenue, Tucson, AZ 85721–0065, USA}
\author{Jura Borissova}
\affil{Departamento de F\'{i}sica y Astronom\'{i}a, Universidad de Valpara\'{i}so, Av. Gran Breta\~{n}a 1111, Playa Ancha, Casilla 5030, Valpara\'{i}so, Chile}
\affil{Millennium Institute of Astrophysics (MAS), Santiago, Chile}
\author{Jos\'{e} G. Fern\'{a}ndez-Trincado}
\affil{Departamento de Astronom\'{i}a, Universidad de Concepci\'{o}n, Casilla 160-C, Concepcio\'{o}n, Chile}
\affil{Institut Utinam, CNRS UMR6213, Univ. Bourgogne Franche-Comt\'e, OSU THETA , Observatoire de Besan\c{c}on, BP 1615, 25010 Besan\c{c}on Cedex, France}
\author[0000-0002-6523-9536]{Adam Burgasser}
\affil{University of California at San Diego, Center for Astrophysics and Space Science, La Jolla, CA 92093, USA}
\author[0000-0002-1693-2721]{D. A. Garc\'{i}a-Hern\'{a}ndez}
\affil{Instituto de Astrof\'{i}sica de Canarias (IAC), E-38205 La Laguna, Tenerife, Spain}
\affil{Universidad de La Laguna (ULL), Departamento de Astrofísica, E-38206 La Laguna, Tenerife, Spain}
\author{Olga Zamora}
\affil{Instituto de Astrof\'{i}sica de Canarias (IAC), E-38205 La Laguna, Tenerife, Spain}
\affil{Universidad de La Laguna (ULL), Departamento de Astrofísica, E-38206 La Laguna, Tenerife, Spain}
\author[0000-0002-2835-2556]{Kaike Pan}
\affil{Apache Point Observatory and New Mexico State, University, P.O. Box 59, Sunspot, NM, 88349-0059, USA}
\author{Christian Nitschelm}
\affil{Unidad de Astronom\'{i}a, Universidad de Antofagasta, Avenida Angamos 601, Antofagasta 1270300, Chile}

\email{marina.kounkel@wwu.edu}

\begin{abstract}
We present an analysis of spectrosopic and astrometric data from APOGEE-2 and \textit{Gaia} DR2 to identify structures towards the Orion Complex. By applying a hierarchical clustering algorithm to the 6-dimensional stellar data, we identify spatially and/or kinematically distinct groups of young stellar objects with ages ranging from 1 to 12 Myr. We also investigate the star forming history within the Orion Complex, and identify peculiar sub-clusters. With this method we reconstruct the older populations in the regions that are presently largely devoid of molecular gas, such as Orion C (which includes the $\sigma$ Ori cluster), and Orion D (the population that traces Ori OB1a, OB1b, and Orion X). We report on the distances, kinematics, and ages of the groups within the Complex. The Orion D groups is in the process of expanding. On the other hand, Orion B is still in the process of contraction. In $\lambda$ Ori the proper motions are consistent with a radial expansion due to an explosion from a supernova; the traceback age from the expansion exceeds the age of the youngest stars formed near the outer edges of the region, and their formation would have been triggered when they were half-way from the cluster center to their current positions. We also present a comparison between the parallax and proper motion solutions obtained by \textit{Gaia} DR2, and those obtained towards star-forming regions by Very Long Baseline Array.
\end{abstract}

\keywords{stars: pre-main sequence, stars: kinematics and dynamics, open clusters and associations: Orion Complex}

\section{Introduction}

\begin{figure}
\epsscale{1}
\plotone{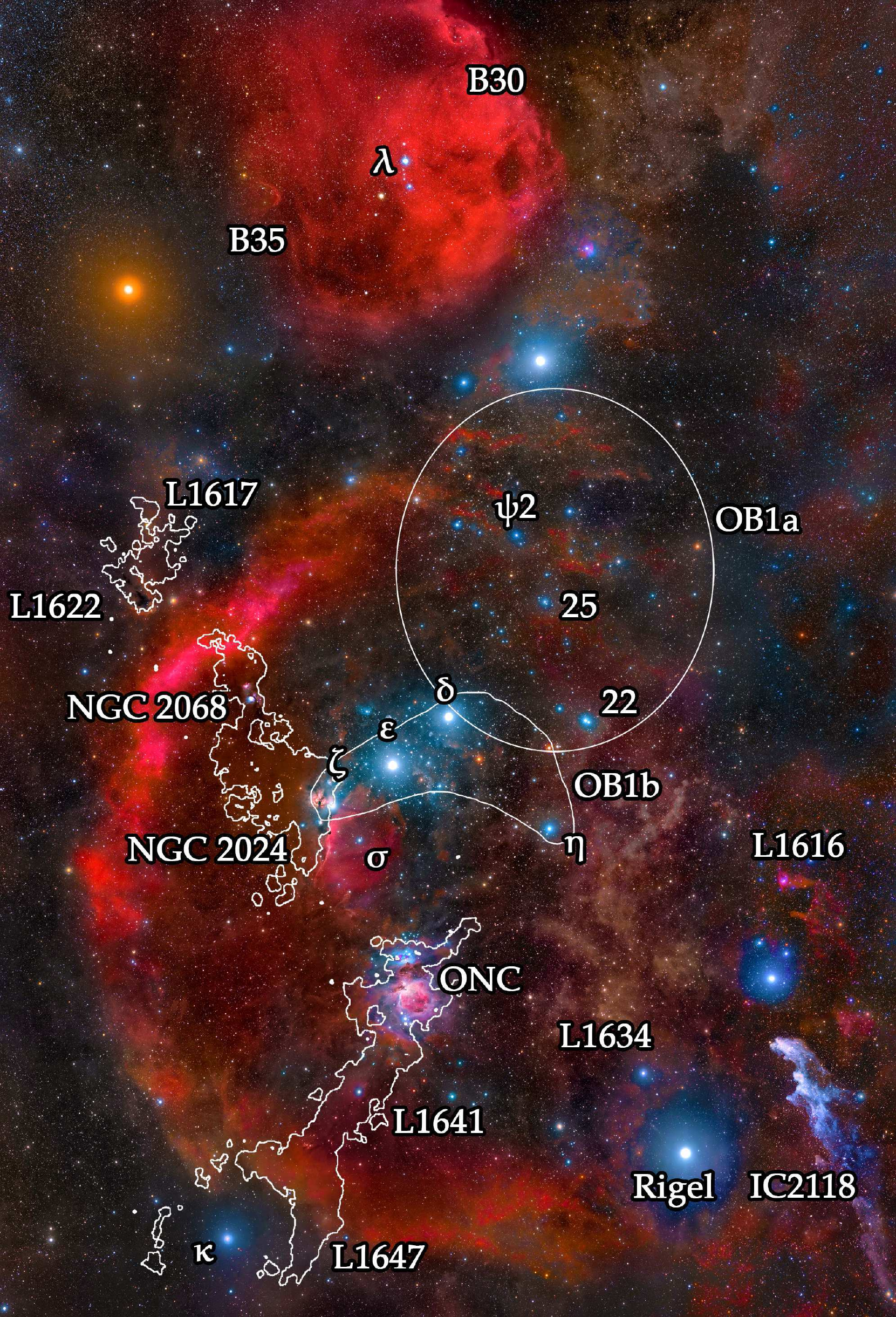}
\caption{Wide-field image of Orion with the identification of the prominent star forming regions. Background image is astrophotography, courtesy of Rogelio Bernal Andreo.
\label{fig:big}}
\end{figure}

Star formation in the Orion Complex has taken place over an extended ($>$10 Myr) period of time, and over a large ($>$100 pc) volume of space, containing multiple stellar populations. \citet{blaauw1964} originally identified the Orion OB1 association at $199^\circ<l<210^\circ$ and $-12^\circ<b<-21^\circ$ (Figure \ref{fig:big}). The current epoch of star formation is confined to the Orion A and B molecular clouds, which contain massive clusters such as the Orion Nebula Cluster (ONC), NGC 2024, and NGC 2068. The stars in these clouds have typical ages of $\sim$1--3 Myr \citep{levine2006,flaherty2008,muench2008,da-rio2010}. Several other populations of young stars that have already dissipated their molecular gas and dust can also be identified. The Orion OB1b region roughly traces the belt stars, with stellar ages of $\sim$5 Myr \citep{briceno2005,caballero2008,bally2008}. A massive cluster centered at $\sigma$ Ori is found towards Ori OB1b, although it is comparatively younger than OB1b itself with an age of $\sim$3 Myr, and the relationship between $\sigma$ Ori and Ori OB1b is still ill-defined \citep{jeffries2006,sherry2008,pena-ramirez2012,hernandez2014}. Ori OB1a has typical stellar ages of $\sim$7--10 Myr; the most well studied group in this region is a cluster near 25 Ori \citep{briceno2007,downes2014}, although many of the properties of this sub-association, including its extent or  membership, still remain highly uncertain. North of Ori OB1 is the $\lambda$ Ori cluster, which has an age of $\sim$6 Myr \citep{dolan2001,bayo2011}. A supernovae occurred near the center of the cluster disassociating nearby molecular gas and sweeping up the remaining material into a ring of dust and gas with a radius of 4-5 degrees \citep{dolan2002,mathieu2008,hernandez2010,bayo2011}. 

Although the Orion Complex is the best studied large scale star forming region, much of its 6 dimensional (6D) structure remains uncertain, compromising our understanding of its star formation history. Located at the average distance of $\sim$400 pc towards the galactic anticenter, it was too far away to have precise parallax or proper motion (PM) estimates by Hipparcos \citep{de-zeeuw1999}. In lieu of direct distance measurements, other studies have used radial velocity (RV), or photometric fitting in order to distinguish structures along the line of sight. \citet{jeffries2006} observed two distinct RV distributions towards $\sigma$ Ori, separated by 7 \kms. They suggested that two populations account for this difference in RVs, one that consists of bona fide members of $\sigma$ Ori, and one that is a foreground population that increases in concentration towards the north of $\sigma$ Ori, consisting of members of the Ori OB1ab sub-associations. Other RV surveys in the region were conducted by \citet{maxted2008}, \citet{sacco2008}, and \citet{hernandez2014}. \citet{caballero2008} and \citet{kubiak2017} have also tried to characterize the population of stars towards Ori OB1b using only photometry. 

\citet{alves2012} and \citet{bouy2014} used photometric observations to argue for an extended foreground population towards the ONC. These findings were supported by \citet{pillitteri2013}. However those conclusions have been questioned by \citet{da-rio2016}, \citet{fang2017}, and \citet{kounkel2017a}.

The first large-scale RV studies with sub-\kms\ resolution towards ONC were conducted by \citet{furesz2008}, followed by work from \citet{tobin2009} and \citet{kounkel2016} to identify a peculiar blueshifted stellar population relative to the molecular gas distribution. 

\citet{da-rio2016,da-rio2017} conducted the first APOGEE survey of Orion A as an SDSS-III Ancillary Science program. Radial velocities from this program allowed the confirmation of new members and the identification of substructures in position-position-velocity space. Stellar parameters inferred from the spectra also provided age diagnostics that agreed well with other methods (e.g., HR diagram placement, disk excess, and association with extinction). The APOGEE Orion data was also used in additional studies of the structure and kinematics of the Orion region. \citet{hacar2016,hacar2018} studied  the distribution of stars and gas in the Orion  A  molecular  cloud  to  identify elongated  strings  that  could  have  formed as  a  result  of  global  cloud  collapse  into  individual  filaments.  \citet{stutz2016} used the APOGEE Orion data in a paper analyzing a mechanism where the regions velocity dispersion reflects the rate at which protostellar cores decouple from the large scale magnetic field.

A few studies of PMs in the region have also been conducted; however, the resulting measurements were either not precise enough to place strong conclusions on the dynamics in the Orion regions \citep{ucac4}, or very limited in scope \citep{dzib2017}. Extensive analysis has also been focused on the region's temporal structure, both in terms of deriving ages for the individual regions (see above), as well as searching for an age gradient within a given cluster \citep[e.g.,][]{beccari2017}. 

Recently, \citet{kounkel2017} used the Very Large Baseline Array to measure parallaxes of 27 non-thermally emitting young stellar objects (YSOs) and create a 3-dimensional model of the region, although this study was confined only to the Orion A and B molecular clouds. However, with the small sample size it was only possible to investigate the overall orientation of the clouds, but not the presence of any substructure along the line of sight. \citet{zari2017} recently used \textit{Gaia} DR1 data to analyze the distribution of young stars in the Orion Complex and identify overdensities corresponding to young clusters. However, the \textit{Gaia} DR1 data only contained 5D astrometric solutions only for the brightest stars, and the uncertainties in distances did not allow \citet{zari2017} to conclusively resolve the populations from each other.

In this paper we present an analysis of structure and star formation history in the extended Orion Complex as determined from stellar positions and kinematics (both PM and RV) from \textit{Gaia} DR2 5D astrometric solutions and APOGEE-2 near-infrared spectra. In Section \ref{sec:data} we discuss the data involved in this study. In Section \ref{sec:algorithm} we describe the hierarchical clustering algorithm used for the identification of stellar groups, including an assessment of its performance and limitations. In Section \ref{sec:analysis} we show the identified groups and discuss their properties. In Section \ref{sec:conclusions} we conclude our results.

\section{Data}\label{sec:data}

\subsection{APOGEE}

\subsubsection{Data products}

Near-infrared high resolution spectral observations of stars towards the Orion Complex were conducted with the Apache Point Observatory Galactic Evolution Experiment (APOGEE) spectrograph, mounted on the 2.5m Sloan Digital Sky Survey (SDSS) telescope \citep{gunn2006,blanton2017}. This instrument covers the spectral range of 1.51--1.7 $\mu$m with $R\sim22,500$ \citep{wilson2010,majewski2017}, and it is capable of observing up to 300 targets simultaneously in a field of view of 1.5$^\circ$. A total of 8991 unique targets have been observed towards the Orion Complex (4259 towards Orion B/Ori OB1ab region, 2991 towards Orion A molecular cloud, and 1741 towards $\lambda$ Ori), most of which were observed repeatedly over the course of 2-3 epochs. These targets were either observed as part of the SDSS-III APOGEE IN-SYNC survey \citep[with the targets selected primarily on the basis of the previous identification of stars as YSOs,][]{da-rio2016,da-rio2017}, or as part of the APOGEE-2 Young Cluster Survey 
\citep[with sources identified based on the observed IR excess, optical variability, or other indicators of membership, more details pertaining to the targeting of the survey are presented in ][]{cottle2018,zasowski2017}. Previously, 43\% of the data used in this work has been included as a part of SDSS DR14 \citep{abolfathi2018}, and 30\% in DR12 \citep{alam2015}.

In addition to the stars in Orion, in the discussion of the data we also include sources observed by APOGEE and APOGEE-2 towards other star forming regions and young clusters. We also analyze spectra from Kepler 21 and NGC 188 fields to test the stability of the stellar parameters we extract from the APOGEE spectra, as both fields have been targeted by APOGEE for more than 20 epochs. However, the exact properties of the targets and the stellar population in these regions are beyond the scope of this work, and we defer the discussion of the non-Orion sources to future papers.

Unfortunately, the gradient in RV remains for the sources with \teff$<$3000 K in the correlation, which offsets RVs by up to 3 \kms\ at 2400 K; however, this affects only a small fraction of the sources in Orion ($\sim$150 stars, which are typically hotter than 2600 K). A more comprehensive analysis of stellar RVs across all the star forming regions (particularly those that are nearby and for which the spectra for the cooler sources are available) would be needed in order model this gradient in the future.

\subsubsection{Stellar parameters}\label{sec:params}

All the spectra are originally processed by the APOGEE Stellar Parameter and Chemical Abundances \citep[ASPCAP,][]{garcia-perez2016} pipeline, providing various stellar properties, including the abundances \citep{holtzman2015}. However, ASPCAP is primarily optimized for giants; given that the evolved YSOs targeted towards Orion are primarily dwarfs, this could introduce potential biases or systematic errors in the derived parameters. It is also a known issue that the uncertainties of various quantities (e.g. RVs) may be significantly underestimated, or that there may be a correlation in RVs for sources with \teff$<$3000 K \citep{nidever2015}.

To account for these and other issues, we processed the spectra using the pipeline developed by \citet{cottaar2014}, which is better suited for the analysis of spectra of YSOs. It fits to each spectrum the effective temperature (\teff), surface gravity (\logg), radial velocity (RV), rotational velocity (\vsini), and veiling ($r_H$), using a synthetic grid and interpolating between the grid points. A Markov Chain Monte Carlo simulation (MCMC) is performed at the end of the fit to measure the uncertainties in all the parameters and validated against the scatter in each parameter across multiple epochs of spectroscopic observations. Average parameters are determined for each star as a weighted average of the parameters measured for each epoch, as presented in Table \ref{tab:apogee}.

\begin{figure}
  \centering
		\gridline{\fig{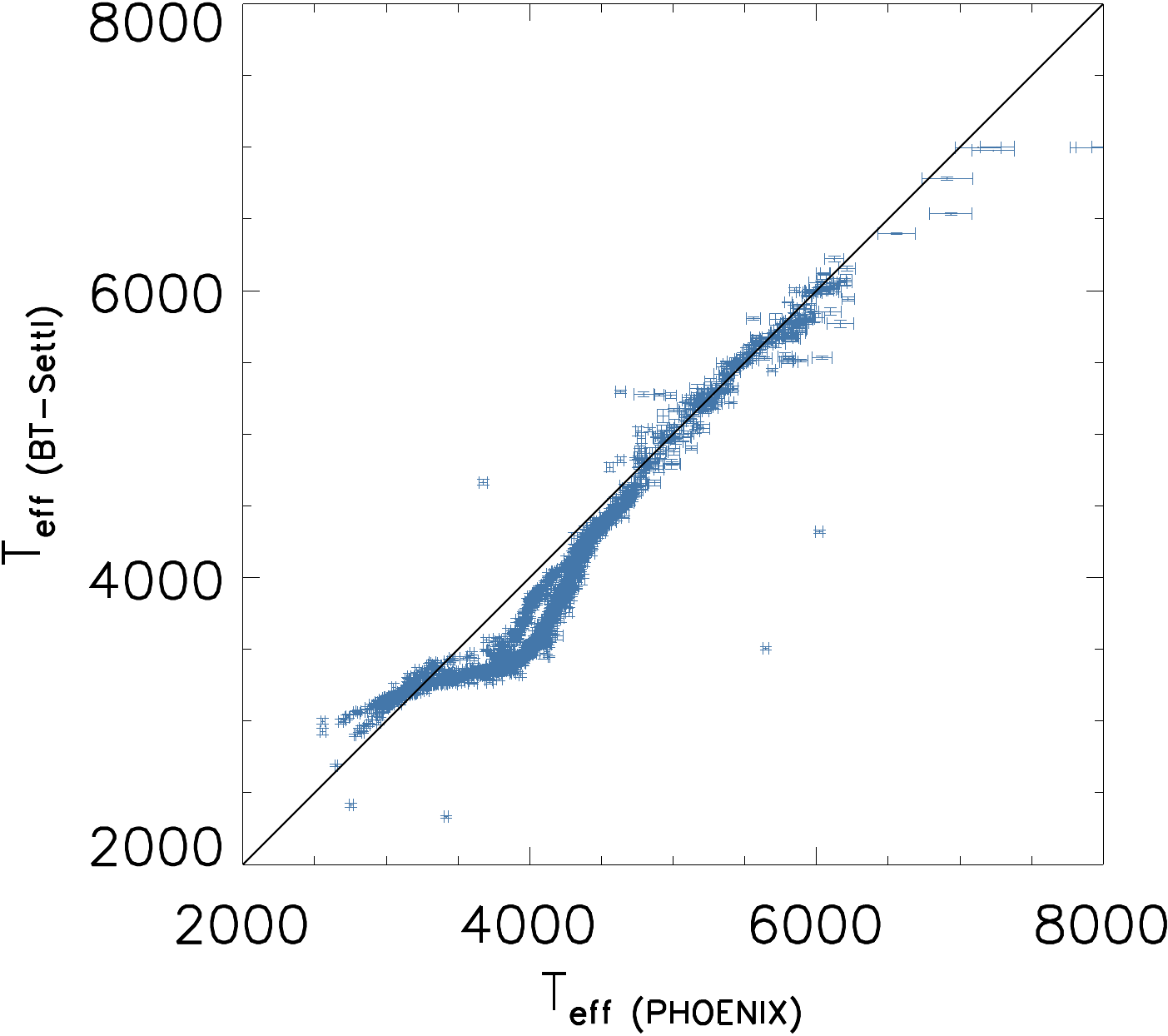}{0.25\textwidth}{}
              \fig{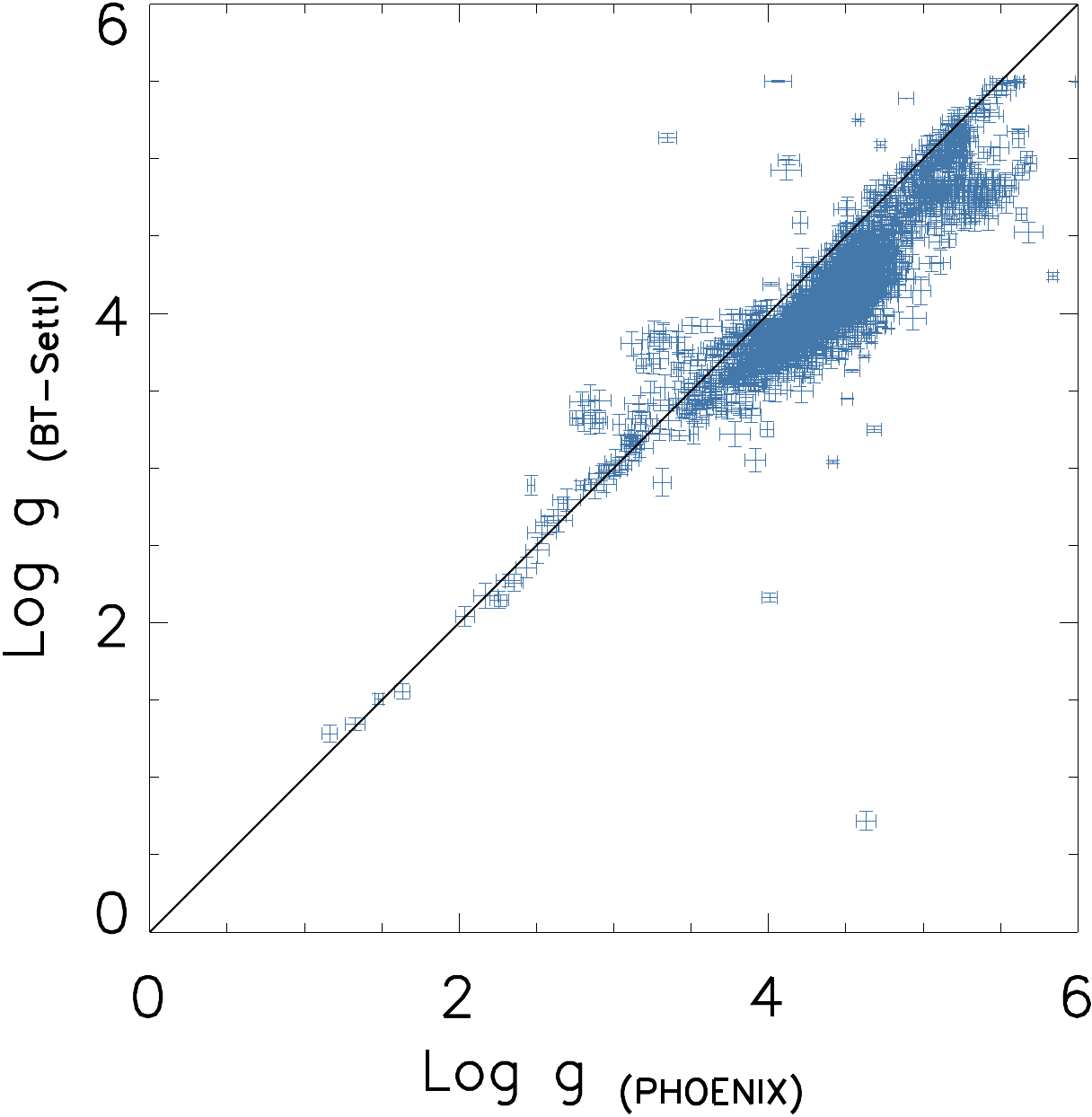}{0.22\textwidth}{}
        }
        \vspace{-0.8cm}
		\gridline{\fig{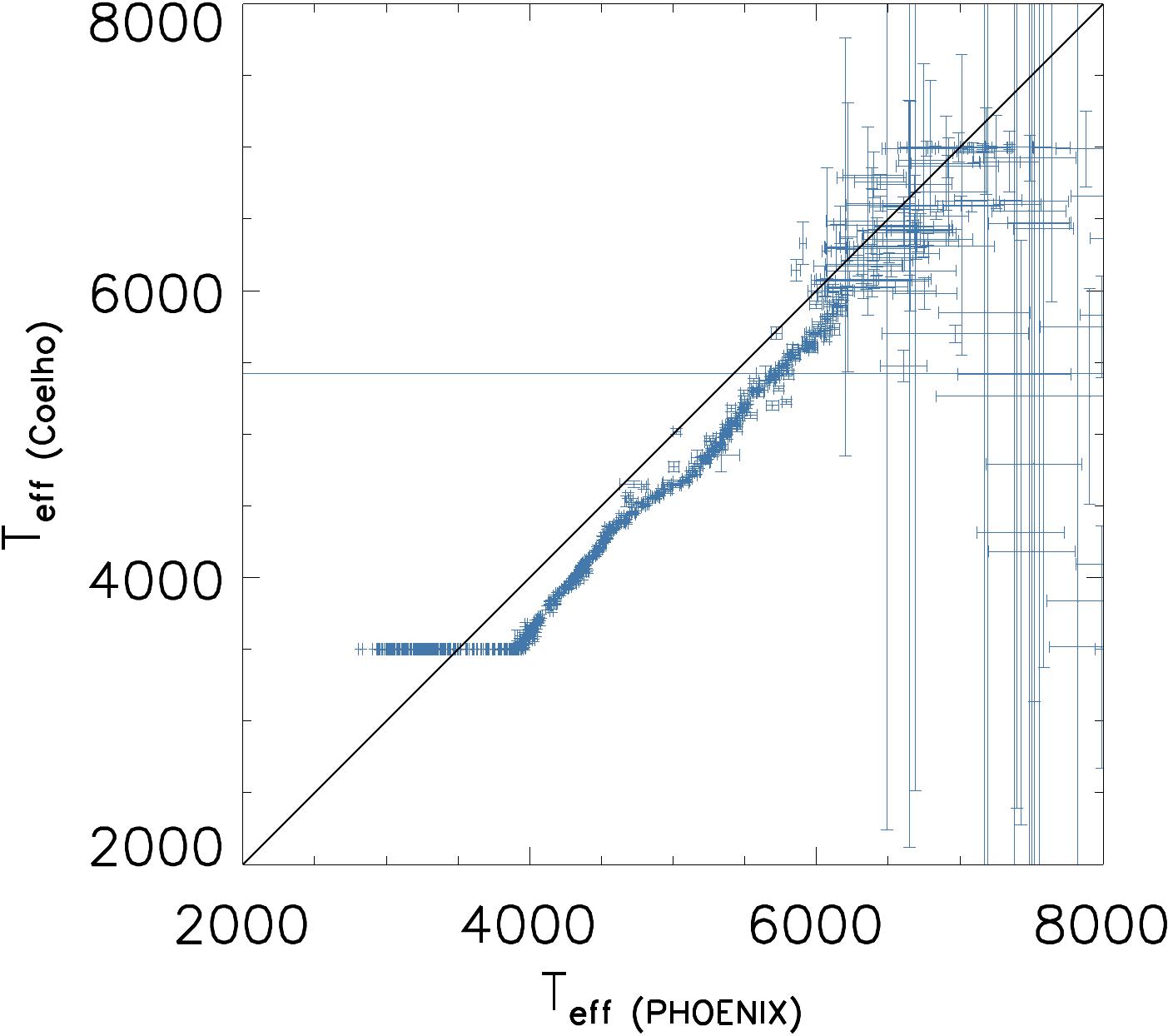}{0.25\textwidth}{}
              \fig{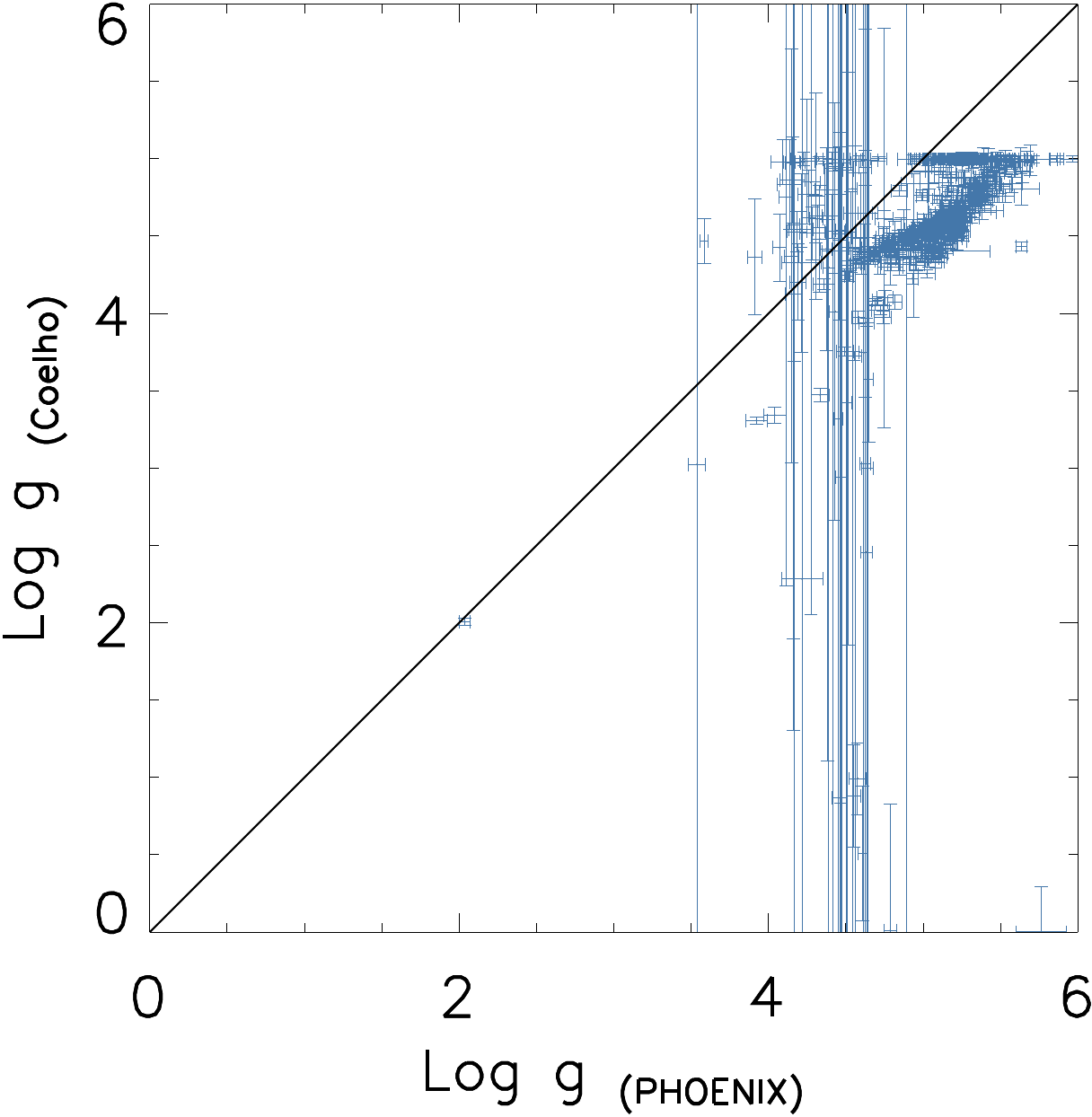}{0.22\textwidth}{}
        }
        \vspace{-0.8cm}
		\gridline{\fig{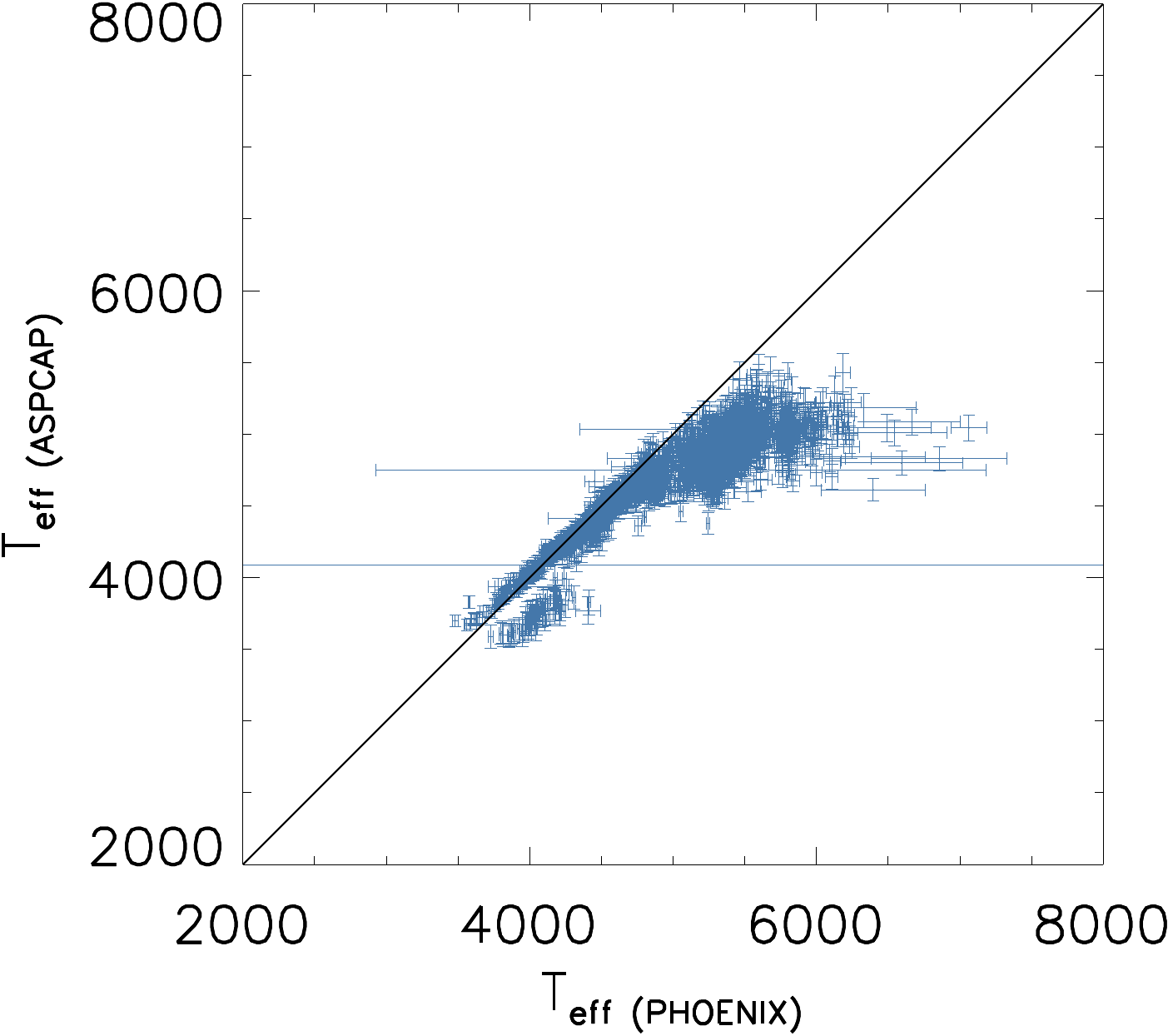}{0.25\textwidth}{}
              \fig{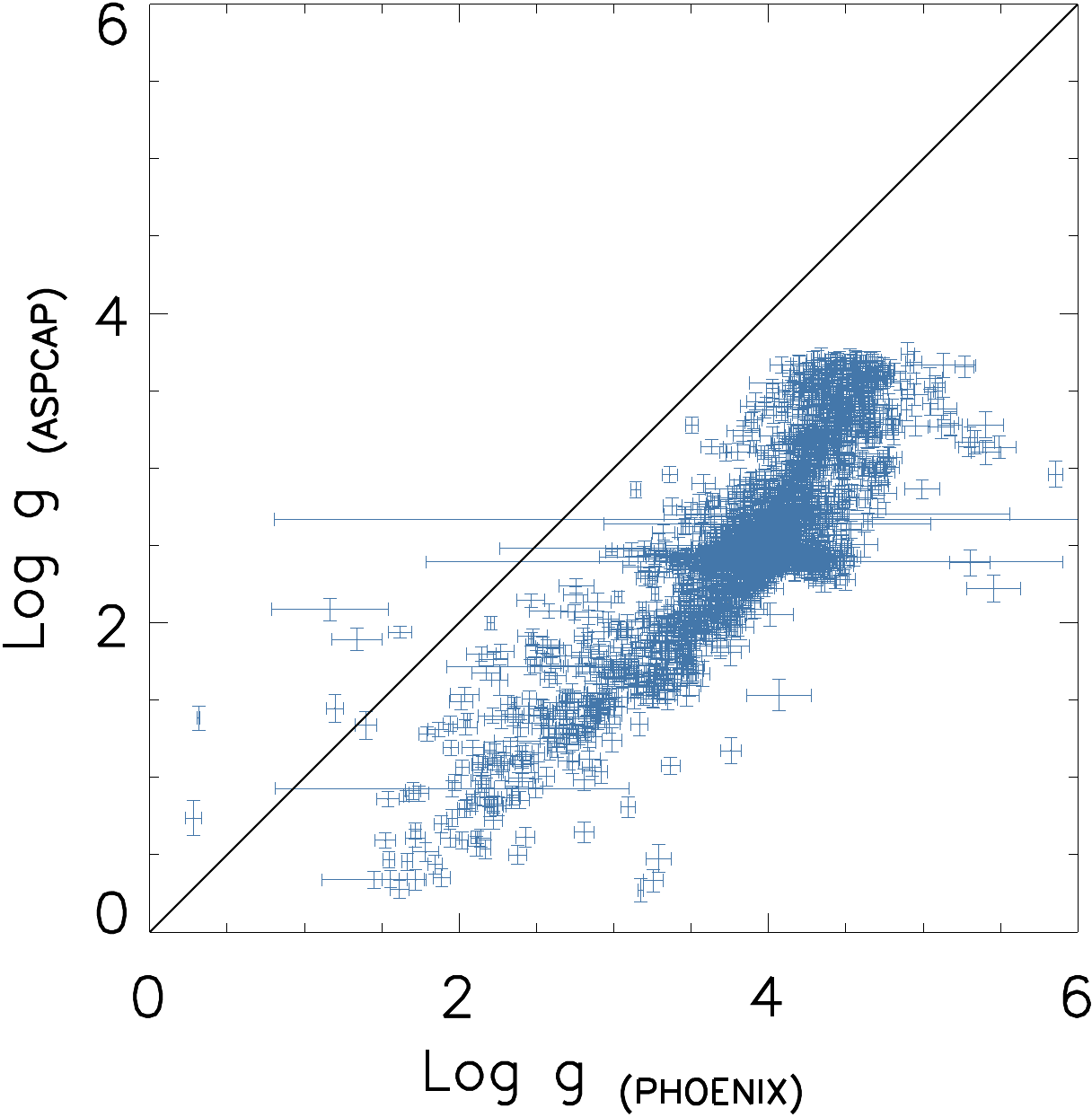}{0.22\textwidth}{}
        }
      \caption{Comparison of \teff\ and \logg\ between the correlations produced with various grids. First row shows comparison with the results from \citet{cottaar2014} and \citet{da-rio2016}, second row is restricted to sources in Pleiades (plateau is from the Coelho grid edges), bottom row includes sources that were observed by APOGEE towards the observed star forming regions that have both \teff\ and \logg\ produced in the ASPCAP catalog (which limits the sample primarily to the background red giants). \label{fig:ltcomp}}
\end{figure}

To test the performance of the pipeline, we processed the APOGEE spectra of the stars observed towards the Pleiades cluster with various synthetic grids, namely the Coelho \citep{coelho2005}, BT-Settl \citep{allard2012}, and PHOENIX \citep{husser2013} spectral libraries, all of which were restricted just to the solar metallicities. A comparison was also made to the parameters previously obtained for the sources in the original IN-SYNC survey \citep{cottaar2014}, which used the BT-Settl grid in spectral fitting. All the grids produced largely comparable solutions, although there were some systematic differences, most likely due to the underlying assumptions in the models (Figure \ref{fig:ltcomp}). The most notable difference is in the absolute value of \logg: while the correlation between all models is largely linear, there is an offset of 0.5 dex offset between the solutions produced with PHOENIX and BT-Settl grids, and a 1.5 dex between the PHOENIX grid and the ASPCAP catalog. There are some differences in the absolute calibration of \teff. The choice of grid has little to no effect to the determination of RV and \vsini. We ultimately adopted the PHOENIX grid for our spectral analysis, as it produced the most self-consistent solutions, and covers the widest range in \teff (2,300 to 15,000 K).

\begin{figure}
\epsscale{0.8}
\plotone{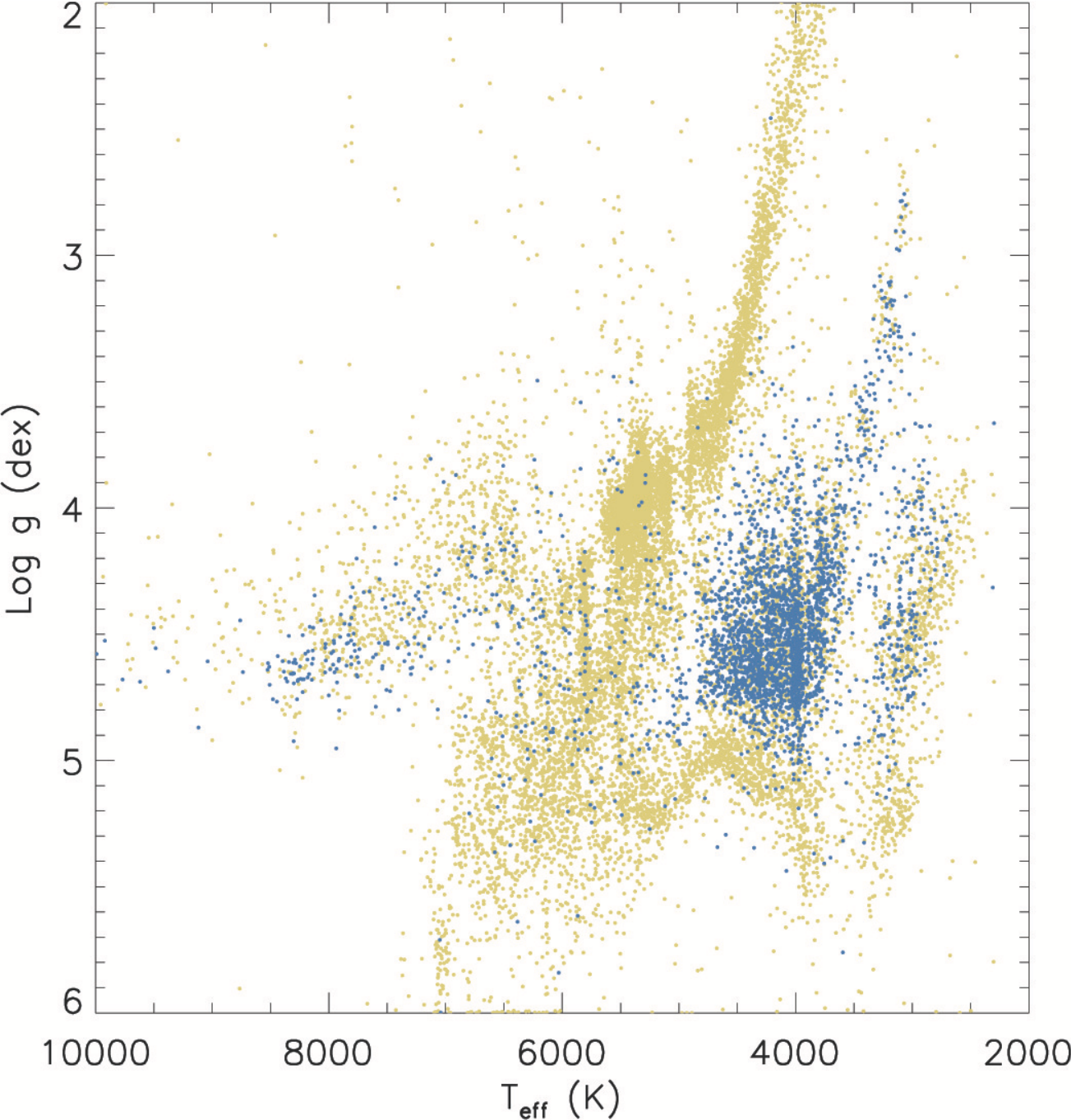}
\caption{Distribution of \teff\ and \logg\ for the sources correlated with the PHOENIX grid. Yellow circles show the entire sample observed by APOGEE towards all the young clusters (including sources in star forming regions other than Orion, as well as the contaminating field population); only sources towards the Orion Complex that pass the imposed membership cuts are shown in blue. \label{fig:teffloggall}}
\end{figure}

While the pipeline does try to interpolate between grid points, it is less successful in some regimes than others (mainly with respect to \teff), and particular values for the parameters are preferred to the exclusion of others. The most prominent zones of avoidance (nodding) for the PHOENIX grid occur at \teff$\sim$3900, 5000, 5200, and 5700 K (Figure \ref{fig:teffloggall}). A similar nodding effect is present in correlations with all grids; although it appears to be most frequent in the BT-Settl grid, and weakest in the Coelho spectral library. Correlations with the PHOENIX grid also show a peculiar absence of sources at \teff$\sim$3500 K, and to a lesser degree at \teff$\sim$5000 K, although this gap does not follow the grid edges. The BT-Settl grid does not appear to show this gap (resulting in a dip deviating from the line of equity in Figure \ref{fig:ltcomp}), and the presence or absence in the Coelho and ASPCAP grids are inconclusive due to the limited temperature range. Comparison with optical surveys \citep[e.g.,][]{fang2018} suggests that most likely this 3500 K gap is due to the PHOENIX model producing NIR spectra that is not entirely representative of the real stars in this temperature regime.

Veiling is a property that is only applicable to the stars that are accreting, and while a number of sources in the sample are disk-bearing accretors, a majority of them are not. However, when present, veiling will weaken spectral lines. Since the spectra were fit with only solar metallicity models, the pipeline uses veiling to address any discrepancies in the absolute line depths. At low \teff, the veiling parameter does systematically rise up to 0.3--0.5, even though most sources are expected to have veiling$\sim$0 (however, the accreting sources with high veiling can be identified above the systematic values). We correlated a small fraction of the stars ($\sim$200 sources in Orion) with a larger PHOENIX grid that does include the effects of metallicity. The systematic effect in veiling at low \teff\ for non-accretors is significantly reduced to only $\sim$0.1. The correspondence in \teff\ between models with and without metallicity variations is mostly 1-to-1 (although there is some scatter beyond 5000 K). Comparing the reduction where metallicity was and was not allowed to vary, the surface gravities are consistent for sources with \logg\ $>4$, but offsets grow for lower \logg. Below \logg$\sim3$, the parameters derived assuming solar metallicity produced \logg\ about 1 dex lower than those where metallicity is allowed to vary (primarily affecting the location of the red giant branch). The effects of nodding and the gaps described above remain unchanged.

The pipeline is capable of two modes of operation: fitting a single epoch of a given star at a time, or processing all the epochs to simultaneously fit a single \teff\ and \logg, and \vsini. However, with a large number of epochs, the latter mode produces increasingly unreliable solutions for the quantities that are allowed to vary between epochs, such as RV and veiling. Testing the pipeline on sources in the NGC 188 and Kepler 21 regions with 20+ epochs, almost all sources had RVs of several epochs with SNR$>$20 were artificially scattered to values far outside the mean RV, compared to the solutions produced in the single epoch fits. This effect was also present in sources with as few as 3 epochs. In extreme cases, an excessive noise from a single low SNR spectrum may affect the invariant parameters as well. For this reason the parameters presented in this paper were derived by fitting each visit spectrum individually, and then computing a weighted average of the visit-specific parameters, as the parameters from the individual visits have a much greater reliability. This is contrary to the approach taken in the IN-SYNC analyses, which adopted simultaneous fits that can result in spurious RVs. Because of this, identifications of spectroscopic binaries made on the basis of the original IN-SYNC multi-epoch RVs may therefore be unreliable.

\subsubsection{Uncertainties}

The formal uncertainties $\sigma$ in all the parameters are calculated by $\sigma=3\sigma_{MCMC}\sqrt{\chi^2_f}$, where $\sigma_{MCMC}$ is the variance in the MCMC output, and $\chi^2_f$ is the reduced \chisq\ of the best fit. These calculations are similar with the approach by \citet{cottaar2014}, and provides a good agreement to the scatter in the data. If multiple observations of the same source are available, weighted averages (and the weighted uncertainties in those averages) are computed for all parameters from spectra with SNR$>20$. It should be noted that these $\sigma$ incorporate only the uncertainties in the fit, and not the systematic differences between various models.

\begin{figure}
\epsscale{1.2}
\plotone{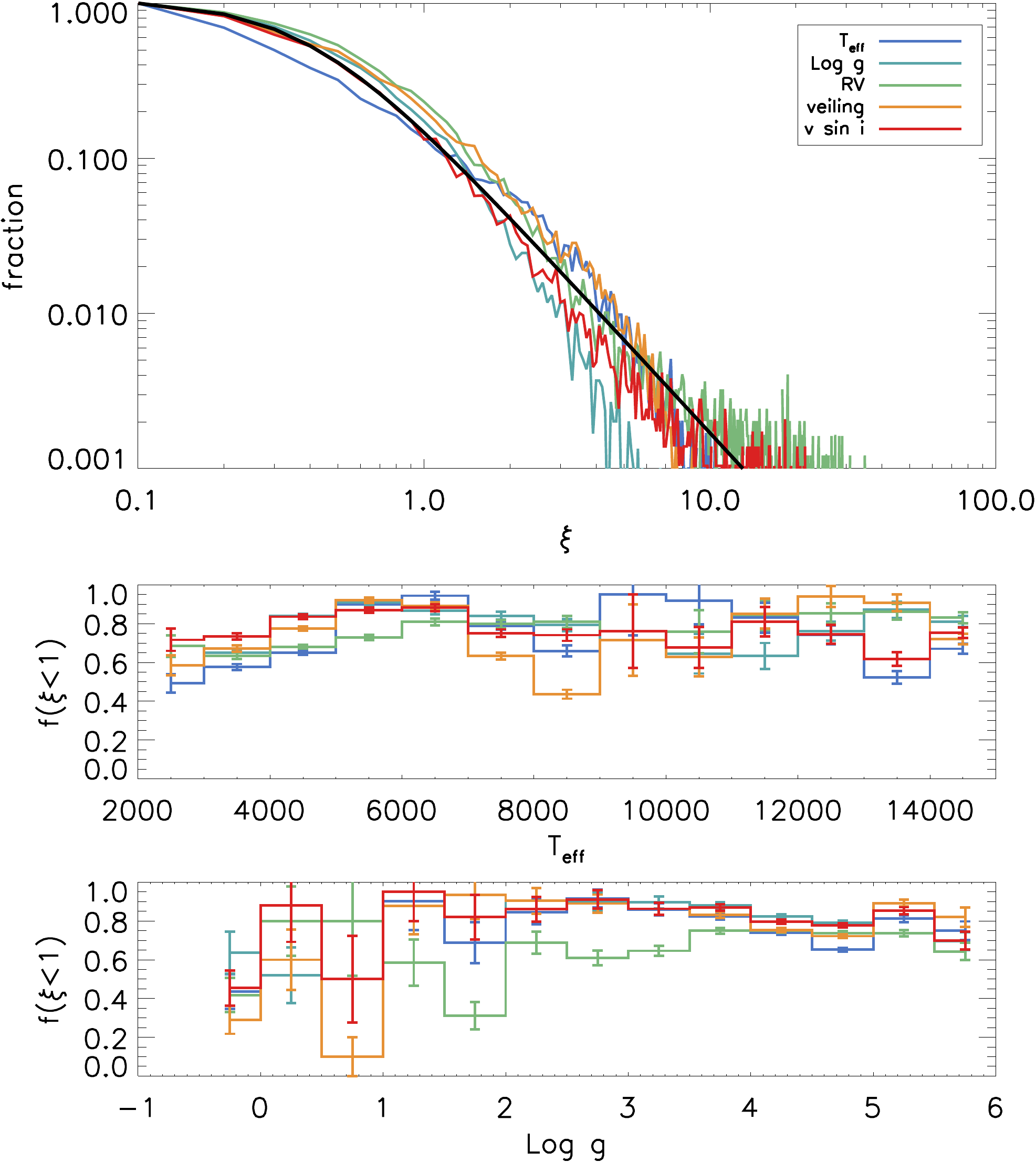}
\caption{Top panel: distribution of $\xi$ values in the entire sample for all parameters (normalized to 1 at $\xi=0.1$). Black line shows the Lorentzian distribution with $\gamma=0.4$. Middle panel: Fraction of sources with $\xi<1$ within \teff\ bins of 1000 K. Corrections for $\sigma$ at \teff$>$6250 K are applied. Bottom panel: Same as above, but for \logg\ bins of 0.5 dex. \label{fig:ksi}}
\end{figure}

For all sources with more than 4 epochs, we compute a degree of scatter $\xi=\Delta/\sigma$, where $\Delta$ is defined as the absolute difference between the weighted average of a parameter and the value of that parameter at a given epoch (Figure \ref{fig:ksi}). The distribution of $\xi$ can be approximated by a Lorentzian $1/(\gamma(1+\xi/\gamma)^2)$ with a width $\gamma=$0.4, and it is comparable for all parameters (the $\xi_{RV}$ distribution does have a long tail, however, largely due to a presence of spectroscopic binaries). It is also largely invariant from the number of epochs available, or from which cluster a source is originated. Approximately 80\% of the scatter is within 1$\sigma$ of the weighted average, $\sim$90\% is within 2$\sigma$, $\sim$95\% is within 3$\sigma$. The median single epoch uncertainties in all the parameters are $\bar\sigma_{RV}$=0.4 \kms , $\bar\sigma_{\log g}$=0.1 dex, $\bar\sigma_{T_{eff}}$=50 K, $\bar\sigma_{v\sin i}$=1.7 \kms, and $\bar\sigma_{veil}$=0.035. These uncertainties should be considered as a lower limit since they do not take into account potential added systematic offsets in the parameters.

There does not appear to be a significant variation in the distribution of $\xi$ of any parameters as a function of temperature or surface gravity at \teff$<$6250 K and \logg$>$1 (Figure \ref{fig:ksi}, although RVs of giants with \logg$<$3.5 may be affected by jitter). Since stars with high \teff\ have only a few spectral features that do not change significantly between different templates, $\sigma_{RV},\sigma_{v\sin i},\sigma_{veil},\sigma_{\log g}$ appear to be underestimated by a factor of 2, and $\sigma_{T_{eff}}$ by a factor of 8 in sources with \teff$>$6250 K. We correct the uncertainties by the corresponding factors. The uncertainties for parameters with \logg$<$1 are somewhat more difficult to correct due to a very low number of the affected sources; however, all of them are on the red giant branch for which more precise fits are available in the ASPCAP catalog.

\subsubsection{Stellar radius via spectral energy distributions}\label{sec:radius}

The observed APOGEE spectra have been fitted by the method discussed by \citet{stassun2016a} and \citet{stassun2017}, in which a star's angular radius, $\Theta$, can be determined empirically through the stellar bolometric flux, \fbol, and effective temperature, \teff, according to
\begin{equation}\label{eq:frt}
\Theta = ( F_{\rm bol} / \sigma_{\rm SB} T_{\rm eff}^4 )^{1/2},
\end{equation}
\noindent where $\sigma_{\rm SB}$ is the Stefan-Boltzmann constant.

\fbol\ is determined empirically 
by fitting stellar atmosphere models to the star's observed spectral energy distribution (SED), assembled from archival broadband photometry over as large a span of wavelength as possible, preferably from the ultraviolet to the mid-infrared. 
\teff\ is measured from the APOGEE spectra, and thus the determination of \fbol\ from the SED involves only the extinction ($A_V$), and an overall normalization as the two free parameters. 

We adopt the broadband photometry available in the literature spanning the wavelength range of approximately 0.15--22~$\mu$m. 
As demonstrated in \citet{stassun2017}, with this wavelength coverage for the constructed SEDs, the resulting \fbol\ are generally determined with an accuracy of a few percent. 
For stars with \teff\ uncertainties of $\lesssim$1\%, the \fbol\ uncertainty is dominated by the SED goodness-of-fit. With the exception of a few outliers, it was shown that one can achieve an uncertainty on \fbol\ of at most 6\% for $\chi_\nu^2 \le 10$, with 95\% of the sample having an \fbol\ uncertainty of less than 5\%. 

Here we apply this methodology to stars for which the estimated veiling in the $H$ band was less than 0.5, suggesting little to no active accretion. Most of the stars should therefore be diskless and the SED fit should be representative of the bare stellar photosphere. However, a number of the stars were found to exhibit significant excess emission in the WISE infrared bands, and in some cases in the 2MASS $K_S$ band as well. Therefore we ran the SED fitting procedure a second time on these stars, now excluding the 2MASS $K_S$ and the WISE photometry. Out of 8991 stars observed towards Orion, 6850 could be fit effectively as pure photospheres with the full wavelength range up to 22 $\mu$m; 298 had somewhat poorer goodness of fit, and 663 can be fit as photospheres up to H band, excluding longer wavelength photometry. The remaining 1180 sources could not be fit due to either high veiling or large infrared excess.

\subsubsection{Sample}

\begin{figure}
\epsscale{0.9}
\plotone{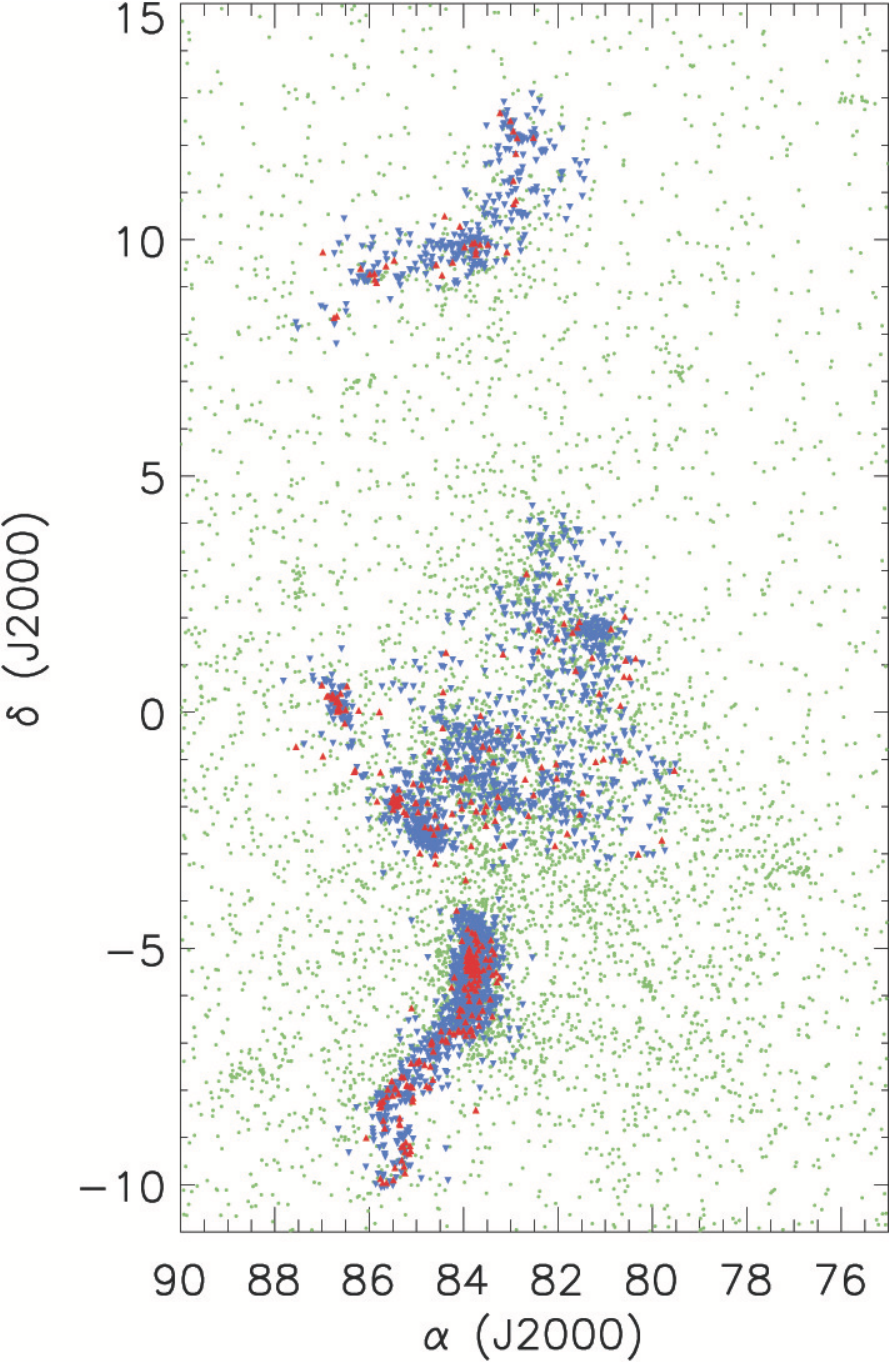}
\caption{Distribution of sources that satisfy the membership cuts towards the Orion Complex. The green symbols show sources that are uniquely detected by \textit{Gaia}, the red - those that are uniquely detected in the APOGEE footprint, and the blue are those that are common between both surveys. \label{fig:ysomap}}
\end{figure}

The sources observed towards the Orion Complex with APOGEE had substantial contamination from the foreground and background field stars. To minimize this contamination in the analysis throughout the paper we excluded sources with RV$<$ 10 \kms\ or RV$>$40 \kms, or those that fail the cuts in parallax, proper motions, or color imposed in Section \ref{sec:gaia}, if the data from \textit{Gaia} are available (Figure \ref{fig:ysomap}). 

\begin{splitdeluxetable*}{ccccccccBccccc}
\tabletypesize{\scriptsize}
\tablewidth{0pt}
\tablecaption{Properties derived from the APOGEE spectra towards the stars in Orion \label{tab:apogee}}
\tablehead{
\colhead{$\alpha$} &\colhead{$\delta$} &\colhead{2MASS ID} &\colhead{\textit{Gaia} DR2} & \colhead{$N_{epoch}$} &\colhead{\teff} &\colhead{\logg} &\colhead{$\overline{RV}$} &\colhead{\fbol}&\colhead{$A_V$}&\colhead{$\Theta$}&\colhead{Flag\tablenotemark{a}}&\colhead{Group}\\
\colhead{(J2000)} &\colhead{(J2000)} &\colhead{} &\colhead{ID} & \colhead{} &\colhead{(K)} &\colhead{(dex)} &\colhead{(\kms)} &\colhead{10$^{-10}$erg s$^{-1}$ cm$^{-1}$}&\colhead{mag}&\colhead{$\mu$as}&\colhead{}&\colhead{}
}
\colnumbers
\startdata
85.1387100 & -10.148064 & 2M05403328-1008530 &    3011876641301608832 &  1 &   4575$\pm$    46 &  3.360$\pm$0.113 &  -12.379$\pm$  0.390 &   2.37$\pm$ 0.10 &   2.40$\pm$ 0.25 & 20.13$\pm$ 0.59 & 0 & field\\
85.5587769 &  -9.938928 & 2M05421410-0956201 &    3011907015310374528 &  1 &   4645$\pm$    31 &  4.326$\pm$0.075 &   21.491$\pm$  0.528 &   6.51$\pm$ 0.43 &   4.05$\pm$ 0.20 & 32.39$\pm$ 1.15 & 2 & l1647-2     \\
85.8625565 &  -9.993770 & 2M05432701-0959375 &    3011892137543646080 &  1 &   5388$\pm$    68 &  4.710$\pm$0.092 &   20.863$\pm$  0.560 &   6.62$\pm$ 1.49 &   2.05$\pm$ 1.00 & 24.27$\pm$ 2.81 & 2 & l1647-2     \\
\enddata
\tablenotetext{}{Only a portion shown here. Full table is available in an electronic form.}
\tablenotetext{a}{0=photosphere, 1=photosphere with poor goodness of fit, 2=photosphere up to H band.}
\end{splitdeluxetable*}

\subsection{\textit{Gaia}}\label{sec:gaia}

\begin{deluxetable*}{ccccccc}
\tabletypesize{\scriptsize}
\tablewidth{0pt}
\tablecaption{Astrometric solutions from \textit{Gaia} utilized in the analysis \label{tab:gaia}}
\tablehead{
\colhead{$\alpha$} &\colhead{$\delta$}  &\colhead{\textit{Gaia} DR2} &\colhead{$\pi$} &\colhead{$\mu_\alpha$} &\colhead{$\mu_\delta$} &\colhead{Group} \\
\colhead{(J2000)} &\colhead{(J2000)} &\colhead{} &\colhead{mas} &\colhead{\masyr} &\colhead{\masyr} &\colhead{} 
}
\colnumbers
\startdata
85.2329562 & -10.639077 &    3010342680846510208 &    2.262$\pm$   0.093 &    0.768$\pm$   0.170 &   -0.601$\pm$   0.158 & l1647-3     \\
87.9103488 & -10.637480 &    3010909277228047744 &    2.385$\pm$   0.075 &   -0.358$\pm$   0.126 &    3.808$\pm$   0.111 & field\\
86.0242576 & -10.557282 &    3011793525094619008 &    2.109$\pm$   0.025 &    0.179$\pm$   0.045 &   -1.225$\pm$   0.043 & l1647-2     \\
\enddata
\tablenotetext{}{Only a portion shown here. Full table is available in an electronic form.}
\end{deluxetable*}

\begin{figure}
  \centering
		\gridline{\fig{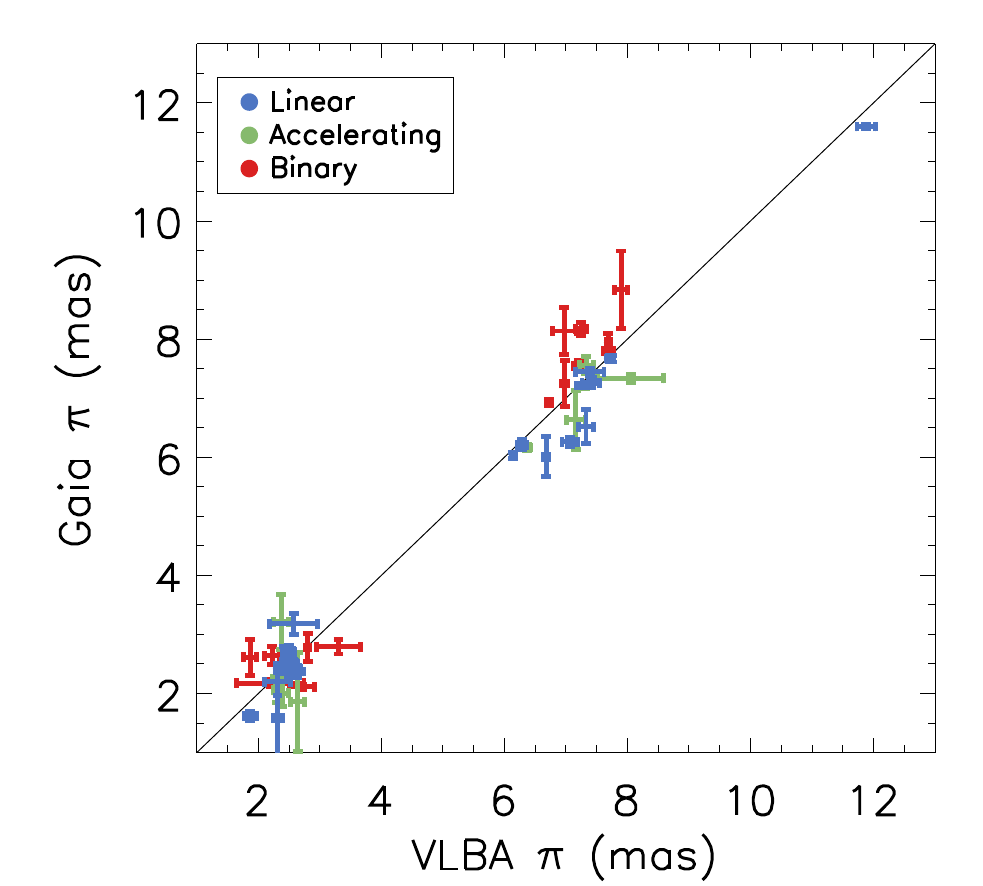}{0.25\textwidth}{}
              \fig{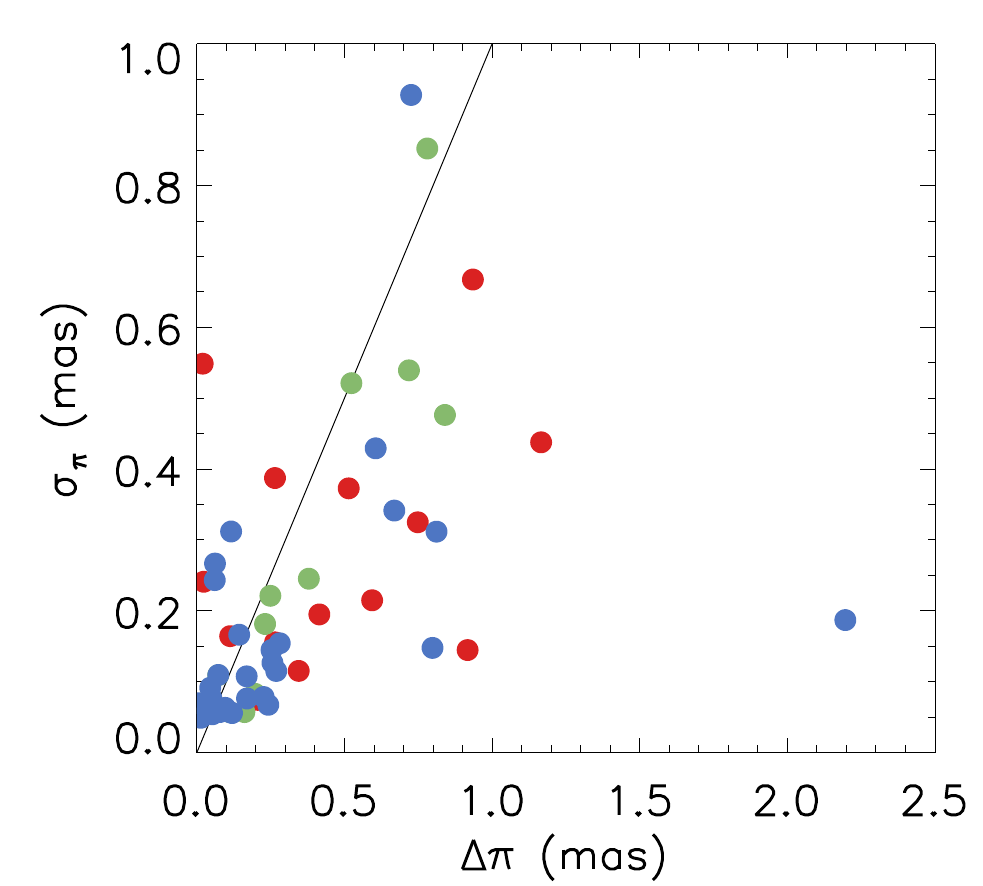}{0.25\textwidth}{}
        }
        \vspace{-0.8cm}
		\gridline{\fig{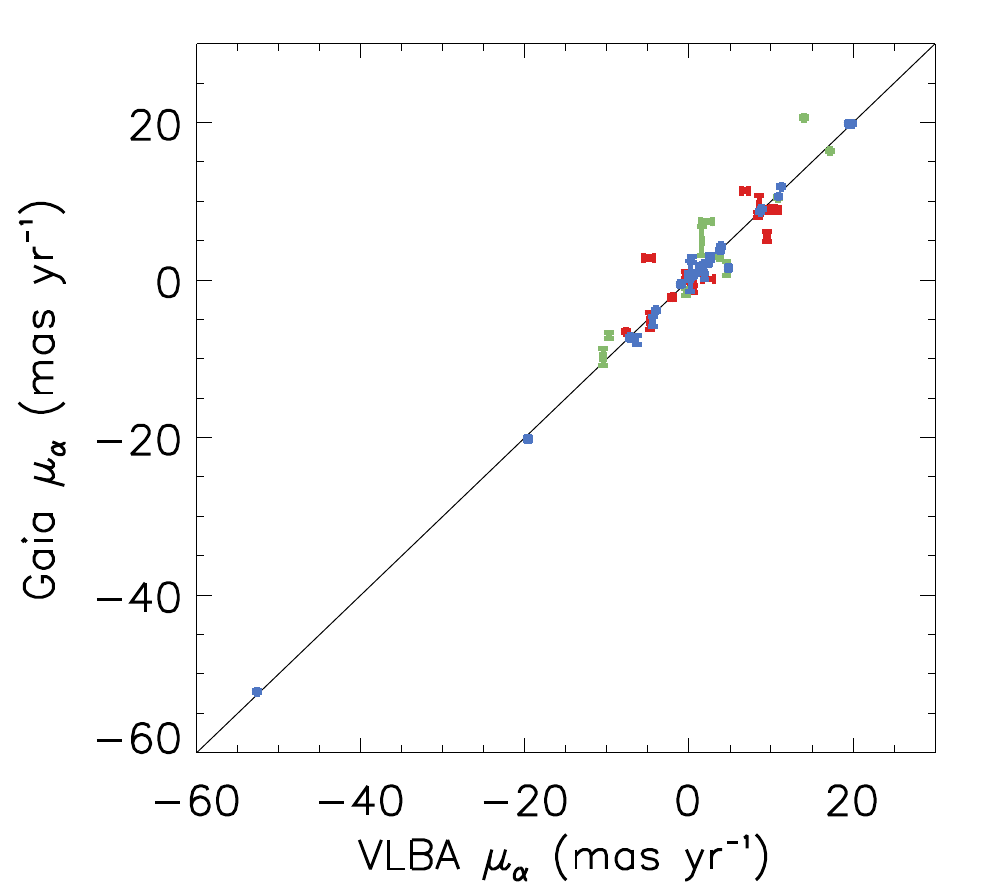}{0.25\textwidth}{}
              \fig{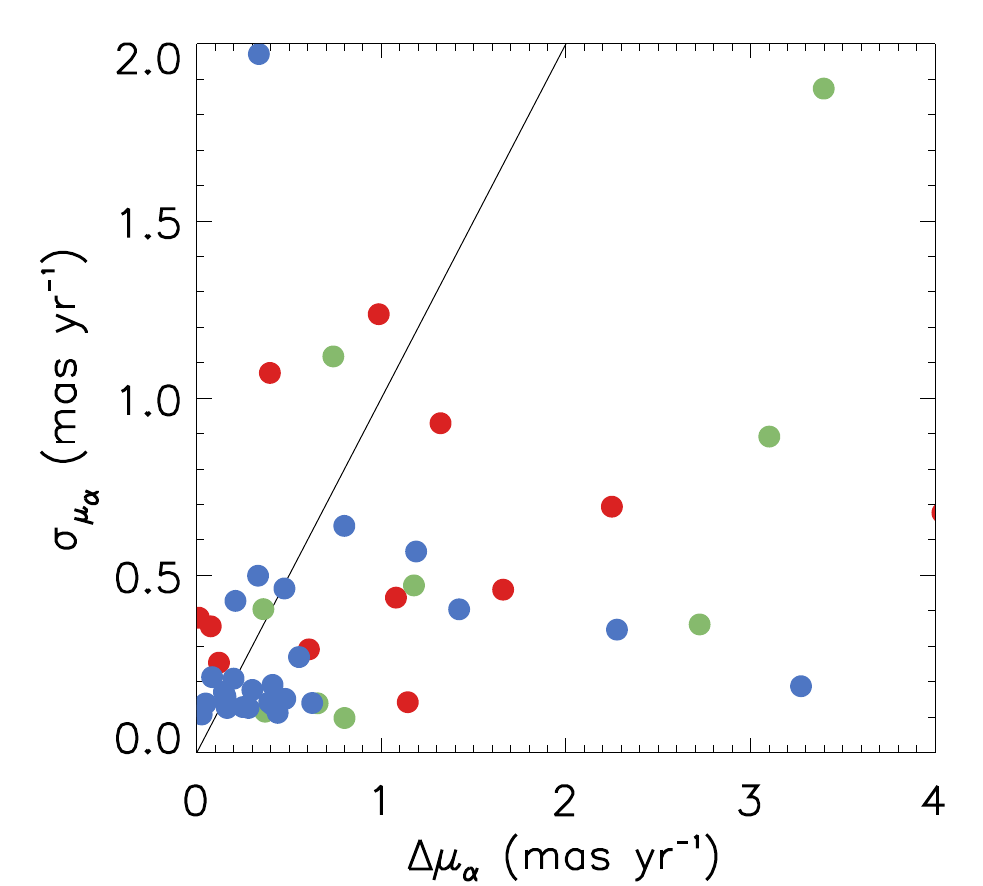}{0.25\textwidth}{}
        }
        \vspace{-0.8cm}
		\gridline{\fig{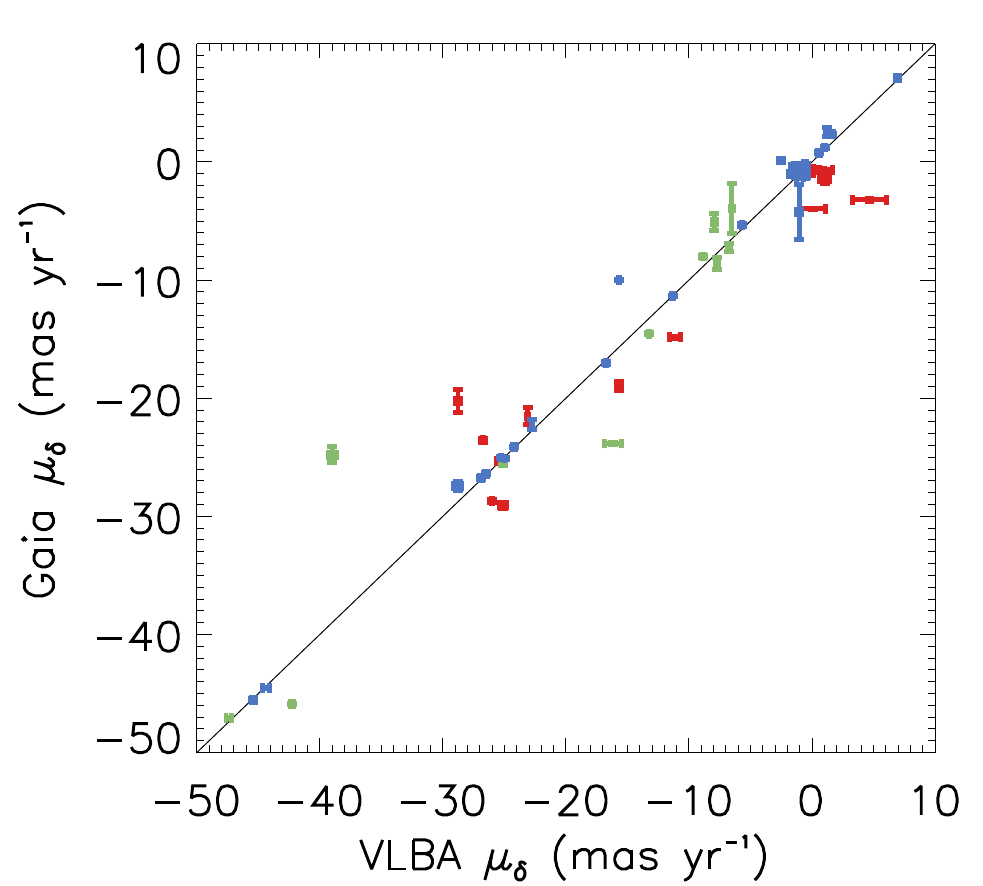}{0.25\textwidth}{}
              \fig{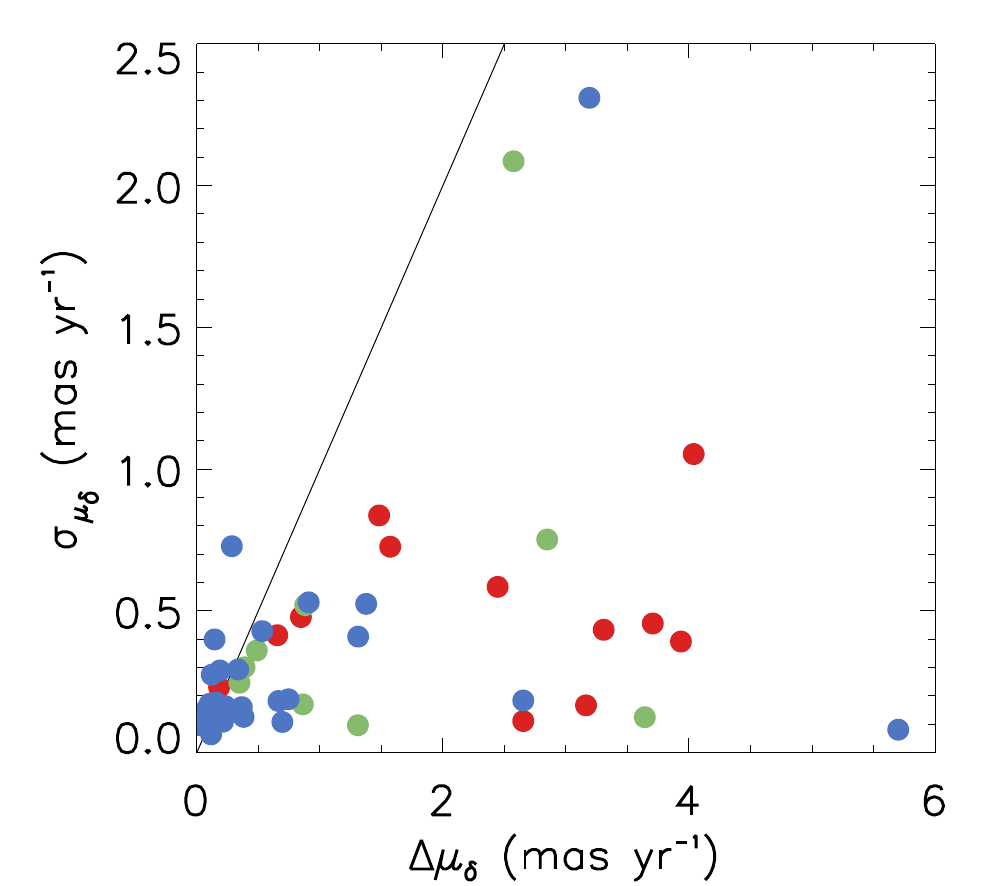}{0.25\textwidth}{}
        }
      \caption{Comparison of the astrometric solutions between \textit{Gaia} and VLBA observations. Blue colors show the sources that can did not show deviation from linearity in the VLBA observations (single stars, spectroscopic binaries, binaries with very long periods), green show the sources that were found to be accelerating over the course of the VLBA observations, and red are the astrometric binaries with periods up to $\sim$15 years. Right column shows the difference between two measurements vs their combined uncertainty. \label{fig:gaiacomp}}
\end{figure}

\textit{Gaia} has recently produced its second data release \citep{gaia-collaboration2018}, which contains astrometric solutions of parallax and PMs for 1.3 billion stars with $G<$21 mag, significantly improving upon the precision of previously available estimates of these parameters from \textit{Gaia} DR1 and Hipparcos. 

Previously, parallax and PMs have been measured for a number of young stars with the Very Long Baseline Array \citep{melis2014,ortiz-leon2017,kounkel2017,ortiz-leon2017a,dzib2018,galli2018}. The precision of these observations is comparable to that of \textit{Gaia} DR2. Currently, a direct comparison of the measured distances between two surveys can be performed for 55 stars.

In general, there does appear to be good agreement between the measurements for single stars, as well as spectroscopic and long period binaries (Figure \ref{fig:gaiacomp}), i.e. the sources that do not strongly deviate from an approximation of a linear proper motion. The astrometric solutions between the two surveys can be described with
\[\pi_{Gaia}=(0.9947\pm0.0066)\pi_{VLBA}-0.073\pm0.034\]
\[\mu_{\alpha_{Gaia}}=(0.9964\pm0.0015)\mu_{\alpha_{VLBA}}-0.030\pm0.073\]
\[\mu_{\delta_{Gaia}}=(0.9960\pm0.0020)\mu_{\delta_{VLBA}}-0.652\pm0.034\]

\noindent The systematic offset between parallax measurements is consistent with the zero point offset of -0.03 mas reported by \citet{lindegren2018}. The large offset in $\mu_\delta$ is driven by a few sources that may be affected by multiplicity that has not been apparent in the timeframe of the VLBA-only observations; this offset is consistent with 0 for the remaining sources. As \textit{Gaia} DR2 still does not have a prescription for astrometric binaries, the systems with a period of a few to a few dozen years present a considerable source of uncertainty that significantly increases the scatter in all parameters. In future data releases, this is expected to be improved.

\begin{figure}
\epsscale{0.8}
\plotone{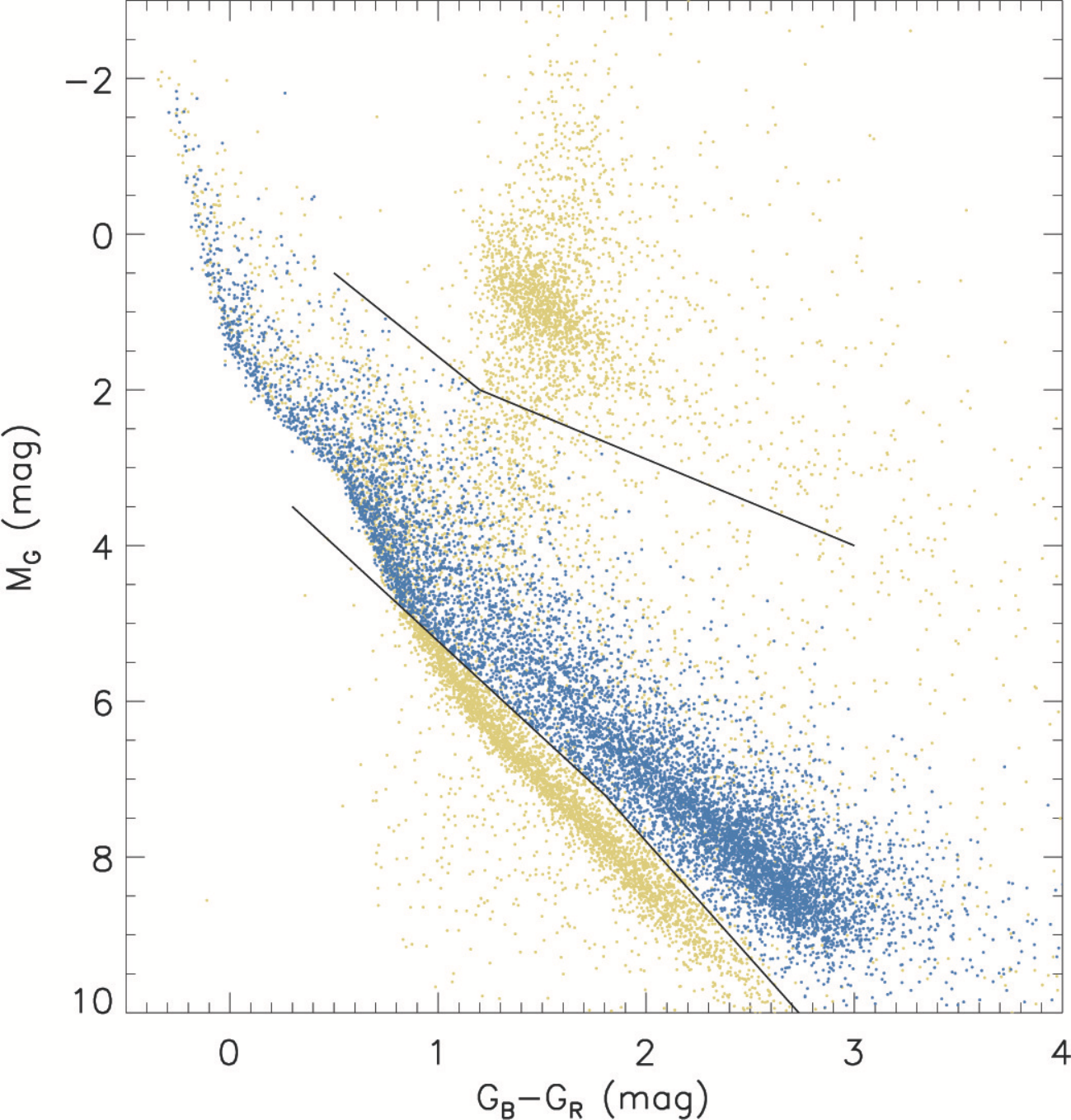}
\caption{Color-magnitude diagram for sources observed with APOGEE and those that satisfy positional and kinematical cuts in the \textit{Gaia} catalog without the APOGEE counterparts. The blue sources show those that pass the color cuts to identify the members of the Orion Complex, the yellow sources are the ones that have been rejected. \label{fig:gaiacolor}}
\end{figure}

To analyze the structure within the Orion Complex, we selected the sources in \textit{Gaia} DR2 with the position on the sky $75<\alpha<95^\circ$ and $-11<\delta<15^\circ$, parallax $2<\pi<5$ mas, and proper motions $-4<\mu_{\alpha}<4$ \masyr\ and $-4<\mu_{\alpha}<4$ \masyr. The ranges were chosen from the visual examination of the data to include all the overdensities that are associated with Orion. We excluded the astrometric measurements for the sources that had $\sigma_\pi>0.1$ mas or $\sigma_\mu>0.2$ mas in order to retain only the sources with precise measurements. Furthermore, discarded the sources that do not satisfy

\[M_G<2.46\times[G_B-G_R]+2.76; 0.3<[G_B-G_R]<1.8\]
\[M_G<2.8\times[G_B-G_R]+2.16; 1.8<[G_B-G_R]\]
\[M_G>2.14\times[G_B-G_R]-0.57; 0.5<[G_B-G_R]<1.2\]
\[M_G>1.11\times[G_B-G_R]+0.66; 1.2<[G_B-G_R]<3\]

\noindent where $M_G=G+5-\log(1000/\pi)$ in order to minimize the contamination from the main sequence and red giant stars (Figure \ref{fig:gaiacolor}).

Throughout the paper, all the calculations are done in the parallax space. Conversion to distances occurs only when we report the averages and to obtain the stellar ages. When converting from parallax to the distance space, we correct for 0.03 mas offset, and assume a systematic error of 0.02 mas \citep{lindegren2018}.

\subsection{Ages derivation}

We estimate the stellar ages using several different methods. One set of estimates is purely photometric. For YSOs without a reliable parallax measurement, we calculate $M_G$ by assigning the average distance of their corresponding group (See Section \ref{sec:analysis}). These calculations of $M_G$ ignore the effect of extinction: outside of the youngest regions that are still associated with the molecular gas, representing most of the footprint of the Orion Complex, the extinction is expected to be relatively small ($A_V\sim$0). We then interpolate $M_G$ and $[G_B-G_R]$ over the PARSEC-COLIBRI grid of isochrones \citep{marigo2017}. Despite the latest bandpass definitions, there appears to be a systematic offset between the data and the grid; to correct for it we offset grid's $M_G$ by +0.2 mag, and $[G_B-G_R]$ by +0.15 mag. We note that the absolute calibration of the resulting ages may be imprecise, but the relative ages across the complex should be largely consistent outside of the regions strongly affected by extinction. We refer to the ages derived with this technique as Age$_{CMD}$.

In addition to the above calculations, we also obtained ages for the sources observed with APOGEE, working with the \teff\ obtained from the spectra (see Section \ref{sec:params}) and \lbol. We work with the extinction values computed as described in Section \ref{sec:radius} as well as those estimated comparing the observed $G-J$ and $G_{BP}-G_{RP}$ colors from Gaia DR2 and 2MASS catalogs with the intrinsic colors obtained interpolating \teff\ in a modified version of Table A5 from \citet{kenyon1995} to allow for the new bandpasses, and then, transforming the color excesses to $A_V$ using the coefficients $C_G$=0.9145, $C_{G_{BP}}$=1.039 and $C_{G_{RP}}$ = 0.601, and a relation of $A_V=2.283((G_{BP}-G_{RP})-(G_{BP}-G_{RP})_o)$. With these extinctions and using the parallaxes from Gaia DR2, we estimated the absolute $M_G$ and $M_H$ magnitudes, which were converted to \lbol\ applying the bolometric corrections from \citet{kenyon1995}. To obtain the ages, we interpolated \lbol\ and \teff\ into the PARSEC-COLIBRI isochrones, as explained in \citet{suarez2017}. When possible, ages obtained from both bands were averaged and we refer to it as Age$_{HR}$. In contrast to Age$_{CMD}$, because of the dependence on \teff\ from APOGEE, Age$_{HR}$ values are not available for the entirety of stars that are identified as members of the Orion Complex. Where they are available, however, they are somewhat more reliable for the regions that are still associated with the molecular gas.

Finally, for the sources observed with APOGEE, we also interpolate the spectroscopically determined \teff\ and \logg\ values directly to estimate the ages. It is difficult to interpolate over these parameters with the PARSEC grid due to strong non-linear offsets between the grid and the data. But the relative distribution of ages for the various populations that we can infer from the spectroscopic parameters is consistent with the photometric distribution.

We do not report age measurements for the individual stars due to the multitude of the utilized methods and the scatter in the measurements, which we plan to study and minimize in future work, but we discuss the population averaged ages (both Age$_{CMD}$, and Age$_{HR}$) in Section \ref{sec:analysis}. In most cases, both are comparable, with the main difference originating from a somewhat different sample size.

\section{Clustering Algorithm} \label{sec:algorithm}

A number of techniques are available for identifying clustering in multi-dimensional data. Three main approaches that are commonly used are nonparametric hierarchical clustering and $k$-means, and parametric mixture modeling. The first method relies on computing a distance matrix between all sources and then grouping sources with a chosen algorithm together with a distance below a certain threshold. $k$-means requires an assumption regarding the number of groups present, and iteratively partitions the data until the centroid within each partition does not change significantly between steps. The mixture modeling fits a population as a collection of parametric distributions, usually Gaussian, and gives a probability to each source of belonging to a particular group based on this distribution. Here the number of clusters emerges as a parameter of the model.

A number of algorithms are available that utilize these techniques. Each one has its own advantages and limitations, pertaining to the ability to reject non-clustered field sources that are not part of any group (such as foreground or background stars), ability to separate populations that are closer to each other than their respective sizes, ability to identify structures of different sizes, shapes, and densities, and dependence on the initial assumptions pertaining to the structure within the data.A review of various methodologies is presented by \citet{everitt2011,feigelson2012}. We chose 'average linkage' hierarchical clustering that gives structures intermediate between elongated (single linkage) and compact (complete linkage).

A major issue that affects most of these algorithms is the difficulty to process sources with incomplete data (such as for sources that have RV but not $\mu/\pi$ and vice versa), making them impractical to use in this analysis. A common statistical treatment in these cases is to either exclude these sources from a clustering analysis entirely or to fill in the missing values with sensible estimates through imputation \citep[e.g.,][]{gelman2006}. While both techniques can work, it is possible for them to introduce strong biases \citep{wagstaff2005,mittag2013}.

In this section we describe a hierarchical clustering procedure for which we developed a prescription for the missing data. We use this code to identify distant structures in the positional and kinematical data provided by \textit{Gaia} and APOGEE towards the Orion Complex. We then test this code on a synthetic population of stars with approximate precision of the APOGEE and the \textit{Gaia} DR2 to determine how reliably it is able to recover clusters. Prior to the release of DR2, the code was also tested with \textit{Gaia} DR1 \citep{gaiadr1} and HSOY \citep{altmann2017}, although due to the significant improvement in the data quality and quantity of DR2, these tests are not described in the text.

Throughout this discussion, all the velocity components are used in the local standard of rest reference frame (lsr), and we will use $v_r$, $v_\alpha$, and $v_\delta$ to explicitly refer to RV$_{lsr}$, $\mu_{\alpha,lsr}$ and $\mu_{\delta,lsr}$. However, multiple definitions of lsr exist, and both are used in the text. One of them can be applied to all three dimensions of motion by first converting them to the galactic reference frame \citep{johnson1987,tycho}, subtracting the stellar motion \citep{schonrich2010}, and converting back to the equatorial reference frame. If one or two of the dimensions of motion is unavailable, they are set to zero for the purposes of the conversion, but they do not have any bearing on the remaining dimensions. The other definition of lsr is used to determine just $v_r$, from assuming the solar motion of 20 \kms\ towards $\alpha=18^h, \delta=30^\circ$ \citep[in B1900 reference frame,][]{gordon1976}. The former definition is used for the analysis of structure and kinematics in Orion, the latter is used for comparisons of the stellar motion to that of the molecular gas, as it is more commonly used in radio astronomy. In Orion, both definition produce $v_r$ comparable to within 1 \kms.

\subsection{Methodology}\label{sec:method}

Hierarchical clustering algorithms require one or more rules (or `constraints') to define where to `cut the tree' so that a small number of physically reasonable clusters are produced. These rules can be algorithmic, as with the choice of critical minimum cluster population in the widely-used single-linkage clustering procedure of \citet{gutermuth2009}. But the constraints can be based on disciplinary knowledge external to the dataset under study. We choose to set four constraints: maximum cluster size, maximum velocity range, an outlier rejection we call `reach', and the minimum number of star. We set the values of these constraints based on a detailed simulation study described in the Section \ref{sec:test}. In statistical parlance, this is an example of `constrained' or `semi-supervised' clustering \citep{basu2008}.  

We identify structures using the 6-dimensional data ($\alpha$,$\delta$,$\pi$,$v_\alpha$,$v_\delta$,$v_r$), first standardizing it by scaling each dimension using the standard deviation of the values within it to avoid a mixture of units between positional and kinematical components and to give all dimensions an equal weight. Then we compute the Euclidean pairwise distance matrix ($L_2$ norm) between all the sources using the IDL routine \texttt{distance\_measure}. Groups that are formed with this metric are invariant to rotations and translations in the p-space \citep{duda01}.

Given that astrometric and spectroscopic catalogs do not have a complete agreement between the sources that are included, APOGEE-only sources have only 3 dimensions of data ($\alpha$,$\delta$,$v_r$), and \textit{Gaia}-only sources have 5 dimensions ($\alpha$,$\delta$,$\pi$,$v_\alpha$,$v_\delta$). 6d distances involving these sources become undefined. To incorporate all the available data, we compute distance matrices on all three permutations (3D, 5D, and 6D) of the catalogs separately. All three matrices have the same size to simplify their merge later on even though certain elements of each matrix may not be defined in cases where data are incomplete. Nonetheless, even though all variables have been normalized, the measure of distance in each matrix are not compatible: those that were measured from the data with the larger number of the available dimensions are systematically larger (e.g. by $\sqrt{5/3}$). Additionally, the difference in source selections may further bias the measured distances. To correct for this each matrix is then normalized by the $2n^{th}$ smallest element in the matrix, where $n$ is the number of sources that have complete data in a given number of dimensions. This is an approximation of the median linking length of a constructed dendrogram for each permutation. The distance matrices are merged into single distance matrix so that a unified clustering calculation can be performed, keeping the value from the permutation with the largest number of available dimensions for each pair of stars.

Then, a modified version of the IDL routine \texttt{cluster\_tree} is used to identify structures in the data using the computed distance matrix, using the pairwise average linkage. Note that average linkage gives more compact and reliable clusters than the more commonly used single linkage (`friends-of-friends') algorithm \citep{everitt2011}. By default, it continues to run until every single source in a catalog is joined into a single group. To minimize excessive merging of unrelated structures or contamination from field stars, we imposed constraints that prevent a star from joining into a group if (1) the resulting group would be bigger than 4$^\circ$ in diameter ($\sim20-30$ pc at the relevant distance range), (2) the absolute velocity difference (in any of the three dimensions) is bigger than 9 \kms, and (3) after a group has begun growing, it cannot accept any new stars if the linking length necessary to do so relative to the median linking length binding all stars to a cluster is greater than the threshold of 1.2 (cluster reach). At the end of the process, each star has a unique group assignment, and groups with fewer than 10 members are rejected. All these parameters are chosen empirically through the testing of performance on a synthetic data set, which is described below.

\subsection{Synthetic tests}\label{sec:test}

\begin{figure}
\epsscale{1}
\plottwo{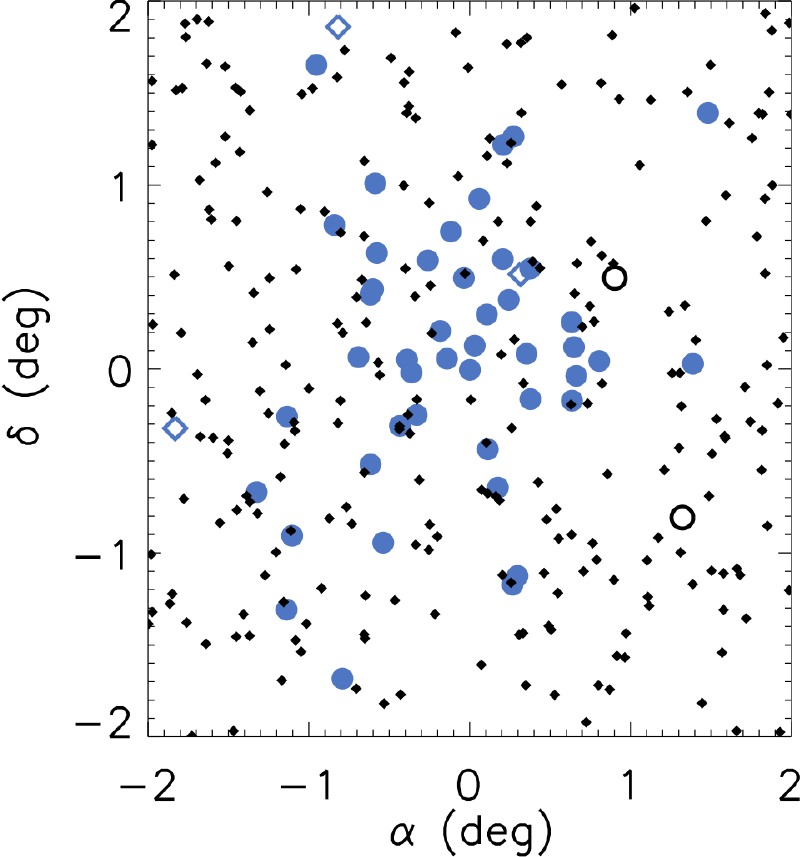}{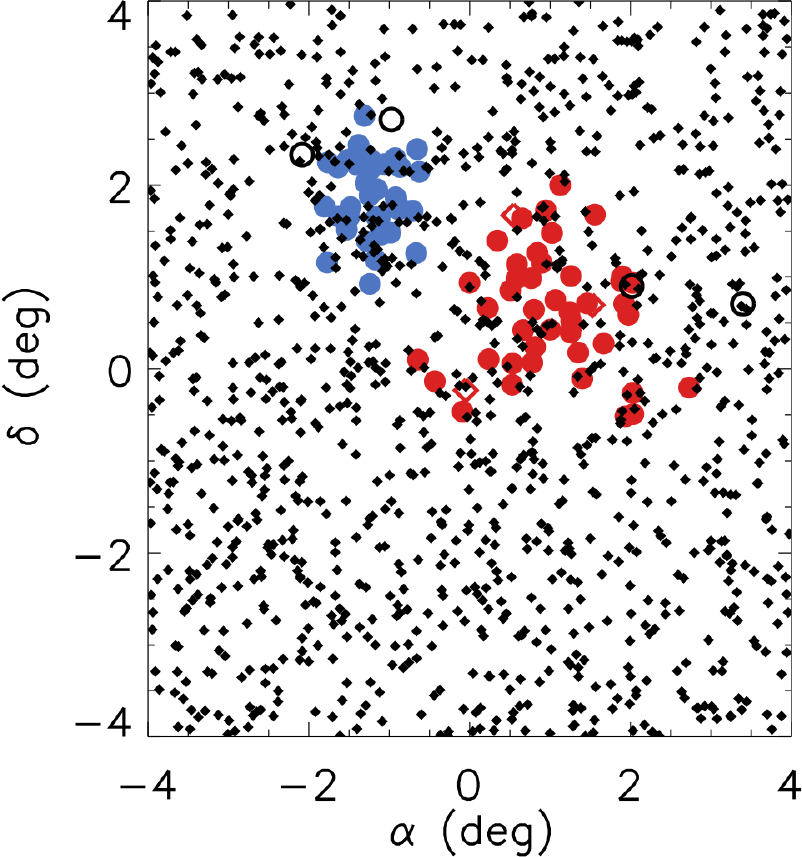}
\caption{Examples of the generated synthetic population (that remains after all the cuts) that is described in the Appendix \ref{sec:synthetic}), both single and multi-cluster runs. Circles show the cluster members, diamonds are the field population. Filled symbols correspond to the souces for which membership was accurately determined; empty symbols to either false positives (empty diamonds) or false negatives (empty circles). Colors are used to distinguish between multiple clusters in the same population. Only projection onto the sky is shown.\label{fig:exsynt}}
\end{figure}

\begin{figure*}
  \centering
		\gridline{\fig{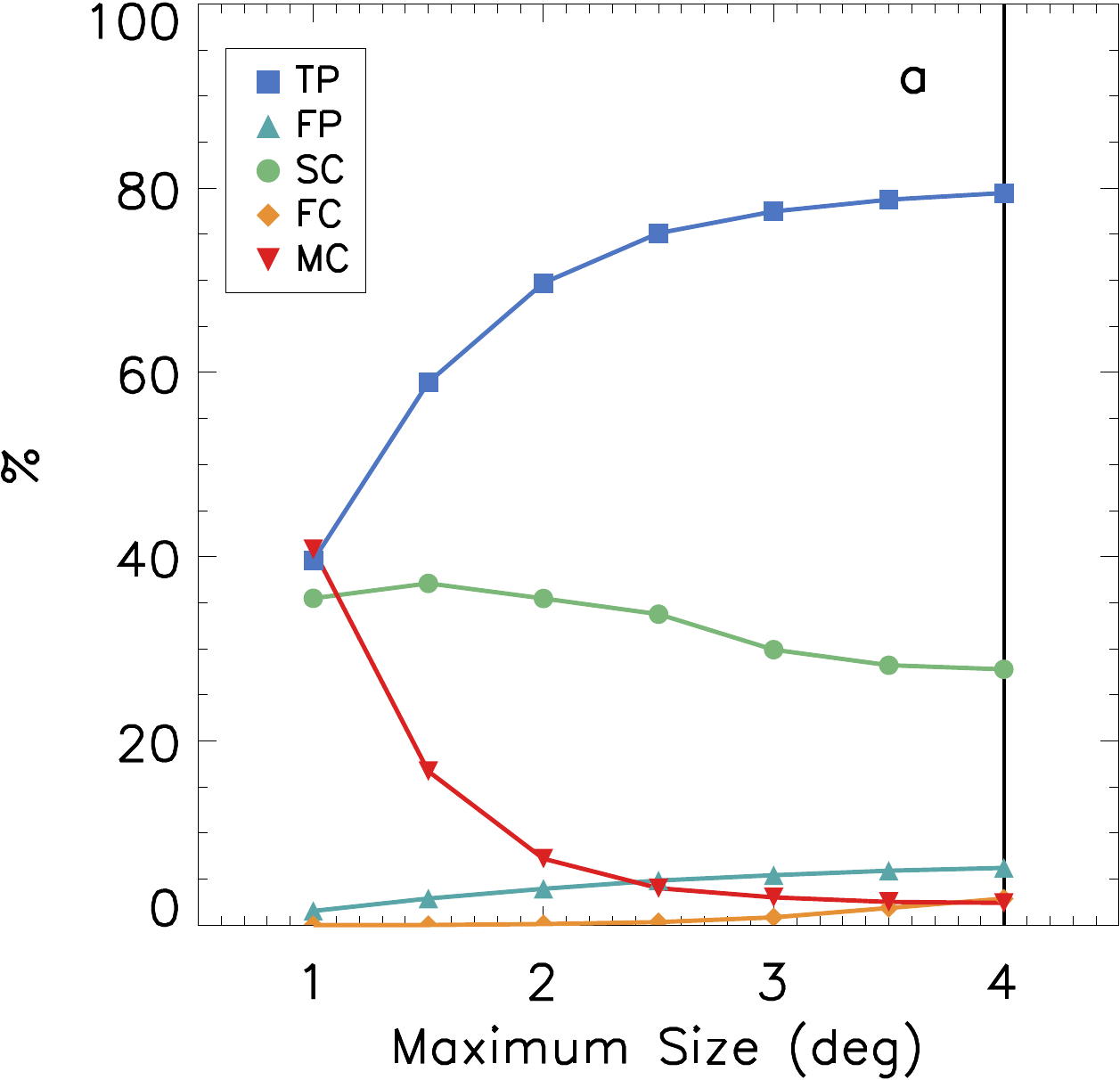}{0.33\textwidth}{}
              \fig{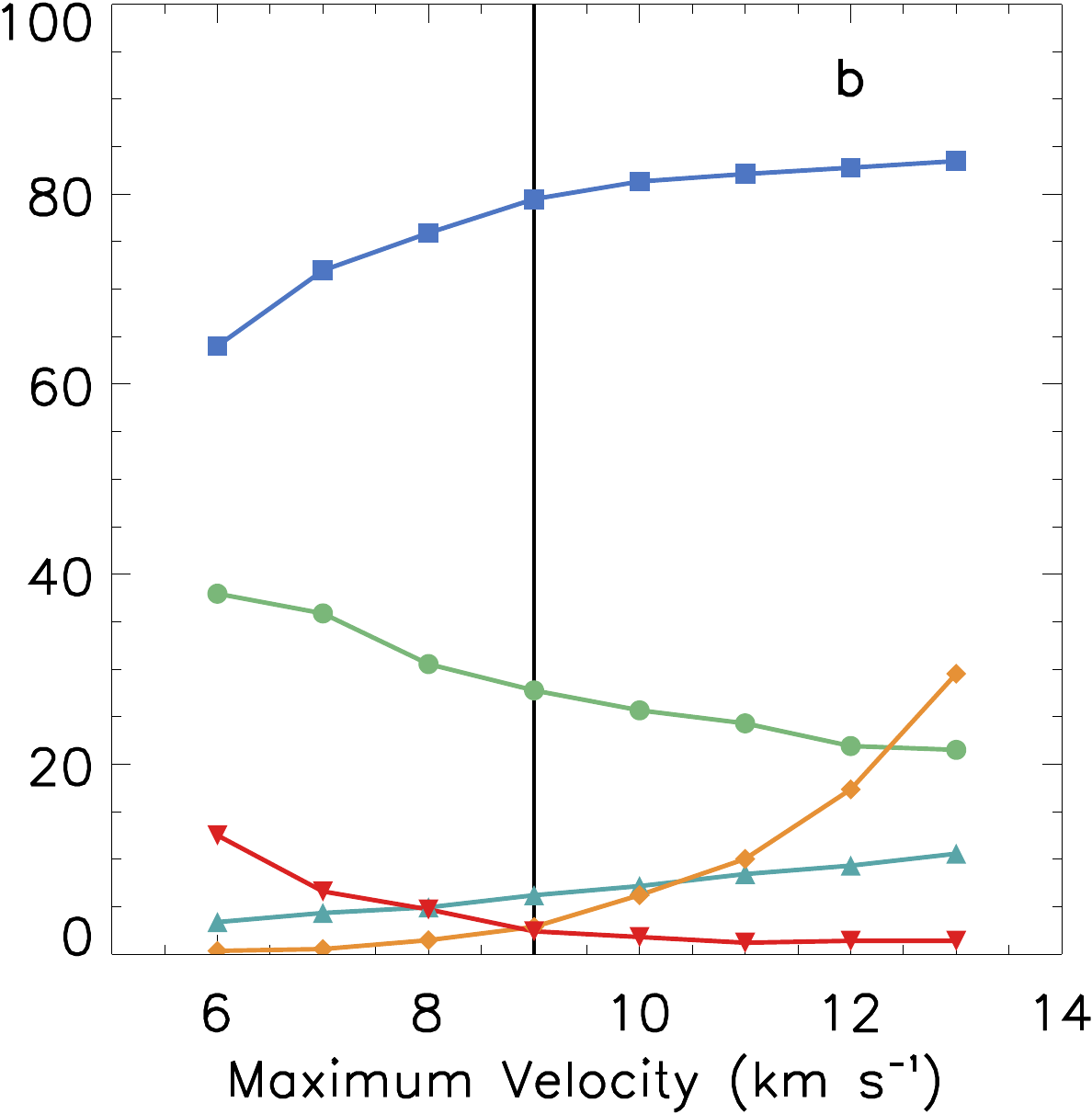}{0.312\textwidth}{}
              \fig{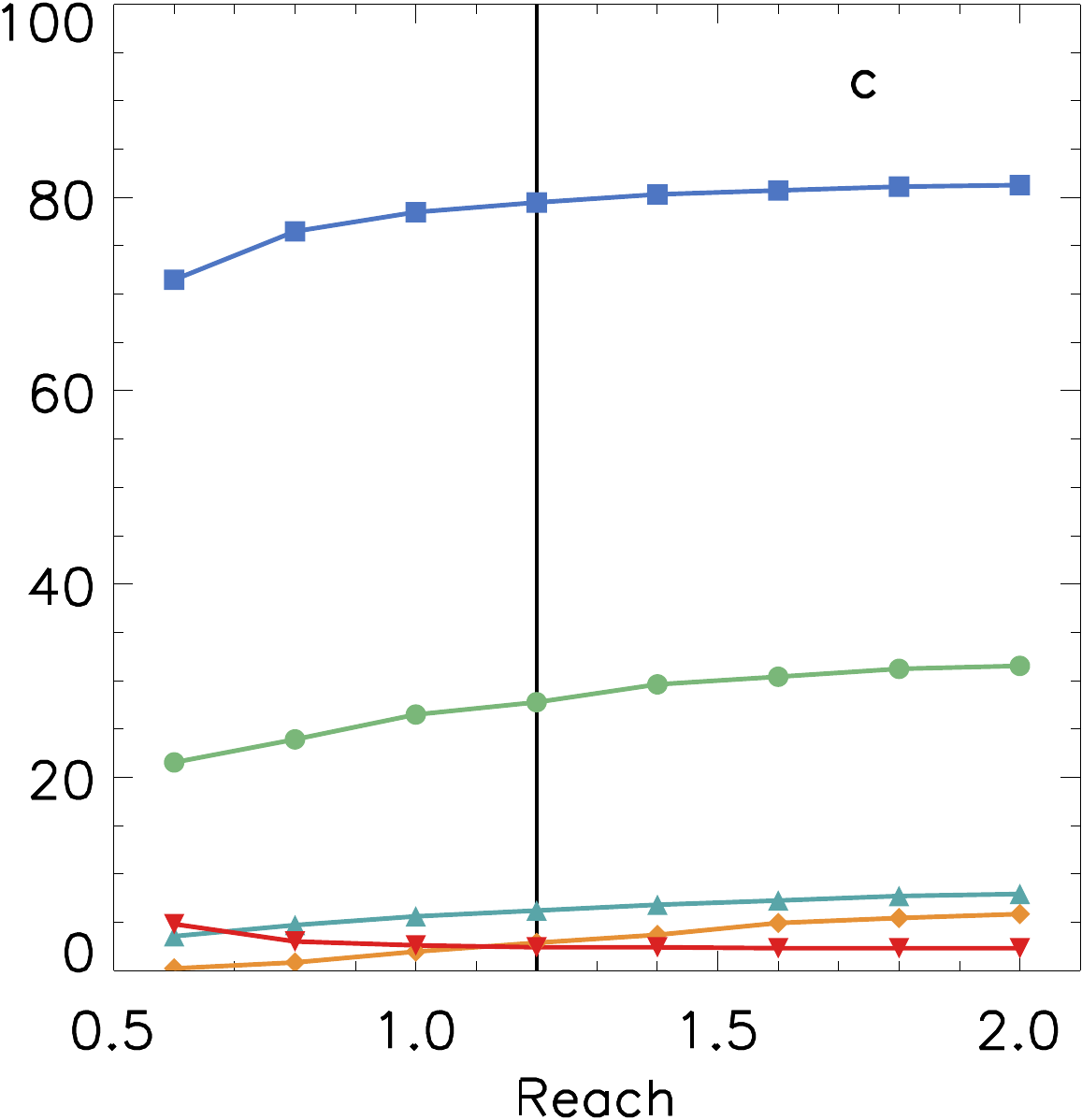}{0.305\textwidth}{}
        }\vspace{-0.5cm}
        \gridline{\fig{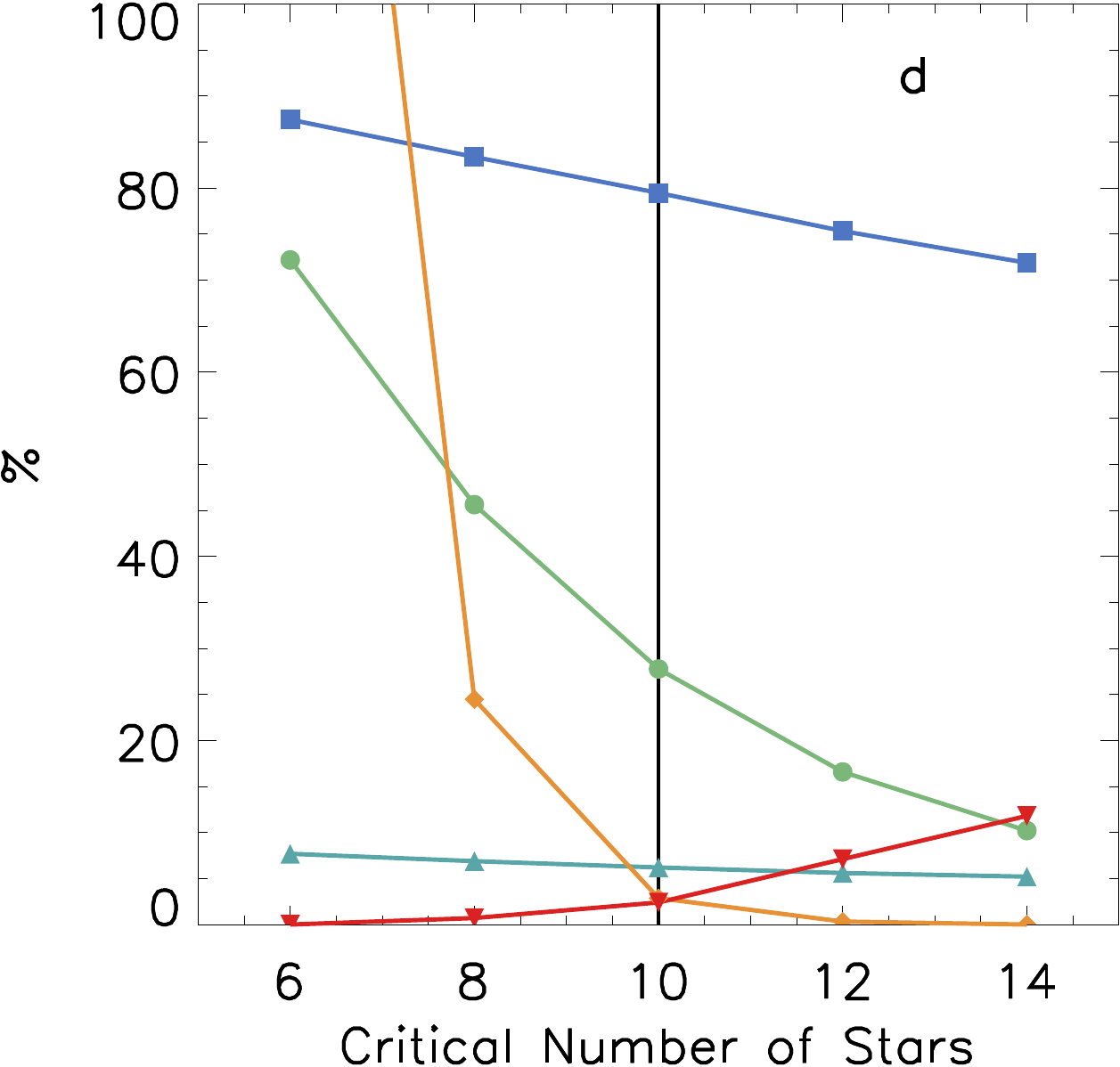}{0.33\textwidth}{}
              \fig{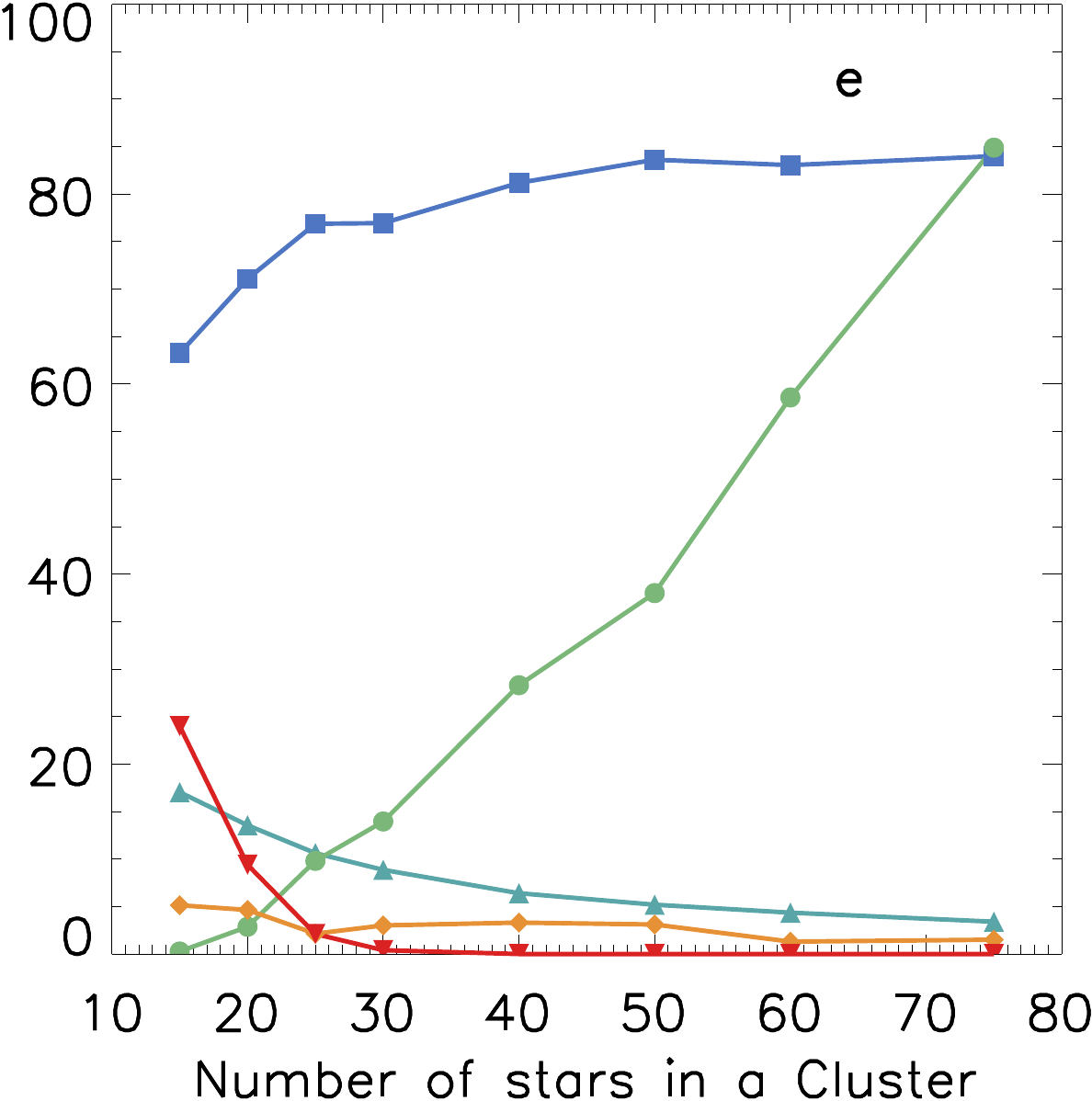}{0.312\textwidth}{}
              \fig{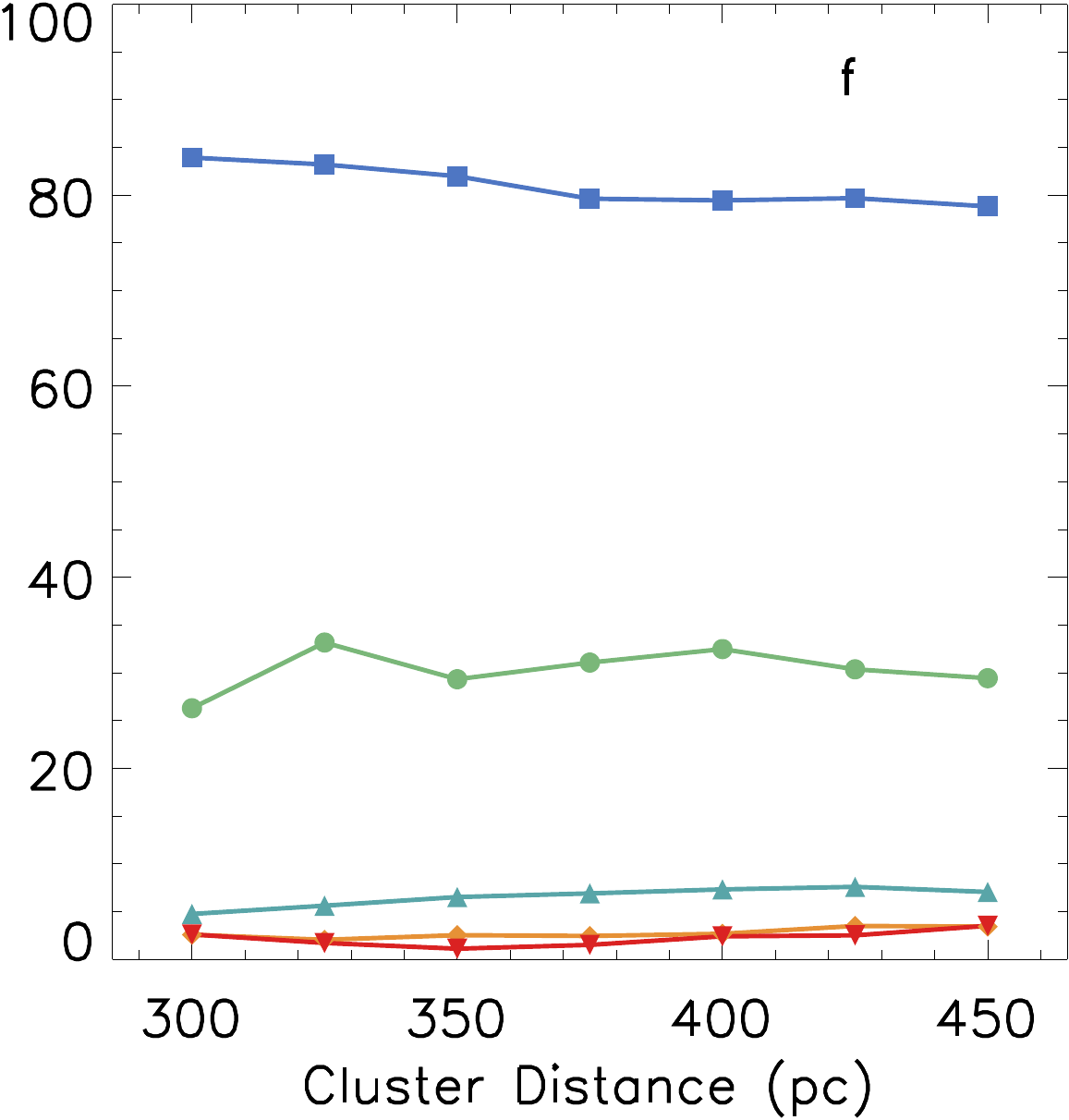}{0.30\textwidth}{}
        }\vspace{-0.5cm}
        \gridline{\fig{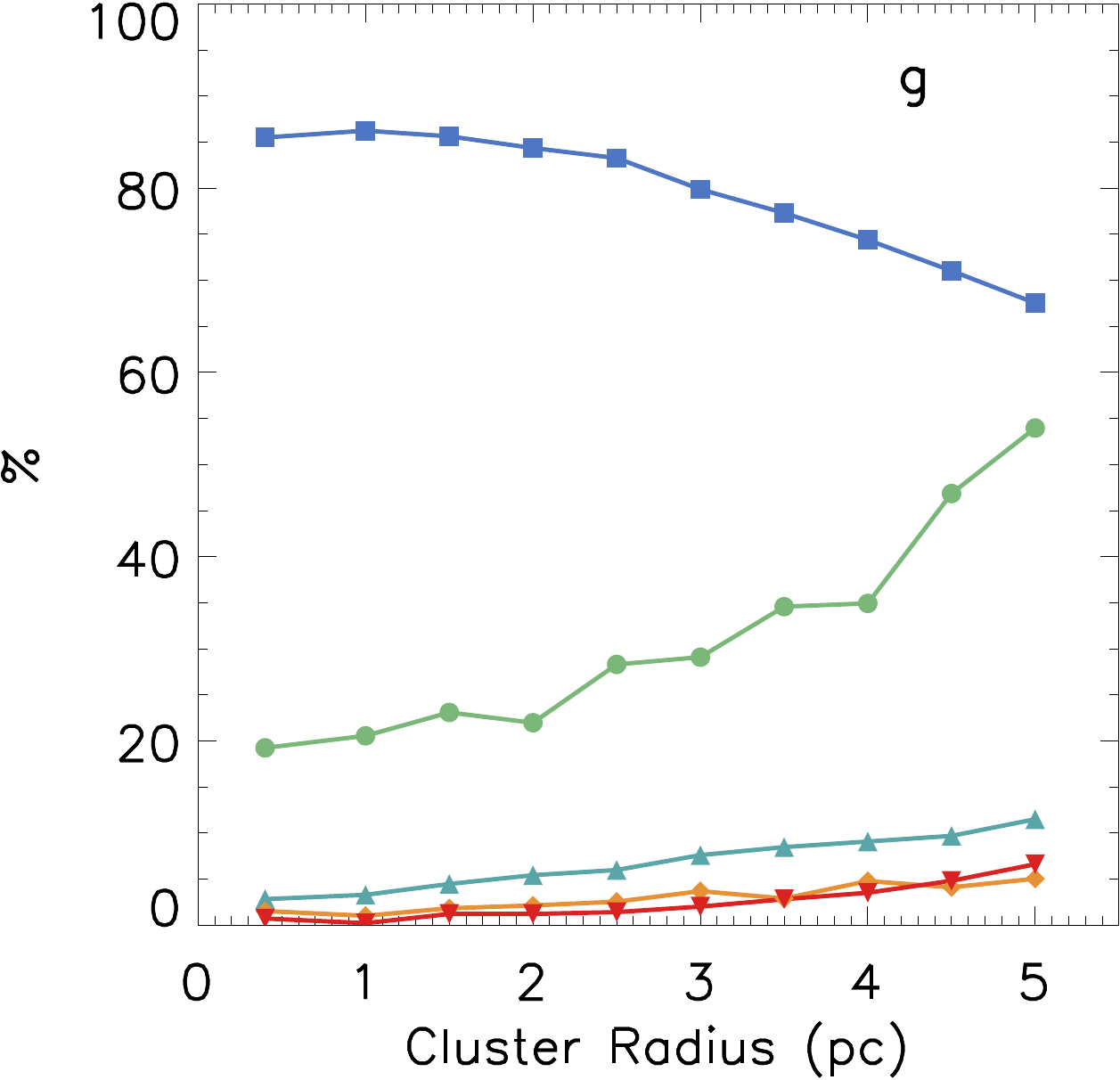}{0.33\textwidth}{}
              \fig{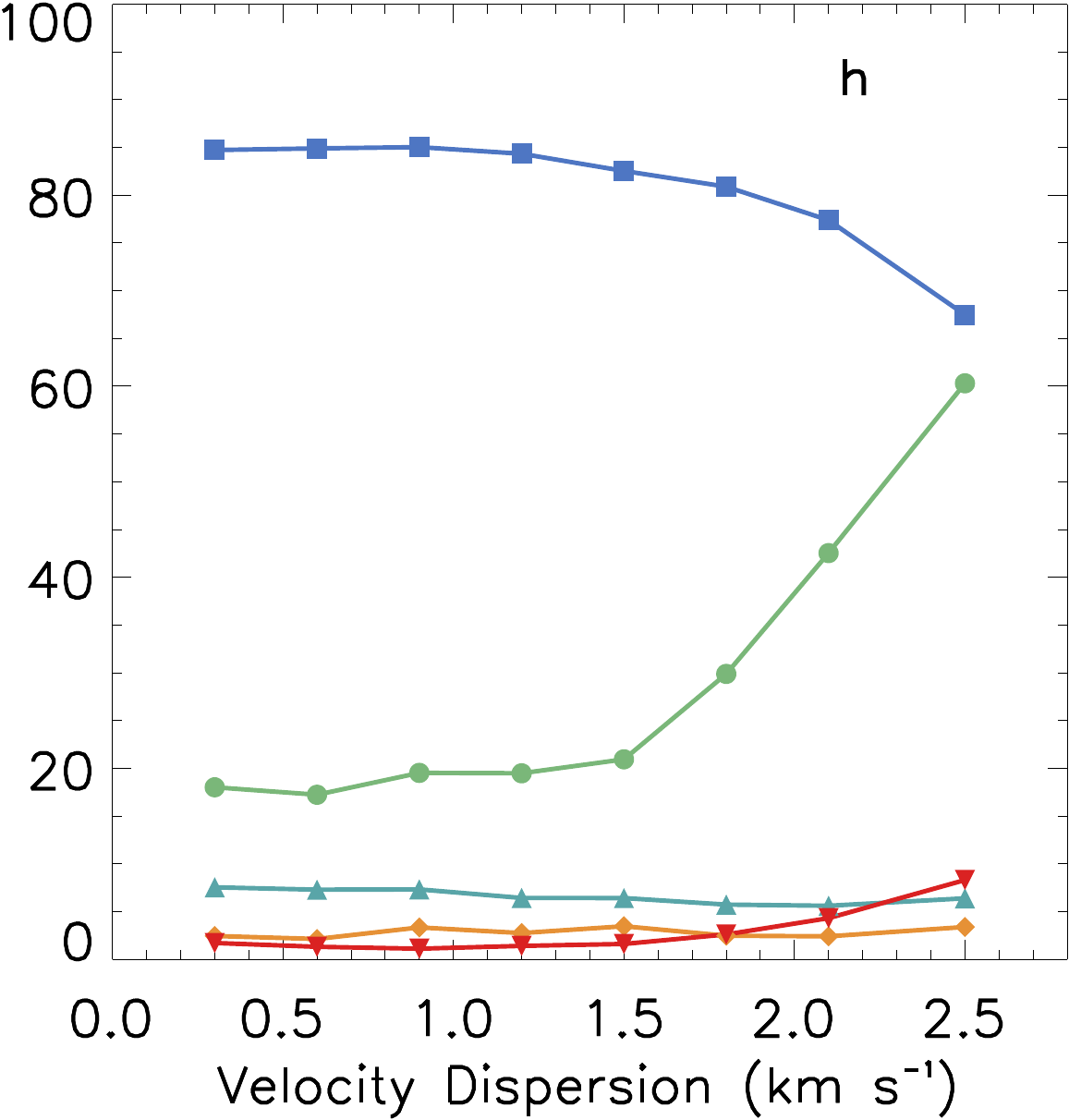}{0.303\textwidth}{}
              \fig{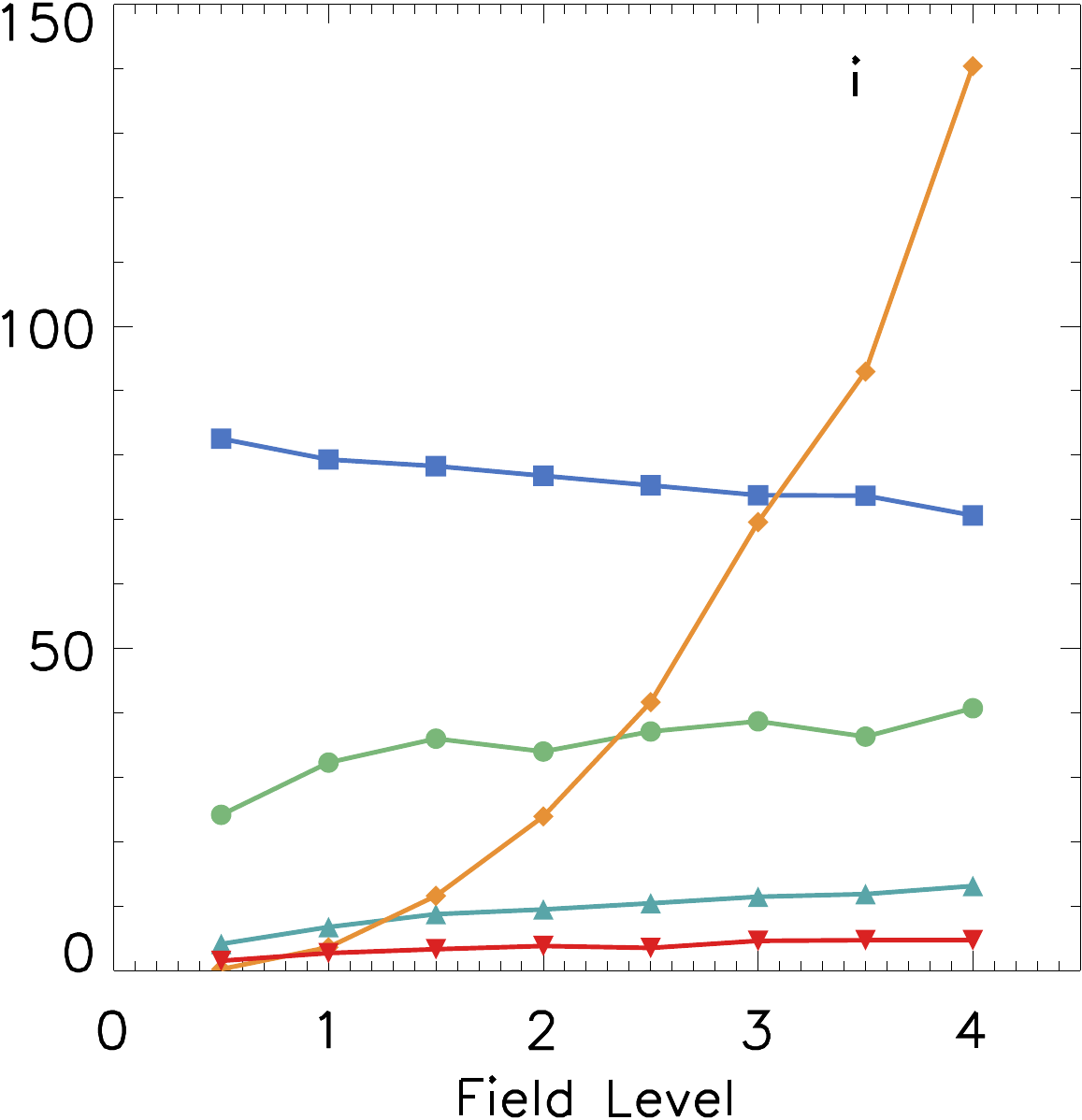}{0.302\textwidth}{}
        }\vspace{-0.5cm}
      \caption{Synthetic cluster recovery performance. Panels a--d refer to the parameters of the algorithm, and the calcualtions were performed on the same set of 1000 clusters. Vertical line shows the default parameters used. Panels e--i refer to the synthetic population properties. \label{fig:complete}}
\end{figure*}

We applied the average linkage clustering algorithm on the synthetic population described in the Appendix \ref{sec:synthetic}. This population consists of a a uniform distribution of field stars into which embedded a randomly generated cluster with the parameters that include distance, characteristic cluster velocity, age, and a number of stars that follow the normal distribution of a given velocity dispersion and the characteristic size. These parameters are varied in a manner that is representative of the ranges appropriate for the young stars towards Orion. While a single cluster cannot compare to the real population in terms of complexity, it can be used to test the performance of the algorithm in the various regimes, and the recovery properties of a single cluster are comparable to the recovery properties of multiple clusters in a single population.

For the purposes of clarity, the term `cluster' will refer to the generated population of YSOs, and the term `group' refers to the output of the algorithm. To test the performance of the algorithm, we analyzed results for 1000 random clusters (each with a unique field contamination) processed by the clustering algorithm under a variety of input parameters (e.g., maximum group size, maximum group velocity, etc). We evaluated the success of the clustering algorithm by computing the following for the resulting groups:

\begin{itemize}
\item False positive fraction (FP) is defined as a number of field stars falsely identified as group members relative to the total number of grouped stars.
\item True positive fraction (TP) is defined as the number of true cluster members in a group relative to the total number of true cluster stars.
\item If FP$>0.5$ then the group is considered as fake. In a given set of permutations, a single TP and FP are calculated, adding the values for all the real groups (i.e., FP$<0.5$) found.
\item Split cluster fraction (SC) is defined as $\frac{\sum N_{real}}{N_{det}}-1$, where $\sum N_{real}$ is the total number of real groups identified in all runs, and $N_{det}$ is the number of runs in which at least one group was detected (which may be $<$1000 as in some configurations of clusters no groups can be identified).
\item Similarly, fake cluster fraction (FC) is defined as $\frac{\sum N_{fake}}{N_{det}}$, where $\sum N_{fake}$ is the total number of fake groups identified in all runs.
\item Missing cluster fraction (MC) refers to the fraction of runs without a single group detection relative to the total number of permutations, i.e. $1-\frac{N_{det}}{1000}$.
\end{itemize}

Prior to determining the exact set of conditions that would be appropriate to use truncate the branches of the cluster tree, we explored the parameter space, varying one property of the algorithm at a time, in order to consider the effect they had (Figure \ref{fig:complete}). The goal of this exercise was to minimize the contamination while maximizing the fraction of true cluster members recovered by the algorithm to determine the appropriate thresholds for all the parameters.

Cluster reach (tolerance relative to the median linking length in a group to reject outliers) has a relatively little effect on all of the parameters. As it relaxes, contamination increases as well (FP increases by 2\% from varying reach parameter from 1 to 2, FC increases by 4\%), and clusters are more likely to be split into several groups (MC increase of 5\% over the same range), while having small gains in TP (TP increases by 2.8\%).

Varying a maximum allowed size of a recovered group has the largest effect on TP and MC: if it is too restricting so as not to encompass the entire cluster, cluster members are more likely to be missed. On the other hand, contamination levels rise relatively modestly with more tolerant thresholds. Varying the maximum size from 2 to 4 degrees increases FP by 2\%, TP by 10\%, and decreases SC by 8\%. This parameter was fixed to 4 degrees, which was the extent of the simulation, and which is almost large enough to encompass an entire molecular cloud in the Orion Complex. 

Maximum velocity cut has a qualitatively similar effect to the maximum size; however, relaxing the threshold beyond 9 \kms\ increases the contamination significantly, with FC rising by 14\% at 12 \kms. Finally, the critical number of stars - defined as the minimum number of stars that can be considered as a group at the end of the output - also has a significant effect. Allowing smaller groups increases TP through producing smaller groups that split from the main cluster, while also exponentially increasing FC.

With the default parameters listed in Section \ref{sec:method} applied to the synthetic dataset TP=79.5 $\pm$ 0.5\% (27648 out of 34784 cluster members identified), FP=6.20 $\pm$ 0.14\% (1830 out of 29524 grouped stars were from the field), SC=27.8 $\pm$ 1.7\% ($\sum N_{real}=1247$; $N_{det}$=976), FC=2.9 $\pm$ 0.5\% ($\sum N_{fake}$=28), MC=2.4 $\pm$ 0.5\%.

Next we tested the effect of the specific cluster properties on the recovery. Holding one cluster parameter at a time fixed to a specific value and allowing others to vary in the manner described in Appendix \ref{sec:synthetic}, we generated 1000 clusters for each iteration. As a number of stars in a cluster becomes larger, the cluster is increasingly more likely to split into several smaller groups: clusters with 75 stars have produced multiple groups in 85\% of iterations. While many of these splits occur along one dimension, some of the resultant groups are significant. While clusters were randomly generated, some correlations between the positional and kinematic components may occur, and multiple groups may portray this correlation. Sometimes, a split may occur just through a single dimension, and a closer examination of the output is necessary to confirm its significance. On the other hand, clusters with fewer than 30 stars are less likely to be recovered in the first place: almost 10\% of clusters with 20 stars have not produced any groups.

Cluster distance has little significant effect on any of the tests, and nor does the cluster age up to 15 Myr, as for clusters older than that would require different photometric cuts, and neglecting any effects of that would increase the dispersion of the older clusters beyond the random generation. Compact clusters with small velocity dispersion are more likely to be recovered, with TP$>$80\% for the typical cluster radii less than 2.5 pc (with the Gaussian distribution of stars in a cluster, a total cluster size is $>$10 pc), and for velocity dispersion less than 2 \kms. Those with velocity dispersion larger than 1.5 \kms\ are likely to run into the edges of the maximum allowed velocity range and be split into multiple groups, with the increase of SC from $\sim$20\% up to 60\% at 2.5 \kms. It should be noted, however, that in the Orion Complex, the ONC is the most massive cluster, the velocity dispersion of which is $\sim$2.5 \kms\ \citep{kounkel2016}; in other regions dispersion on the order of 1 \kms\ is not uncommon \citep[e.g.,]{briceno2007,nishimura2015}.

Then, we considered the effect of the field contamination, by increasing the number of field stars by a multiple factor to what has been previously adopted in Appendix \ref{sec:synthetic}. If the field population is underestimated in the generation of the synthetic population (or, if the photometric cuts are modified to include YSOs older than 15 Myr), then TP does decrease somewhat (dropping to $\sim$74\% if the field population is 3$\times4.5\times10^4$ stars), with a slight increase in FP (up to 11\% at the same level), but the main difference is the significant rise in FC (almost every run produces one or more fake groups), which requires additional vetting.

Lastly, a test was performed for the confusion in assignment of the stars in a multi-cluster population. Two clusters were placed at a random position in a sky in a field population covering 8$\times$8 deg$^2$. Other properties were allowed to vary as above. A group was identified as being associated with a particular cluster if more than 2/3 of all stars identified in the group originated from that cluster. In a run consisting of 10,000 simulations, we recorded the fraction of stars from a neighboring cluster that made it into a wrong group, which was 0.6$\pm$0.01\%. We also tracked as a fraction of the permutations in which at least one of the identified groups contained a mix of stars such that no clear cluster assignment could take place (with at least a third of sources in a group originating from one cluster, and a third form the second cluster). This fraction was somewhat higher at at 2.04$\pm$0.14\%, where 60\% of the affected groups were split off from the main groups associated with each cluster, 20\% had confident grouping of one of the clusters and confusion regarding the assignment of the second one due to the contamination from the first, and in 10\% of the cases both clusters are merged into a single group. Unfortunately, due to a multitude of the dimensions in which parameters could vary it is difficult to provide the specific conditions of the minimum distance between two stellar populations before confusion would occur.

\section{Results} \label{sec:analysis}
\begin{splitdeluxetable*}{ccccccccccccBcccccccccc}
\tabletypesize{\scriptsize}
\tablewidth{0pt}
\tablecaption{Identified groups and their properties\label{tab:clusters}}
\tablehead{
\colhead{Cluster ID} &\colhead{$\overline{\alpha}$} & \colhead{$\overline{\delta}$} & \colhead{$\overline{v_r}$\tablenotemark{a}} & \colhead{$\overline{v_\alpha}$\tablenotemark{a}} & \colhead{$\overline{v_\delta}$\tablenotemark{a}}& \colhead{$\overline{RV}$\tablenotemark{b}} & \colhead{$\overline{\mu_\alpha}$\tablenotemark{b}} & \colhead{$\overline{\mu_\delta}$\tablenotemark{b}} & \colhead{$\overline{\pi}$} & \colhead{$\overline{Age_{CMD}}$} & \colhead{$\overline{Age_{HR}}$}& \colhead{${\sigma_\alpha}$} & \colhead{${\sigma_\delta}$} & \colhead{${\sigma_{v_r}}$} & \colhead{${\sigma_{v_\alpha}}$} & \colhead{${\sigma_{v_\delta}}$} & \colhead{${\sigma_\pi}$} & \colhead{${\sigma_{Age_{CMD}}}$} & \colhead{${\sigma_{Age_{HR}}}$} & \colhead{N$_{in group}$}& \colhead{N$_{APOGEE}$}\\
\colhead{ } &\colhead{(J2000)} & \colhead{(J2000)} & \colhead{(\kms)} & \colhead{(\masyr)} & \colhead{(\masyr)}& \colhead{(\kms)} & \colhead{(\masyr)} & \colhead{(\masyr)} & \colhead{(mas)} & \colhead{(Myr)} & \colhead{(Myr)} & \colhead{(deg)} & \colhead{(deg)} & \colhead{(\kms)} & \colhead{(\masyr)} & \colhead{(\masyr)} & \colhead{(mas)} & \colhead{(Myr)} & \colhead{(Myr)} & \colhead{}& \colhead{}
}
\colnumbers
\startdata
l1641N-1     & 83.70 & -6.76 &   7.42 &   0.36 &   2.58 &  24.71 &   0.92 &  -0.18 &    2.572 &   3.5 &   2.2 & 0.54 & 0.67 & 1.44 & 0.29 & 0.35 &    0.086 & 1.9 & 1.6 & 104 &  29 \\ 
l1641N-2     & 84.61 & -6.97 &   4.24 &  -0.53 &   1.99 &  21.67 &  -0.19 &  -0.54 &    2.436 &   0.8 &   2.1 & 0.47 & 0.51 & 1.60 & 0.37 & 0.23 &    0.084 & 1.1 & 1.1 &  25 &  15 \\ 
l1641N-3     & 83.85 & -6.45 &   3.79 &   0.18 &   3.38 &  21.09 &   0.61 &   0.61 &    2.539 &   3.0 &   5.2 & 0.37 & 0.65 & 2.85 & 0.18 & 0.16 &    0.061 & 1.9 & 1.1 &  16 &   6 \\ 
l1641N-4     & 83.89 & -6.20 &   9.21 &   0.81 &   3.33 &  26.48 &   1.34 &   0.72 &    2.503 &   2.0 &   0.7 & 0.09 & 0.22 & 0.19 & 0.72 & 0.38 &    0.196 & 1.6 & 1.9 &  11 &  11 \\ 
\enddata
\tablenotetext{}{Only a portion shown here. Full table is available in an electronic form.}
\tablenotetext{a}{lsr}
\tablenotetext{b}{Heliocentric}
\end{splitdeluxetable*}

\begin{figure}
\epsscale{1}
\plottwo{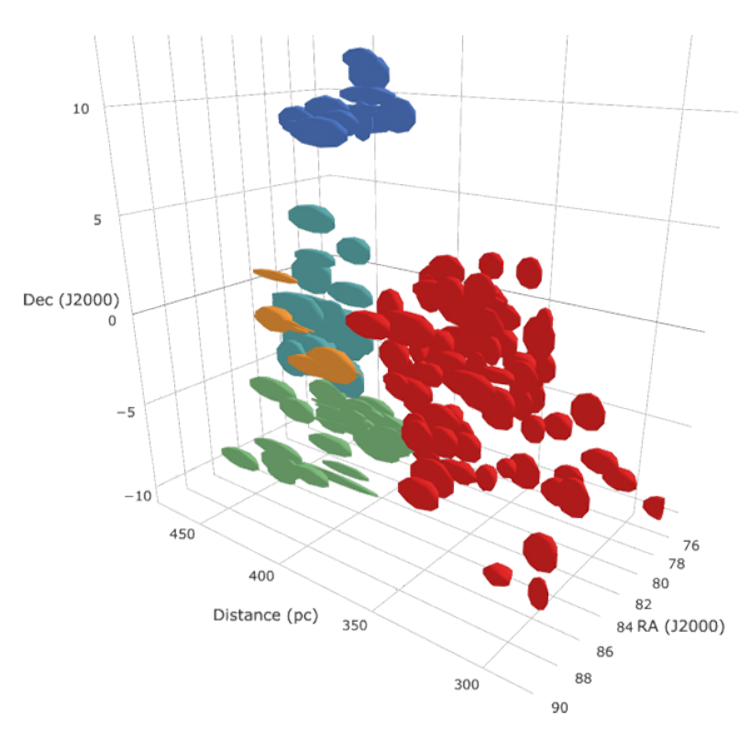}{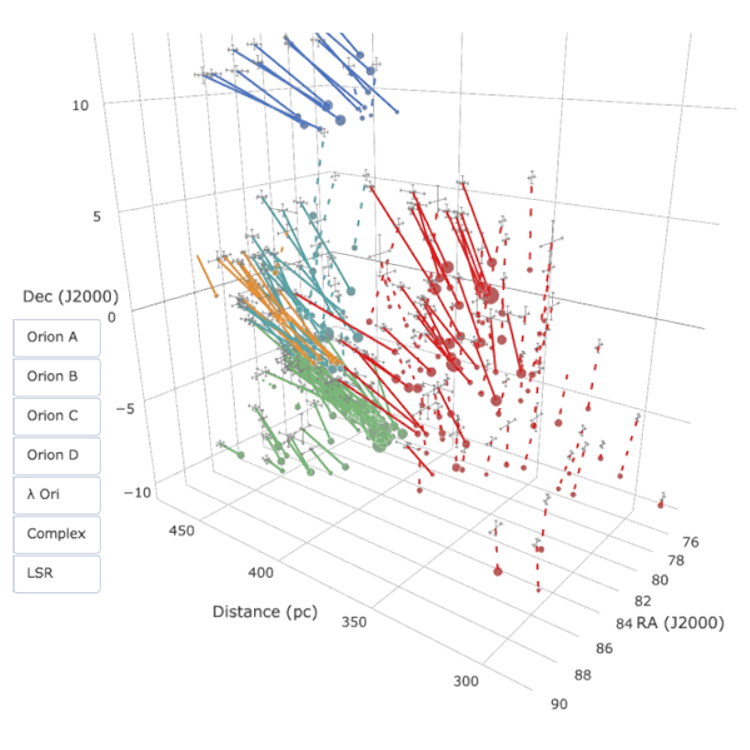}
\caption{Left: Structure of the Orion Complex as traced by the characteristic sizes of the identified groups. Colors correspond to the different populations: Orion A (green), Orion B (orange), Orion C (cyan), Orion D (red), and $\lambda$ Ori (blue). Right: Typical kinematics of the groups. Dashed lines show the proper motion of those groups in which no RV information is available. The sizes of the points at the beginning of the lines correlates to the number of stars in a group. By default, lsr kinematics are displayed, buttons convert the kinematics to the reference frame of a given population. Interactive figure is available in the online version.\label{fig:interactive}}
\end{figure}

\begin{figure*}
\epsscale{0.8}
		\gridline{\fig{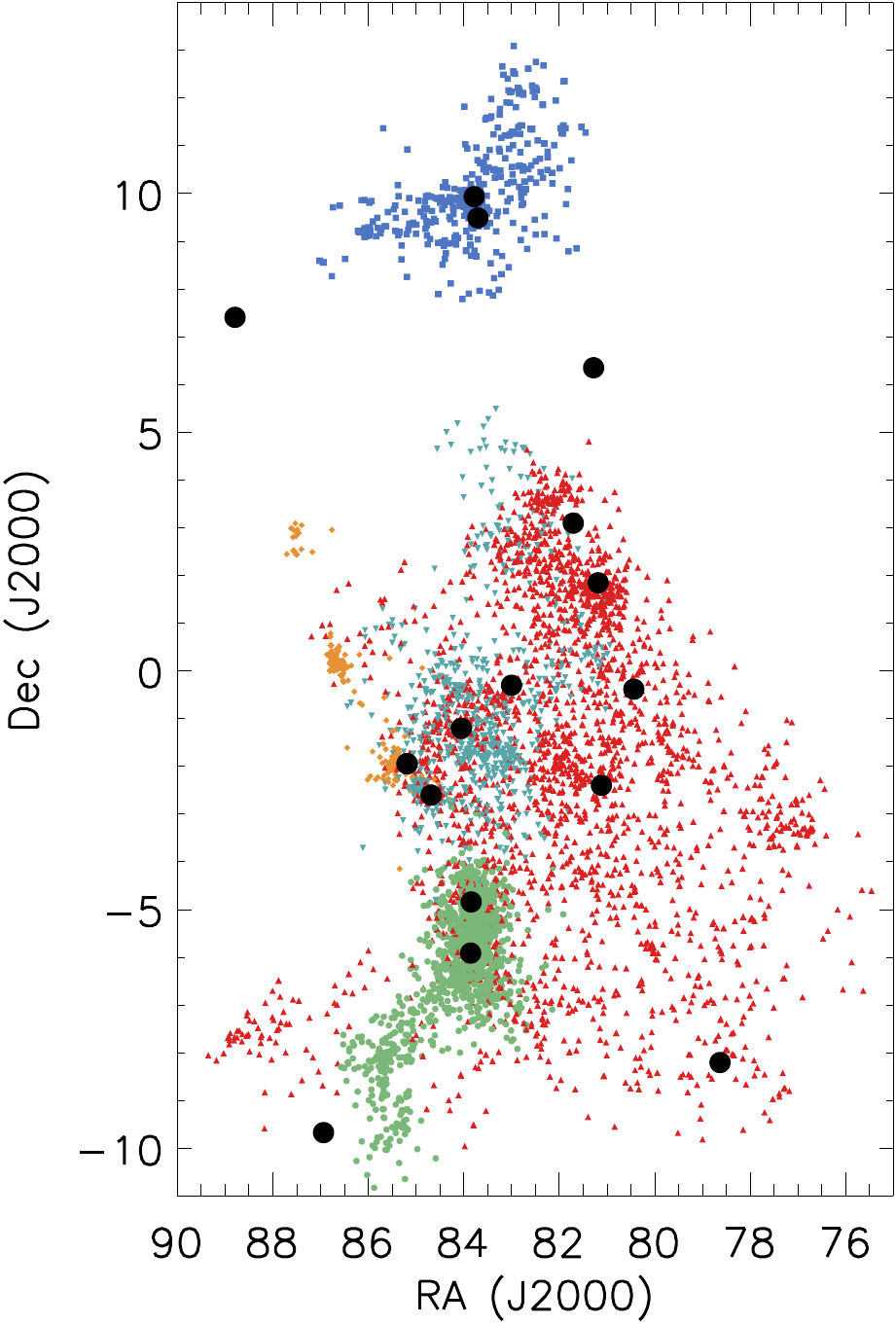}{0.38\textwidth}{}
              \fig{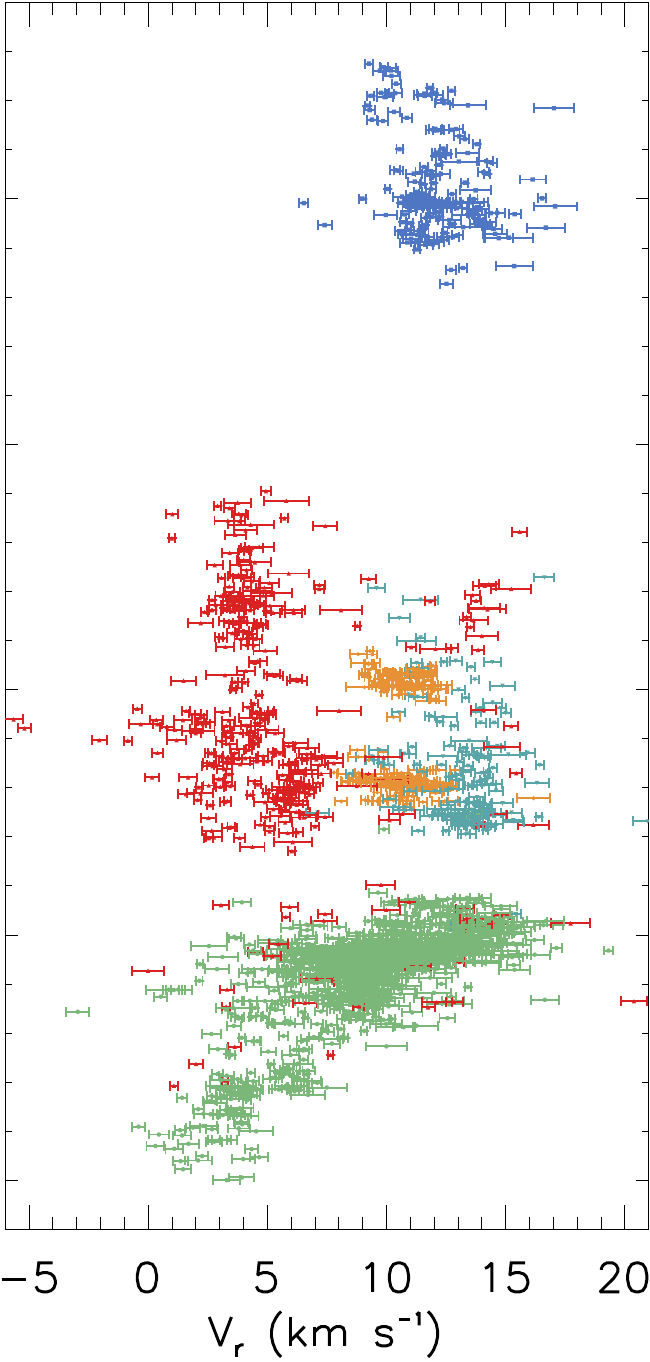}{0.267\textwidth}{}
              \fig{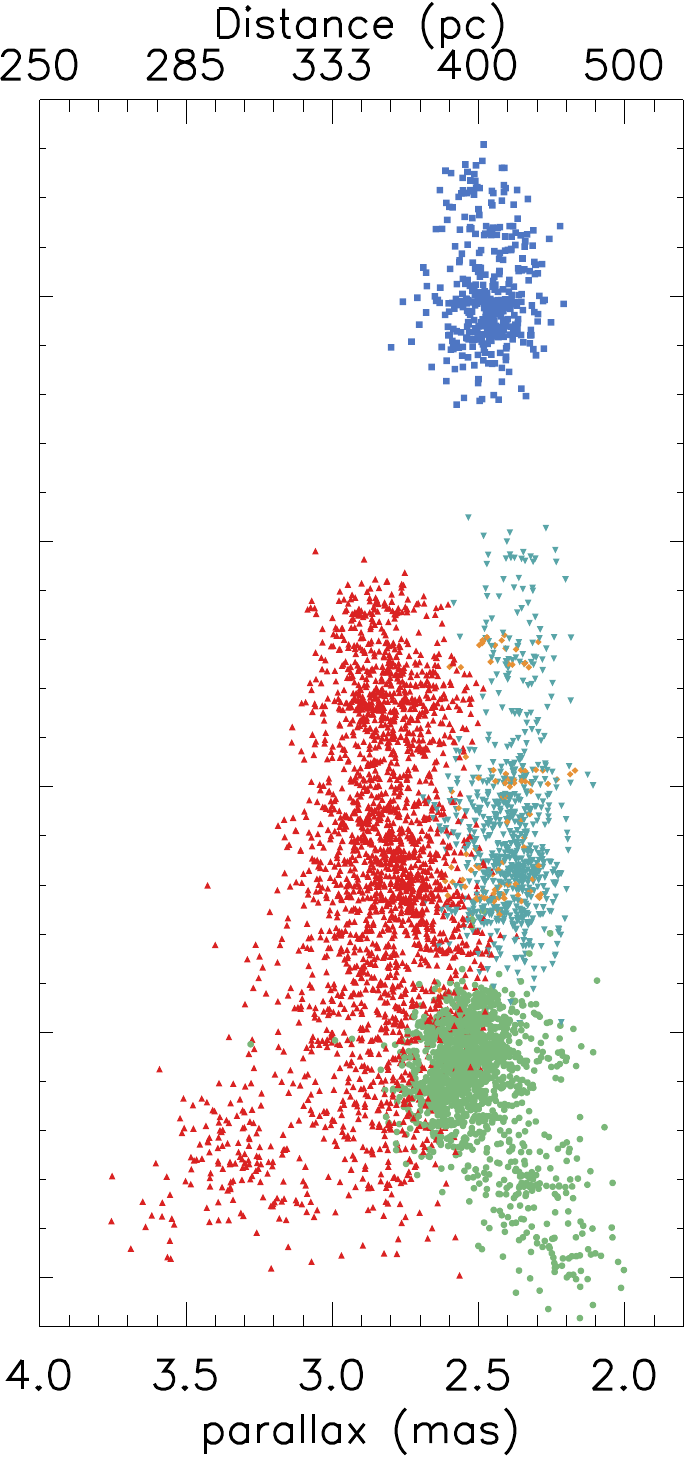}{0.28\textwidth}{}
        }
\caption{Distribution of stars identified as members Orion A (green), Orion B (orange), Orion C (cyan), Orion D (red), and $\lambda$ Ori (blue). In the first panel, black dots show the position of the major bright stars in Orion. \label{fig:largerv}}
\end{figure*}

\begin{figure*}
\epsscale{1}
		\gridline{\fig{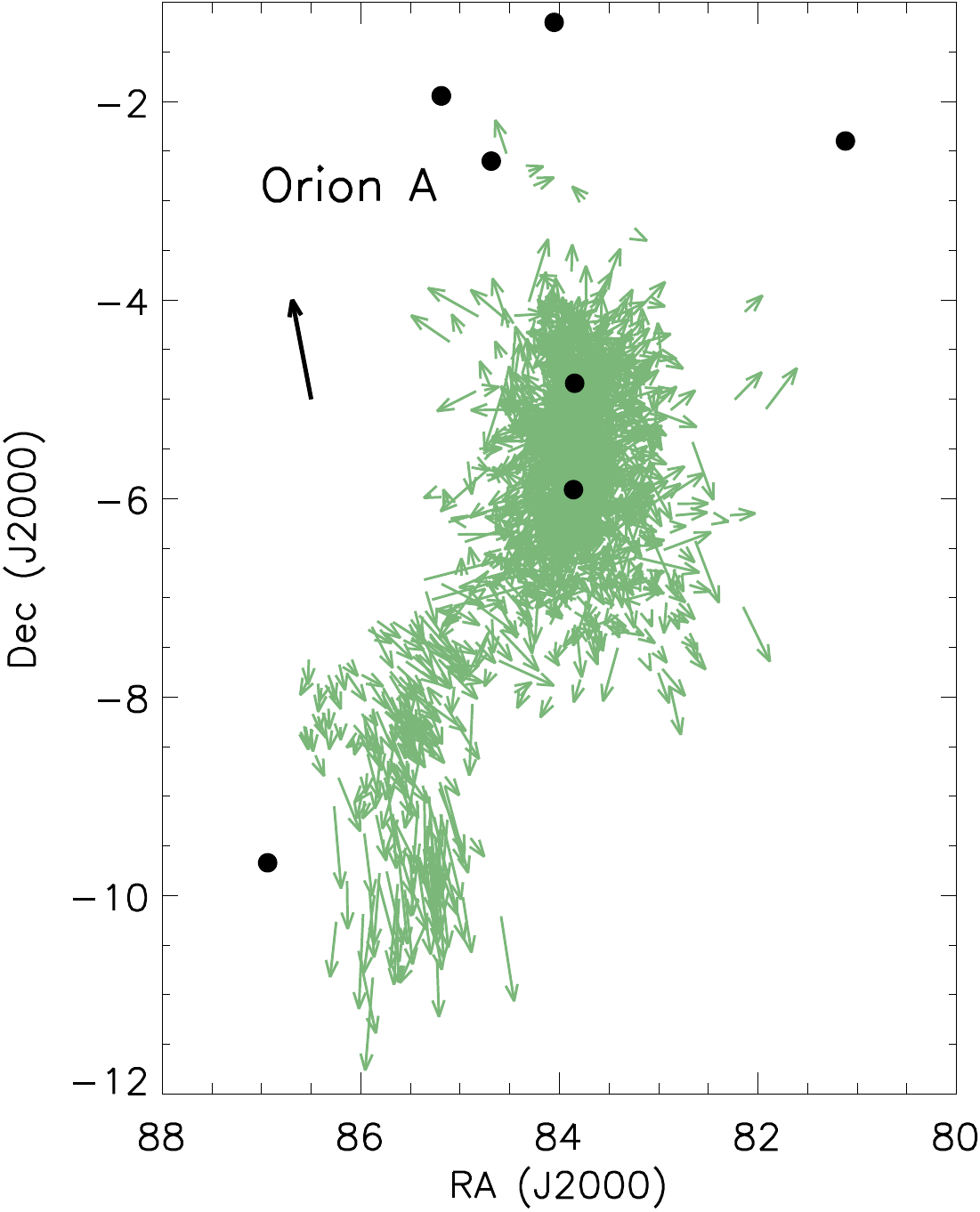}{0.36\textwidth}{}
              \fig{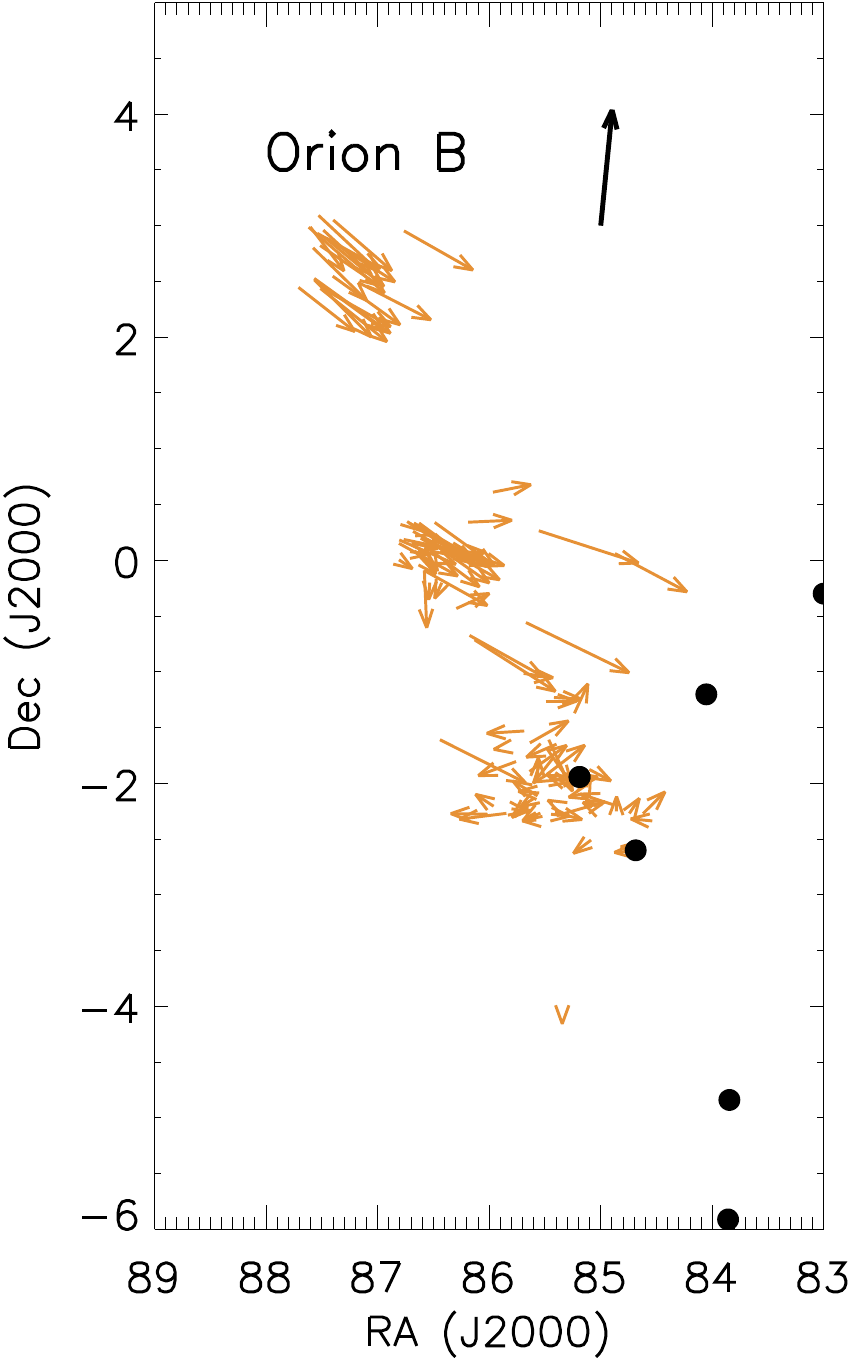}{0.28\textwidth}{}
              \fig{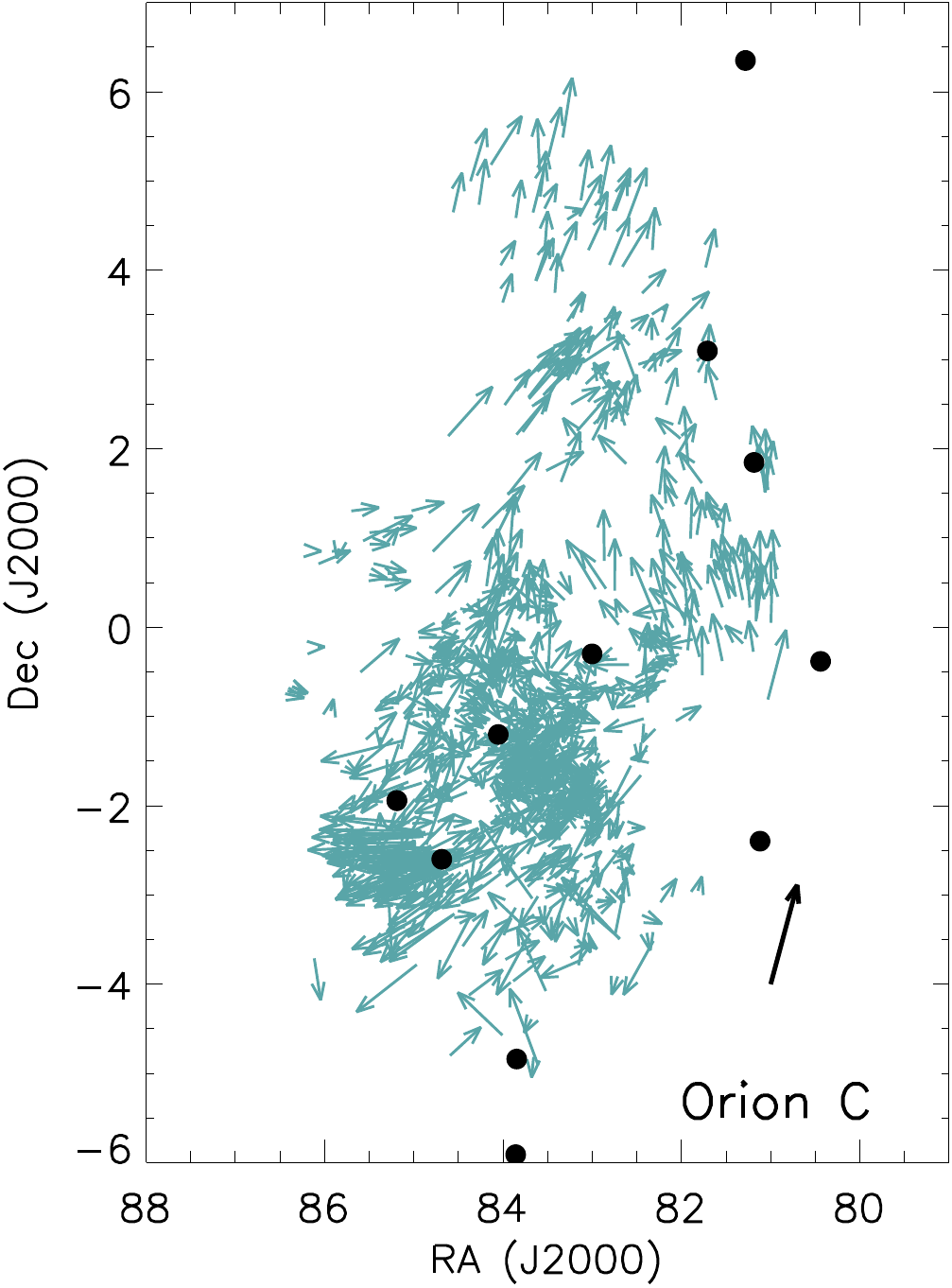}{0.33\textwidth}{}
        }
        	\gridline{\fig{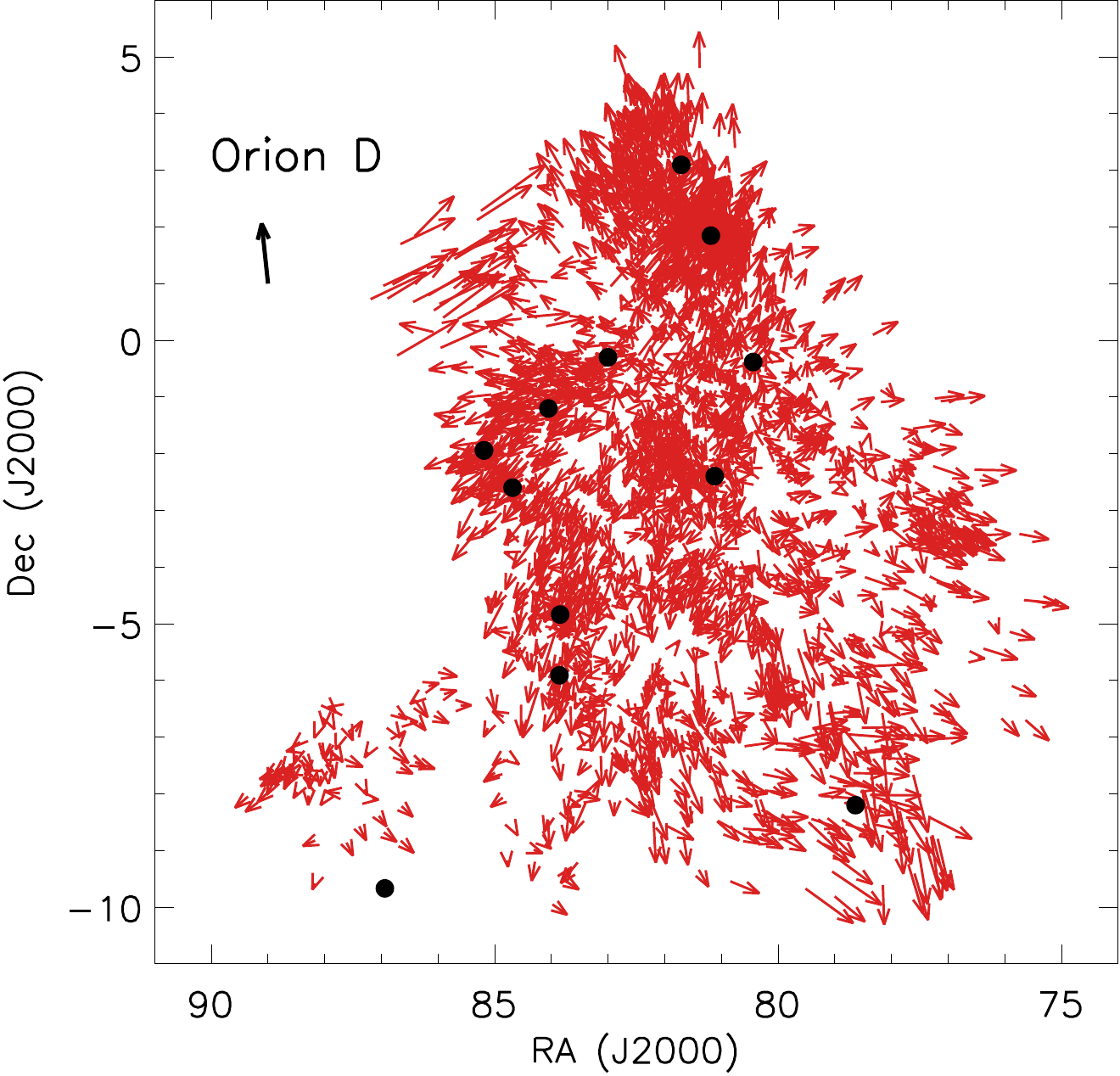}{0.4\textwidth}{}
              \fig{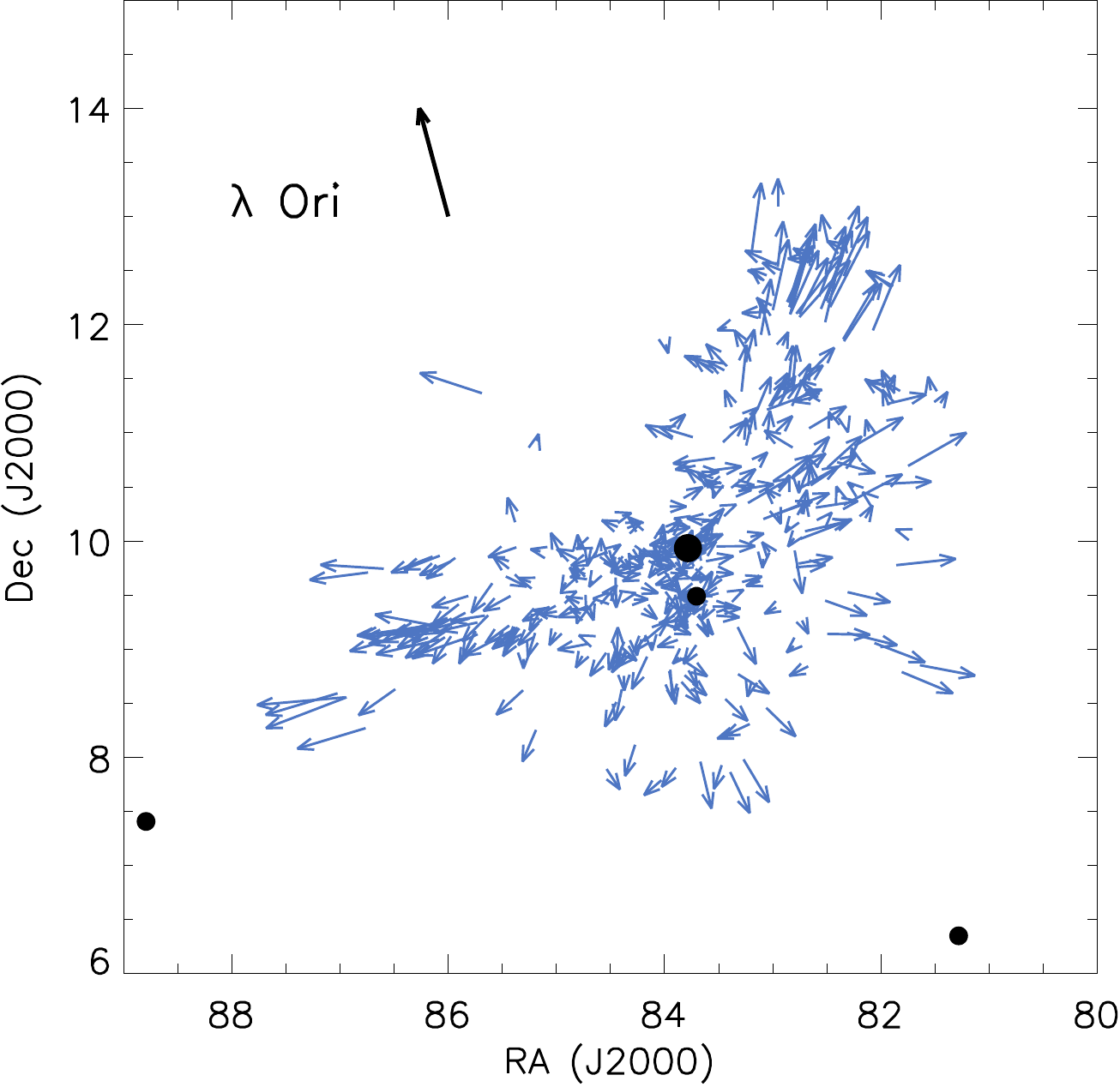}{0.4\textwidth}{}
        }
\caption{Proper motions of the stars identified as members of the complex, relative to the average (lsr) proper motion in each structure. The black arrow shows the subtracted average motion, which has the magnitude of $\sim$3.1, 3.4, 3.7, 2.9, and 3.4 \masyr in Orion A, B, C, D, and $\lambda$ Ori respectively. The length of the vectors correspond to the motion of over 1.2 Myr. Note that the scale is not consistent across all panels. Black dots show the position of the major bright stars in Orion; in the last panel the bigger dot is $\lambda$ Ori.\label{fig:pms}}
\end{figure*}
\begin{figure}
\epsscale{1.2}
\plotone{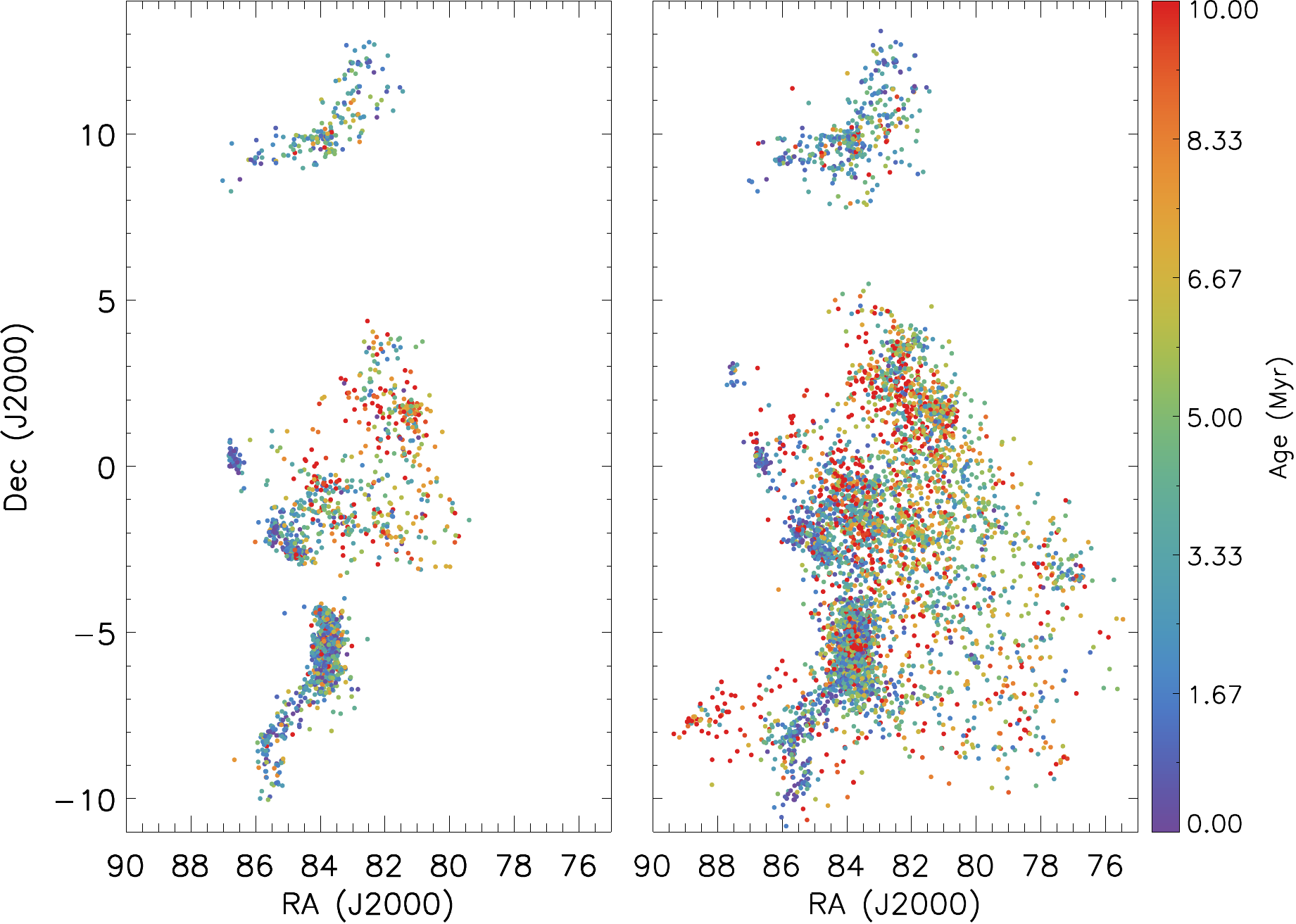}
\caption{Estimation of stellar ages for the sources in the Orion Complex. Left: Age$_{HR}$, right: Age$_{CMD}$. \label{fig:age}}
\end{figure}

In this section we present the results of the clustering algorithm (Figure \ref{fig:interactive}), which was applied to the 10248 stars, of which 890 came solely from the APOGEE sample, 7081 were unique to \textit{Gaia}, and 2277 appeared in both datasets with complete 6d characterization. In total, we identified 190 groups throughout the Orion Complex, their average properties are listed in Table \ref{tab:clusters}. These groups were manually combined into 5 larger structures that are described below (Figures \ref{fig:largerv},\ref{fig:pms},\ref{fig:age}), and when feasible cataloged according to the closest major object near them, such as a star visible to the naked eye, known cluster, or molecular gas region.

There are 6 groups identified as spurious detections as they are unlikely to correspond to any larger structure; they are also listed in the table, for completeness. 

In many cases, the identified groups may not necessarily correspond to distinct subclusters. In some cases, the split may occur only along one principle axis with the artificial split smaller than the dispersion in the cluster (e.g in NGC 2024 groups are separated in $v_r$ by $\sim$0.5 \kms, with the total dispersion velocity of 1.3 \kms). This is most apparent in the massive clusters: the ONC alone is associated with more than 30 groups. This occurs because the algorithm is calibrated to recover less numerous, less concentrated populations that occurs in the other parts of the Complex. There may be a few clustering algorithms that might have an improved performance when confronted with such vast differences in scale through a mixture of several techniques \citep[e.g.,][]{zhang1996,karypis1999}. Here, however, we use the identified groups to trace the distribution of the larger structure and look for strong significant deviations that are attributable to real subclusters. Some confusion may also occur when the sources from two separate large structures may have some overlap in one of the dimensions, if they have incomplete data in other dimensions. For example, there are a few groups associated with Orion D where only one or two sources have RV information which is may be more consistent with the Orion C population. This affects a relatively small fraction of sources, and when reporting on the averages we require at least 3 measurements in a given dimension.

\subsection{Orion C and D}\label{sec:oric}

\begin{figure*}
\epsscale{1}
\plotone{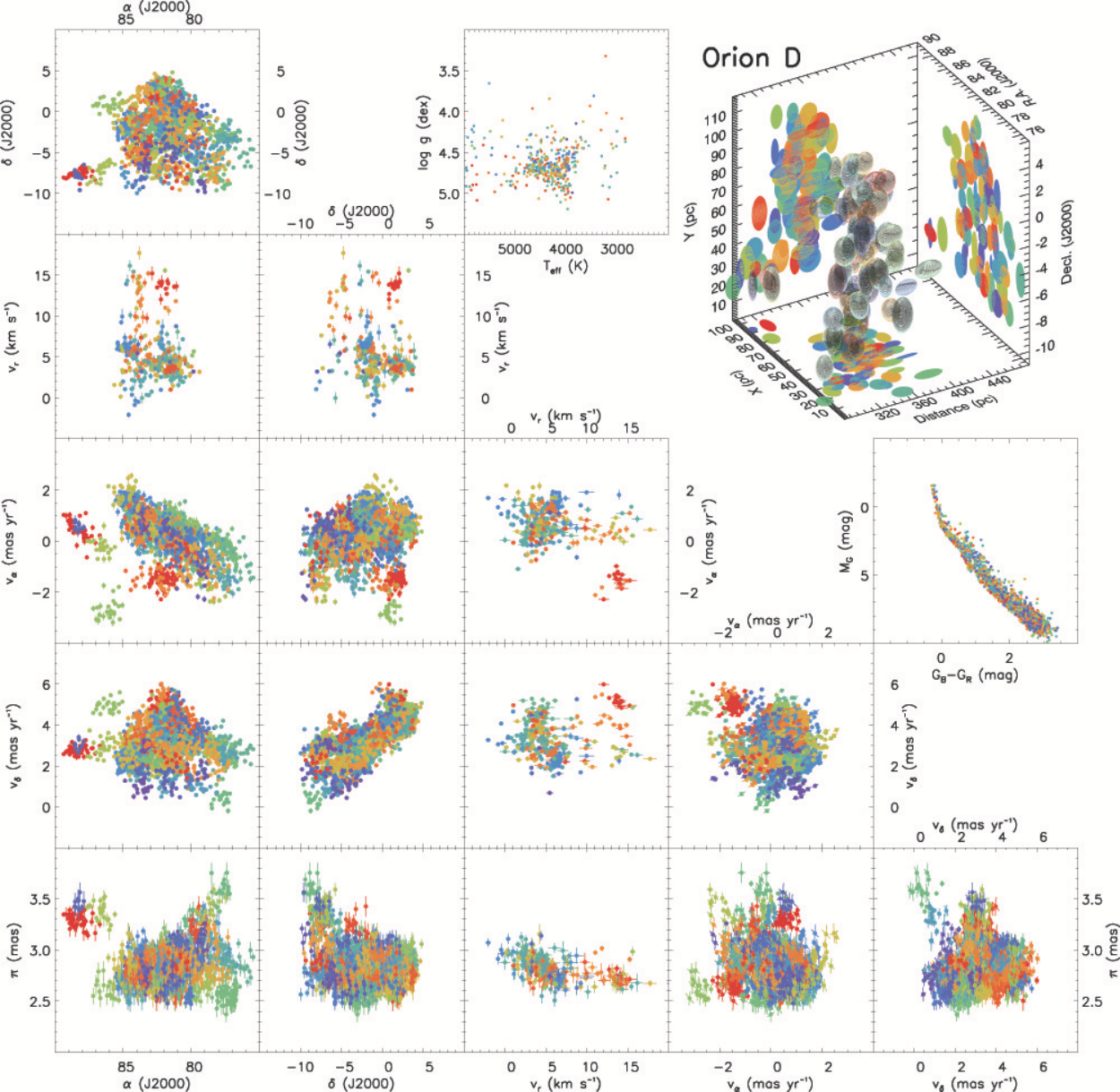}
\caption{Structure and kinematics of all of the identified groups towards the Orion Complex that show the projection of the individual measurements in all six dimensions. Orientation of the uncertainties in parallax and proper motions include the correlations from \textit{Gaia}. Other panels include the distribution \teff\ and \logg\, color-magnitude diagram, and a three-dimensional rendering of the structure. Colors are randomly assigned to distinguish the groups from each other in all panels. Orion D is shown here; the complete figureset (5 images) is available in the online journal.\label{fig:6d}}
\end{figure*}

%
%
%
%
%

In the observations of RV towards $\sigma$ Ori, \citet{jeffries2006} have observed two distinct velocity components separated by 7 \kms, one of which was interpreted to originate from the population of stars associated with $\sigma$ Ori, whereas the other component as attributed to Ori OB1ab. We obtain similar results in the RV observations with APOGEE. The two components appear to extend northwest significantly beyond the area originally covered by \citet{jeffries2006}. The component centered at $v_r\sim$3--5 \kms\ (Figure \ref{fig:6d}.1) traces the population of YSOs surrounding the belt stars, as well as the OB stars such as $\eta$ Ori, 22 Ori, 25 Ori, and $\psi^2$ Ori. This population includes stars associated with Orion OB1a and OB1b. It stretches further south beyond the area covered by the APOGEE footprint towards Rigel, which includes a few of the Orion outlying clouds \citep{alcala2008,biazzo2014}, and some authors recently referred to the region as Orion X \citep{bouy2015,zari2017}. On the other hand, the component centered at $v_r\sim$13 \kms\ (Figure \ref{fig:largerv},\ref{fig:6d}.2) consists of the $\sigma$ Ori cluster, and it stretches diagonally towards an area east of 25 Ori, concluding northeast of $\psi^2$ Ori, largely excluding any other notably bright OB stars. The two populations are also found at different distances; this has also been recently found by \citet{briceno2018}. This demonstrates that $\sigma$ Ori originated from a different cloud than the rest of the OB1ab regions. We extend the naming convention of the Orion A and B molecular clouds, and we will refer to this progenitor as the (former) Orion C molecular cloud. Collectively, we will also refer to the entire region encompassing OB1a, OB1b, and Ori X as Orion D. We note that the original definition of the OB1 sub-associations is largely based on the area on the sky - not necessarily representative of the underlying structure. As two populations are largely superimposed onto each other, caution must be exercised in comparison with the literature, especially in vicinity of OB1b. The motivation behind grouping it with Ori D is due to the tighter correlation of the foreground population with the belt stars, but many of the stars that have been associated with OB1b are also associated with the Orion C region.

At the present day, the gas from the Orion C has been almost completely dissipated. There does appear to be a clear age gradient in the stars along the filamentary structure from $\alpha=85^\circ, \delta=-3^\circ$ to $\alpha=82^\circ, \delta=3^\circ$ (Figure \ref{fig:age}). Group C-North is the oldest, having a wide distribution in ages around a median of $\sim$7.5 Myr, C-Central has an equally broad distribution, with a median of $\sim$5.5 Myr, but also a peak at $\sim$3 Myr, and the $\sigma$ Ori cluster is the youngest population with the age of 1.9$\pm$1.6 Myr. Both Age$_{HR}$ and Age$_{CMD}$ show a very similar distribution. Correcting for systematics, we measure the weighted average distances of 406$\pm$4 pc for $\sigma$ Ori, 413$\pm$4 for Ori C-C, and 416$\pm$4 for Ori C-N. Orion C also appears to be moving uniformly radially, however, it is expanding along the plane of the sky, with $\sigma$ Ori and Ori C-N moving away in opposite directions from Ori C-C.

On the other hand, it is difficult to conclusively determine whether all parts of the Orion D population shares the same progenitor or not, as the population containing them is largely continuous and mostly dispersed with few spacial kinematical differences. (In the case of OB1a and OB1b, this contradicts what has been previously observed by \citet{briceno2007}; however, the sample of stars that they used as representative members of OB1b have actually originated from the Orion C population.) Orion D has a comparable size in the sky to the entirety of Orion A, B, and C put together, and in another few Myr it is possible that those A, B, and C regions may become largely indistinguishable from each other. On the other hand, it is possible that Orion D is all part of a single expanding population, and most of the proper motion vectors are consistent with the process of expansion, they dispersing most likely due to the loss of the molecular gas, \citep[e.g.,][]{tutukov1978}.

Orion D has very little molecular gas remaining. Projected in part onto Orion B, the tail end of D corresponds to the very diffuse gas that can be seen in Figure \ref{fig:oribrv} that has a gradient in $v_r$ from 6 to 3 \kms. L1622 also is most likely associated with Orion D (see Section \ref{sec:orib}). Few diffuse clouds are found in the southwest, namely L1616, L1634, and IC 2118. While most of the population is largely dissipated, few concentrated clusters still are apparent, such as 25 Ori at 354$\pm$3 pc \citep[Age$_{CMD}$=6.2$\pm$2.3 Myr, Age$_{HR}$=7.4$\pm$2.0 Myr), consistent with][]{briceno2018}, $\psi^2$ Ori at 347$\pm$3 pc (Age$_{CMD}$=5.5$\pm$1.7 Myr, Age$_{HR}$=4.9$\pm$1.5 Myr), $\eta$ Ori at 347$\pm$3 pc (Age$_{CMD}$=5.7$\pm$1.4 Myr, Age$_{HR}$=6.4$\pm$1.9 Myr), OB1b at 357$\pm$3 pc (Age$_{CMD}$=3.7$\pm$1.8 Myr, Age$_{HR}$=4.1$\pm$1.6 Myr), and L1616 at 360$\pm$3 pc (Age$_{CMD}$ =5.0$\pm$1.7 Myr). The southern portion does deviate in distance from the rest of Ori D - population that is associated with IC 2118 is located at 291$\pm$2 pc (Age$_{CMD}$=7.9$\pm$2.5 Myr).

A small group of stars is found north of $\kappa$ Ori that has a distance similar to IC 2118 at 302$\pm$2.5 pc. This group has not been cataloged previously, although a few stars we identify have been previously confirmed as YSOs by \citet{alcala2000}. We will refer to this group as Orion Y. Further tests will be necessary to confirm the properties and membership of this group. This group is spatially discontinuous from Ori D, and older (Age$_{CMD}$=10.5$\pm$2.5 Myr). However, it is does have similar proper motions, therefore we group it together with the Orion D on all the plots.

\citet{pillitteri2017} have also observed a nearby group of stars surrounding V1818 Ori for which they estimated a distance of 270 pc. It is not clear if it is associated with Orion Y, as we do not recover it, and nearly all sources identified as YSOs by \citet{pillitteri2017} have $\pi<$2 mas.

One group identified by the algorithm is kinematically peculiar. 25Ori-2 \citep[it was recently identified by][under the name of HR 1833]{briceno2018} is kinematically distinct by almost 10 \kms\ in $v_r$ and 1.8 \masyr\ in $v_\alpha$ from the main group. It also appears to be significantly older than 25 Ori, with Age$_{CMD}$=15.1$\pm$3.4 Myr (Age$_{HR}$=12.9$\pm$2.8 Myr). Its kinematics are similar to those of Orion C, however, its distance is more consistent with the Orion D population. Since it is closer than Orion C, but its $v_r$ is redshifted, it is unlikely that these stars have originated in Orion C, but may instead be part of an earlier wave of star formation in Orion D.

\subsection{Orion A}

The clustering algorithm used in this work has a relatively poor performance when it is applied on very numerous and extended stellar populations with a large velocity dispersion, creating many largely artificial distinctions between separate regions of the same cluster. As such, the ONC creates a particular challenge to disentangle in terms of its true internal structure (Figure \ref{fig:6d}.3). The high degree of extinction towards the cluster exacerbates the problem further. The ONC has been the subject of a number of studies to analyze its structure using methods that may be more suited to such an environment, including the work done by \citet{hillenbrand1998}, \citet{kuhn2014}, \citet{megeath2016}, and \citet{hacar2016}, using 2d and 3d data. Here, we present the identified groups mainly as a method to trace the extremes of the stellar population in a 6d space, with the caveat that the groups themselves may not be physical in nature, particularly compared to the results observed in the other regions in this paper.

In previous RV studies towards the ONC, a population of stars were identified that are blueshifted to the molecular gas from which these stars have formed, although no obvious correlations have been found with any of the stellar properties or the position of those stars on the sky \citep{tobin2009,kounkel2016,da-rio2016}. Unfortunately, stellar proper motion information cannot be compared directly to the kinematics of the molecular gas. With the addition of the astrometric solutions from \textit{Gaia}, the mystery of the blueshifted population remains, showing no correlation with the distribution of distances or proper motions when applied to the optical observations from \citep{kounkel2016}. Although it is notable that the APOGEE sample does show significantly better agreement between the stellar RVs and those of the molecular gas.

In general, regarding the 3d structure of the Orion A, there is a good agreement with the model from \citet{kounkel2017}. Using \textit{Gaia} astrometry, the average parallax towards the ONC is 2.540$\pm$0.001 mas; corrected for systematics it results in a distance of 389$\pm$3 pc. The southern end of L1641 is found at $\pi=2.364\pm0.004$ mas, or 417$\pm$4 pc, with relatively smooth continuity between them, and L1647 at $\pi=2.227\pm0.006$ mas, or 443$\pm$5 pc. 

Similarly good agreement is found for the individual measurements of kinematics \citep[however it should be noted that there was an error in conversion from $\mu_{\alpha,\delta}$ to $v_{\alpha,\delta}$ in][]{kounkel2017}, but now it is possible to reconstruct the full map of the PMs along the cloud. The motions in the ONC appear to be mostly random, with slight preference for expansion near the outer edges. PMs in L1641 are preferentially oriented perpendicular to the filament in the plane of the sky (most of the grouped sources are located near the northern edge of the molecular cloud, their proper motions are oriented towards the gas), in L1647, stars are primarily moving away from the main filament in the plane of the sky.

We estimate Age$_{HR}$=1.6$\pm$1.5 Myr (Age$_{CMD}$=3.2$\pm$2.2 Myr, which is an overestimate due to extinction). This is consistent with the estimates by \citet{getman2014}. Similar distributions are present in the northern and southern parts of L1641 with the ages of 2.0$\pm$1.5 and 1.9$\pm$1.4 Myr (Age$_{CMD}$=2.4$\pm$1.8 Myr and 2.1$\pm$1.7 Myr) respectively, consistent with the measurements from \citet{hsu2013}. L1647 is the youngest region, with the Age$_{CMD}$=1.3$\pm$1.3, Age$_{HR}$=1.9$\pm$0.9 Myr.

There is one group towards the ONC that is kinematically peculiar from the rest (ONC-22), as it is offset from the main population in $v_\alpha$ by 2.6 \masyr. Unfortunately no RV measurements are available, but it does not significantly deviate in other dimensions, though it may be found closer to the back of the cluster.

A particular interest has been given in the past to NGC 1980, a population which exhibits relatively little extinction compared to the rest of Orion A. Several studies have suggested that this region is more evolved compared to the ONC, and that it may be located somewhat in the foreground \citep{alves2012,pillitteri2013,bouy2014}. However, there has also been some doubt to these claims \citep{da-rio2016,fang2017,kounkel2017a}. We do find a diffuse distribution of YSOs in the foreground as part of the Ori D structure, and some of it does coincide spatially with Orion A. It does not appear to be particularly concentrated near NGC 1980 region, and only $\sim$8\% of the total that is found towards it ($\sim$30 stars) are likely to be associated with the foregorund population, but it is possible that they may have contributed to the confusion surrounding the issue.

Confusion with Orion D towards Orion A is most notable towards the north of the cluster, near NGC 1981, which is another population that has been previously observed to be older than the rest of the ONC \citep[e.g.][]{maia2010}. There are a number of groups that are observed to have a distance more consistent with that of the Orion D, at 357$\pm$3 pc, with Age$_{CMD}$=5.0$\pm$1.8 Myr/ Age$_{HR}$=2.8$\pm$1.7 Myr. Most of these groups do not have reliable $v_r$ measurements, but those that do have a good agreement with that of the ONC - somewhat surprising considering the peculiar RV structure of the Orion A north of $\delta<-5^\circ$ that is not representative of Orion D. While we associate these groups with Orion D, this assignment may not necessarily be accurate.

\subsection{Orion B}\label{sec:orib}

\begin{figure}
\epsscale{0.8}
		\gridline{\fig{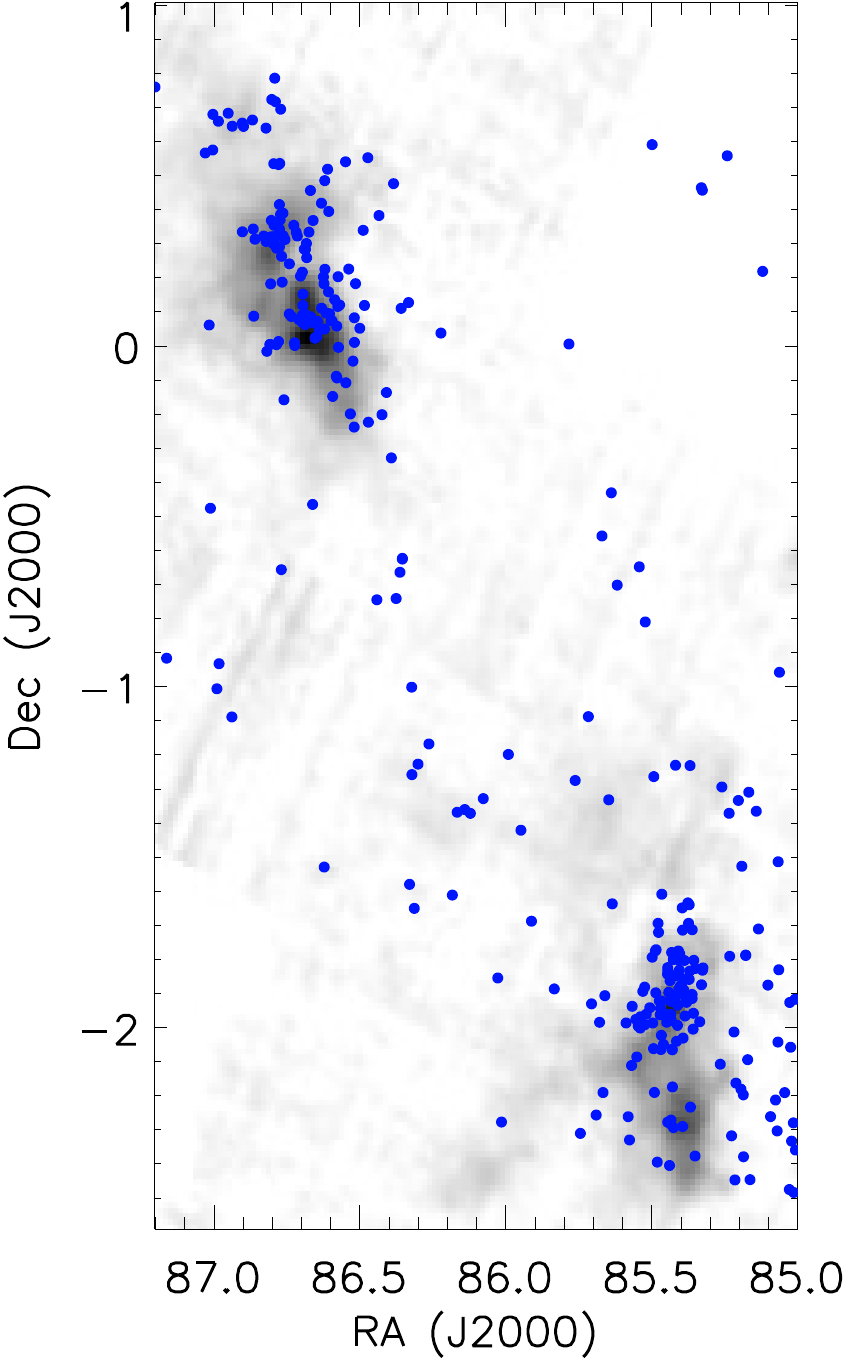}{0.25\textwidth}{}
              \fig{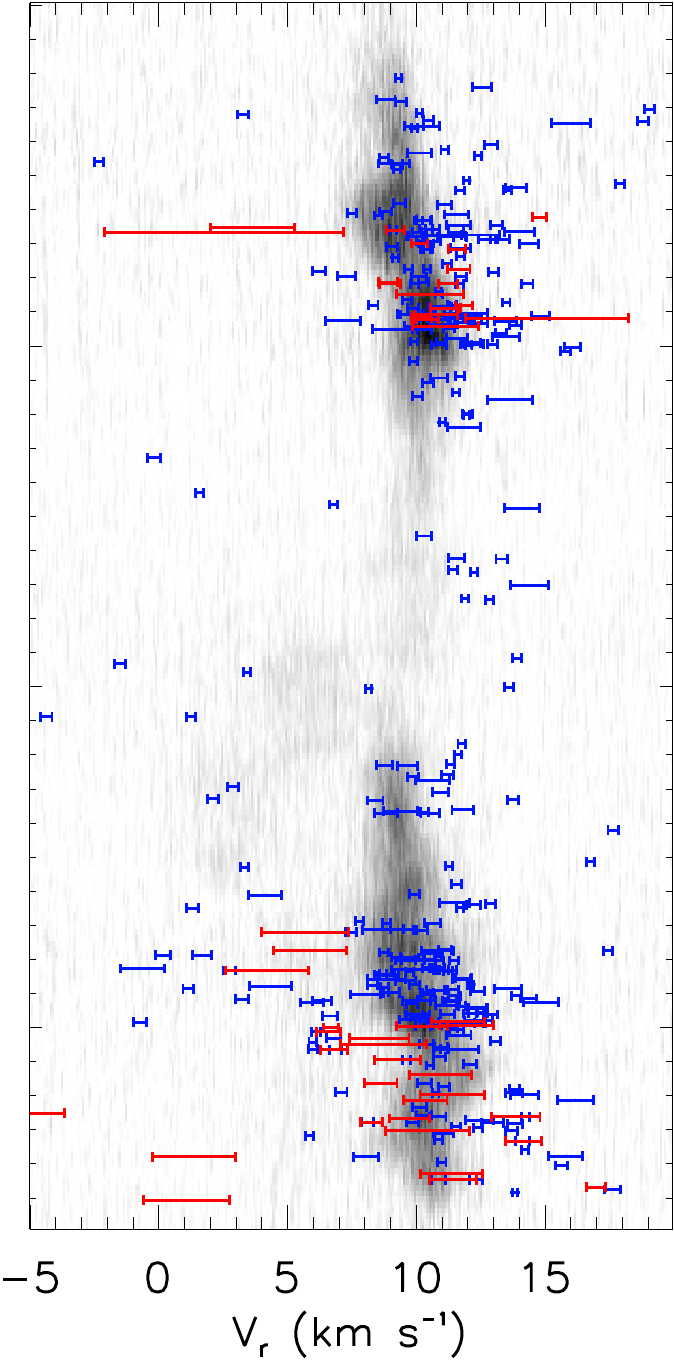}{0.20\textwidth}{}
        }
\caption{RVs of stars towards the Orion B molecular cloud. Blue symbols show measurements from APOGEE, red those uniquely detected by \citet{kounkel2017b}. The grayscale background shows the $^{13}$CO molecular gas map from \citet{nishimura2015}. \label{fig:oribrv}}
\end{figure}

Previously \citet{kounkel2017b} have studied the RV structure of the Orion B molecular cloud. They identified a peculiar RV structure towards the NGC 2024 region that is largely asymmetric from the RV structure of the molecular gas. Although due to a high degree of extinction towards it, their optical sample contained relatively few sources. With the expanded sample of the APOGEE observations, the agreement between the kinematics of the stars and the molecular gas is significantly improved. The blueshifted clump can be resolved into a kinematically distinct population which is centered at $v_r\sim6$ \kms\ (Figure \ref{fig:oribrv}), which appears to be associated with Ori D. On the other hand, the red-shifted clump is traced by the sources that originate from $\sigma$ Ori (see Section \ref{sec:oric}).

Towards the NGC 2068 cluster the agreement between the stellar and molecular kinematics remains good. There is some hint of a slight excess of stars somewhat redshifted relative to the gas, but it is difficult to conclusively state how significant it is.

We do recover a population that is located east of L1617, in the gas-free region close to the edge of the molecular cloud. We will refer to this population as the Orion B-North group. All three clusters are found at comparable distance of 404$\pm$5, 417$\pm$5 and 403$\pm$4 pc for Ori B-N group, NGC 2068, and NGC 2024 respectively (Figure \ref{fig:6d}.4). There is a deviation between the distances measured by \textit{Gaia} and those measured by \citet{kounkel2017} towards these regions; this could be attributed to the small sample size and multiplicity. In terms of proper motions, both Ori B-N and NGC 2068 appear to be moving towards NGC 2024.

L1622 is found outside of the footprint covered by APOGEE, and we do not recover it due to limited number of sources in \textit{Gaia} associated with it because of the high extinction. From 7 Class II stars that have been identified by \citet{megeath2012} as YSOs in L1622, we measure average astrometric parameters of $\pi=2.871\pm0.026$ mas ($d=345\pm$5.5 pc), $v_\alpha=5.343\pm0.045$ \masyr, and $v_\alpha=4.271\pm0.040$ \masyr. Combined with the typical $v_r=1.17$ \kms\ \citet{kun2008}, we conclude that L1622 is not associated with the Orion B molecular cloud, but may possibly have been related to Orion D. On the other hand, while L1622 and L1617 appear to be projected in the similar area of the sky, it is probable that the two are unrelated. While no parallaxes are available for the sources associated with the molecular gas in L1617, is more likely to be a part of Orion B given its $v_r$ \citep{reipurth2008}.

We estimate Age$_{HR}$ of 1.1$\pm$1.0 and 1.0$\pm$0.5 Myr (Age$_{CMD}$ of 1.9$\pm$2.0 and 1.0$\pm$1.4 Myr), for NGC 2024 and NGC 2068 respectively. Ori B-N group has Age$_{CMD}$=2.2$\pm$1.0 Myr.

\subsection{$\lambda$ Ori}

\begin{figure}
\epsscale{0.9}
\plotone{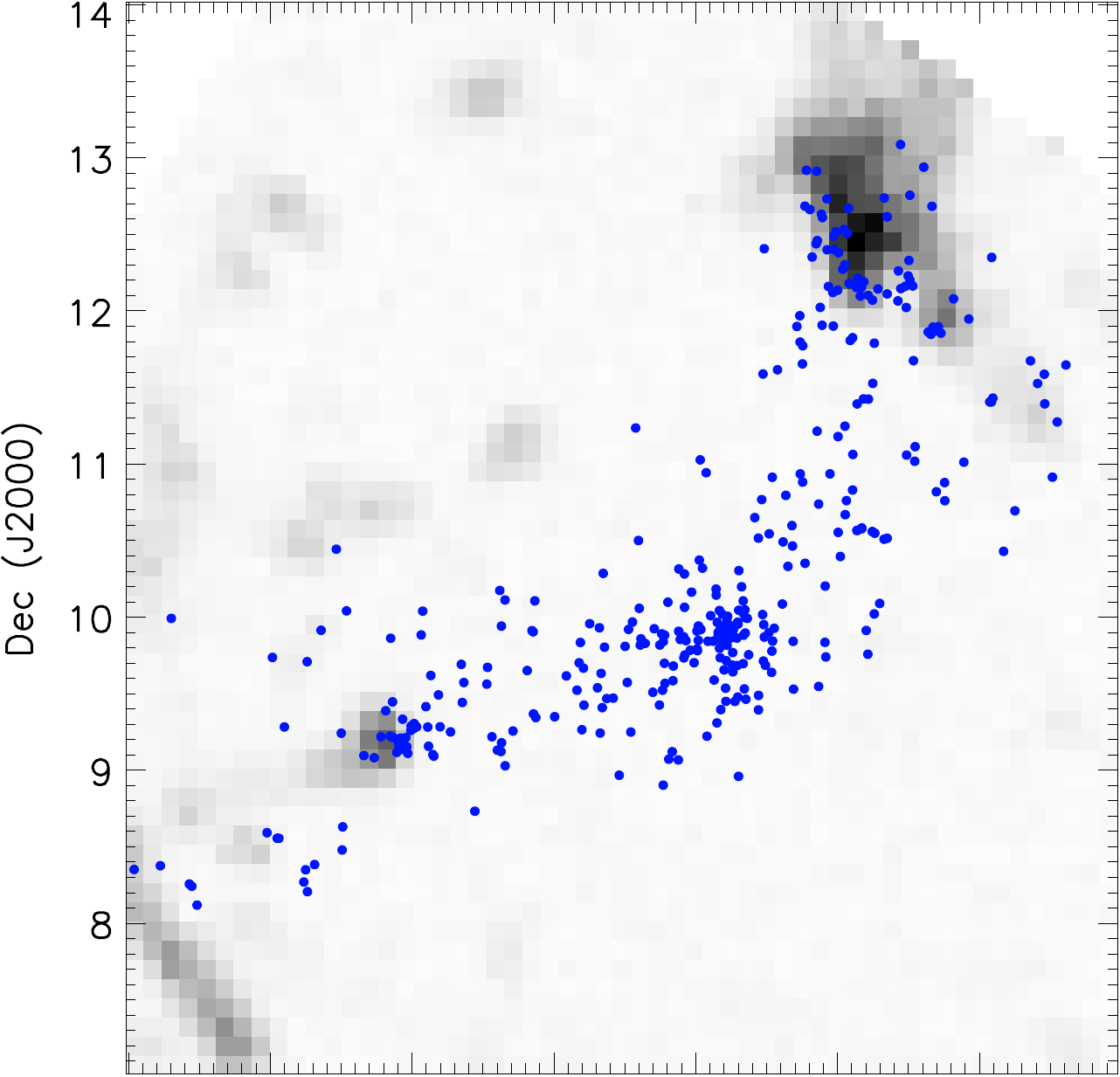}
\hspace*{0.05mm}\plotone{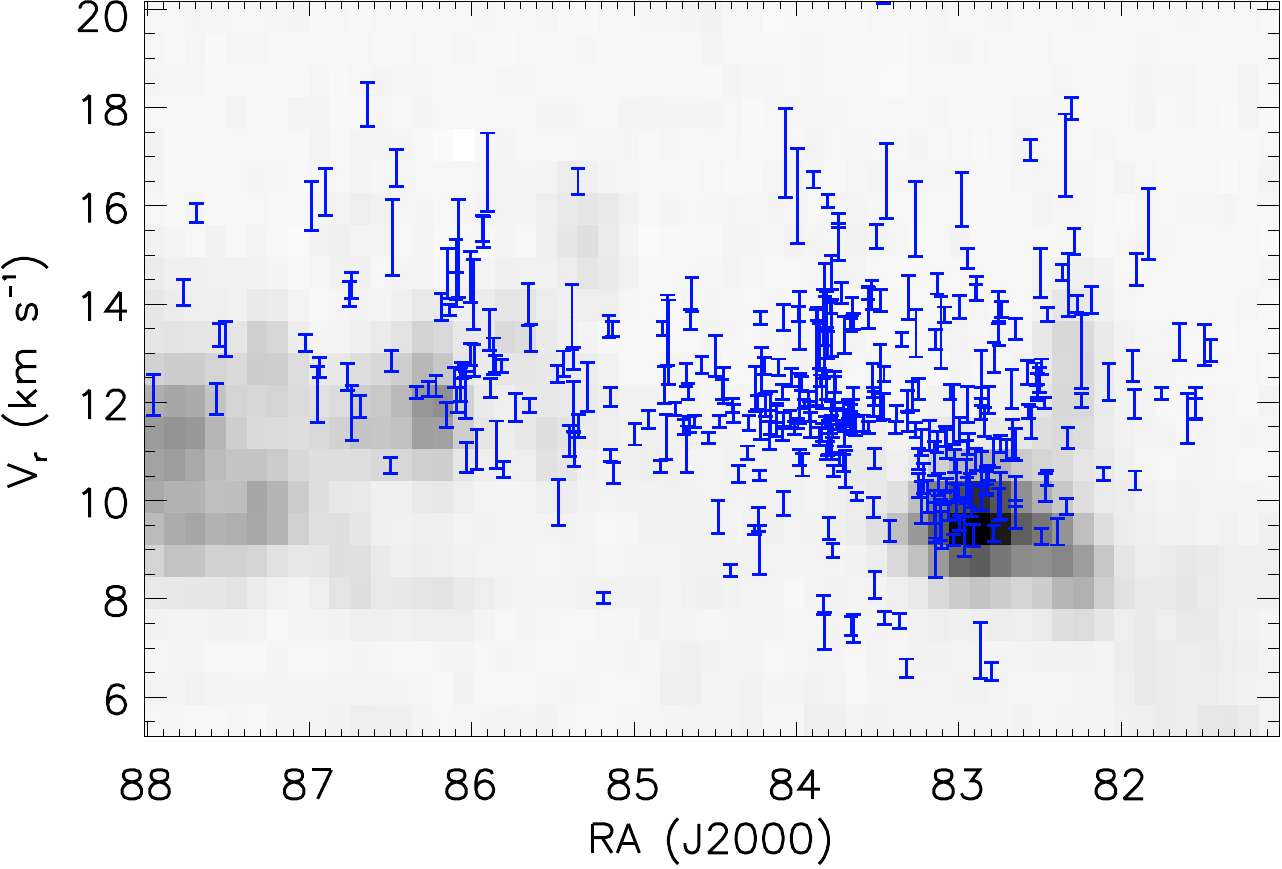}
\caption{RV measurements of stars towards $\lambda$ Ori. The grayscale background shows the CO molecular gas map from \citet{dame2001}. \label{fig:larv}}
\end{figure}

\begin{figure}
\epsscale{1.2}
\plotone{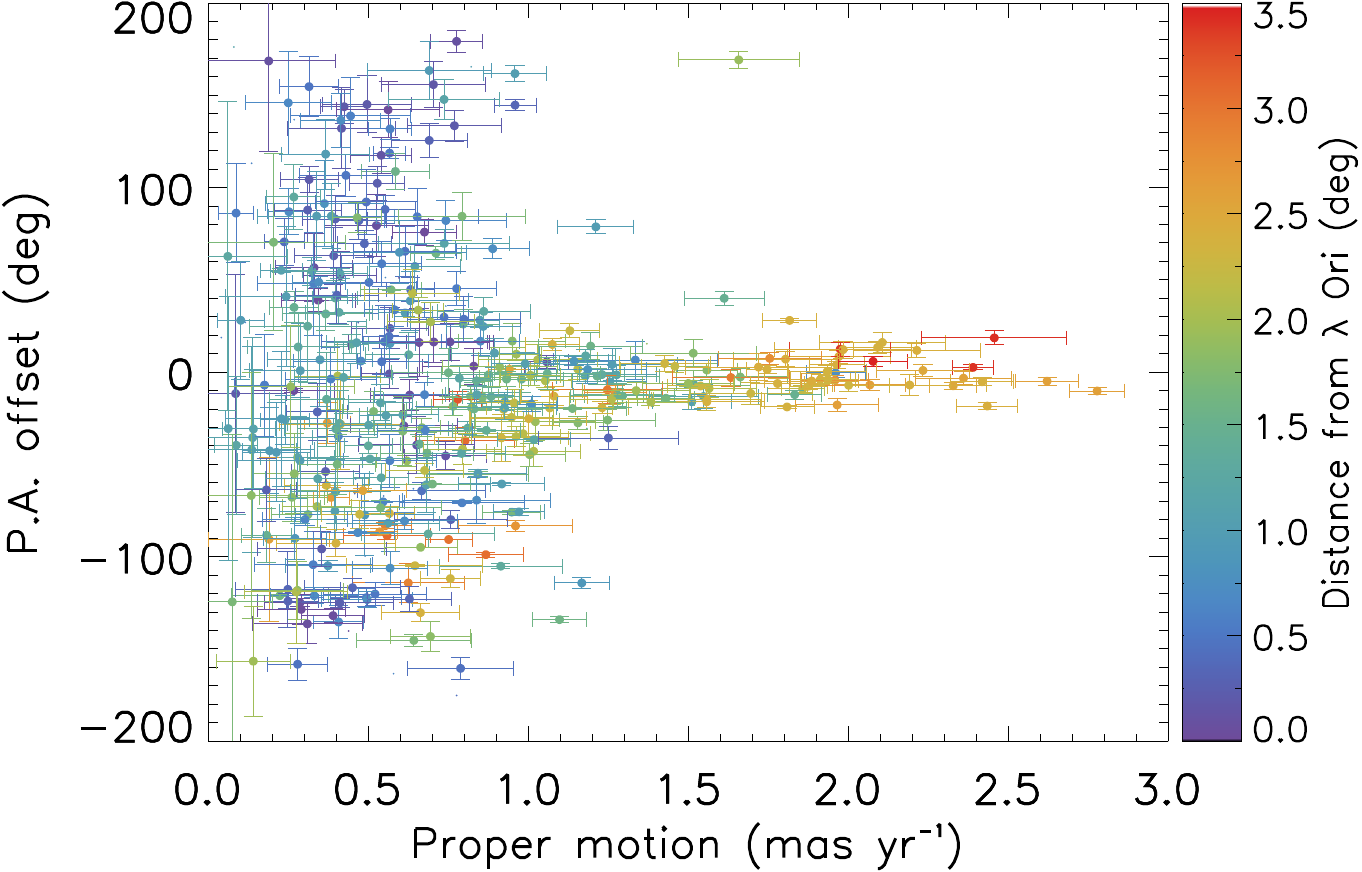}
\caption{Magnitude of the proper motion vector of stars towards $\lambda$ Ori (corrected for the average motion of the cluster) versus the offset between the position angle of the proper motions and the position angle of the star relative to the center of the cluster. Sources are color-coded according to their distance from the center of the cluster.  \label{fig:lapm}}
\end{figure}

Most of the molecular gas towards $\lambda$ Ori has already dissipated by a supernova; however, a few clouds within the ionized bubble remain: namely, B30 in the northwest, and B35 in the southeast. YSOs observed towards this region are primarily located alongside a filamentary structure stretching between these clouds. Stellar RVs are largely consistent with the RVs of the molecular gas, although a direct comparison is no longer possible (Figure \ref{fig:larv}).

In the cluster centered on $\lambda$ Ori, though, there do appear to be two distinct kinematical components (Figure \ref{fig:6d}.5): the main one at $v_r\sim$12 \kms, and a secondary sub-cluster (lambdaOri-2) at $v_r\sim$14 \kms. The separation is also apparent in proper motions, but not in parallax (average distance to the cluster of 404$\pm$4 pc) or in the spatial position on the sky. It appears lambdaOri-2 favors somewhat older ages than lambdaOri-1, with Age$_{HR}$=3.7$\pm$1.0 Myr vs 5.1$\pm$1.1 Myr (Age$_{CMD}$=3.5$\pm$0.9 Myr vs 4.4$\pm$1.4 Myr). Few groups (lambdaOri-3 and -4) also have somewhat peculiar proper motions, although they lack a reliable number of the RV measurements.

YSOs near B30 (located at a distance of 396$\pm$4 pc) are younger, 2.4$\pm$1.3 Myr, with a near-uniform distribution of ages from 2 to 5 Myr, with the northermost sources ($\delta>12^\circ$) near the cloud having Age$_{CMD}$=1.2$\pm$0.8 Myr (Age$_{HR}$=2.1$\pm$0.8 Myr). These YSOs are clumped preferentially near the inner edge of the molecular cloud, with a common $v_r\sim$10 \kms. A similar distribution is found towards B35, with the typical age of 2.6$\pm$1.3 Myr (distance of 397$\pm$4 pc), with the eastern-most sources becoming somewhat preferentially younger with the age of Age$_{CMD}$=1.8$\pm$1.5 Myr (Age$_{HR}$=2.0$\pm$1.0 Myr). 

A supernova was produced near the cluster center a few Myr ago. Nearly all proper motions larger than 1 \masyr relative to the average cluster motion are moving radially away from $\lambda$ Ori, and the further the sources are away from the center of the cluster, the faster they tend to move (Figure \ref{fig:lapm}). This is consistent with a single trigger expansion, which has the travel time from the cluster center of $\sim$4.8 Myr. On the other hand, sources within 1.5$\circ$ from the cluster appear to be mostly relaxed.

Based on their age, B30 and B35 would have formed half-way from $\lambda$ Ori to their current position, and would not have been accelerated due to the sudden mass loss of the molecular gas from the cluster. The fastest moving stars have proper motions of up to 6 \kms\ relative to the average motion of the cluster. Considering that 
\begin{itemize}
\item observationally, kinematics of YSOs usually closely mirror kinematics of the gas from which they form,
\item it is easier for an expanding shockwave to influence the velocity of the molecular gas compared to the already forming YSOs,
\item outlying stars are significantly younger than those in the cluster center, 
\end{itemize}
\noindent it is possible that the formation of a number of those stars have been triggered by the supernova, similarly to the scenario described by \citet{mathieu2008}. However, it should be noted that from simulations it appears to be difficult to distinguish between stars formation of which has been truly triggered, and those that are in process of formation regardless of any feedback \citep{dale2015}. Further modeling of the cluster dynamics will be needed to demonstrate the role that triggering has played in the formation of the outlying stars.

There may be additional structure near the outskirts of the $\lambda$ Ori that was suggested by \citet[such as L1588, for example]{zari2017}, although we do not recover it.

\section{Conclusions} \label{sec:conclusions}

\begin{itemize}
\item For the first time, we map the full extent of the population of YSOs found in the regions towards the portions of the Orion Complex that are presently devoid of the molecular gas and trace their 3d kinematics. Using these data we can now reconstruct the properties of the dispersed molecular clouds that produced them, and they represent an example of potential evolution of their younger counterparts.
\item Most notably, we identify two separate populations towards the Ori OB1ab sub-associations that are coherently separated in RV and distance space. These two populations are projected on top of one another, both spanning several deg$^2$. One of them, to which we refer as Orion C, and has 3 distinct epochs of star formation, ranging from 2 to 10 Myr. This population is the progenitor of the $\sigma$ Ori cluster. The other, which we refer to as Orion D, spans the full extent of OB1ab, tracing most of the brightest OB stars in the region, and extends further south towards Rigel encompassing the Orion X region. We also recover a previously uncatalogued population north of $\kappa$ Ori that we refer to as Orion Y. 
\item We identify several peculiar groups that may represent kinematically distinct subgroups. Particularly notable are those located within $\lambda$ Ori, ONC, and 25 Ori. 
\item We measure average distances of 386$\pm$3 pc for the ONC, extending up to 443$\pm$4 at the southern end of Orion A, 407$\pm$4 pc for Orion B, 412$\pm$4 pc for Orion C, 350$\pm$3 pc for Orion D, 400$\pm$4 pc for $\lambda$ Ori, and we also trace the kinematics and the distribution of ages within them. Large structures that are devoid of molecular gas are preferentially expanding (although further investigation will be necessary to confirm whether individual clusters in these structures are bound or not). In $\lambda$ Ori the expansion is largely radial and attributable to a supernova explosion, which could be modeled in the future. In Orion A, the kinematics are preferentially perpendicular to the cloud, and Orion B appears various groups are preferentially moving towards NGC 2024.
\item Together these data represent a major step forward in terms of understanding the dynamical evolution within the young star-forming regions. This will be instrumental for the future modeling efforts of the assembly and dispersal of molecular clouds and the stars that form within them, as well as determining whether any of the subclusters will remain bound, and analyzing the difference in the kinematical population as a function of age.

\end{itemize}

\appendix
\section{Synthetic population properties}\label{sec:synthetic}

\begin{figure*}
  \centering
		\gridline{\fig{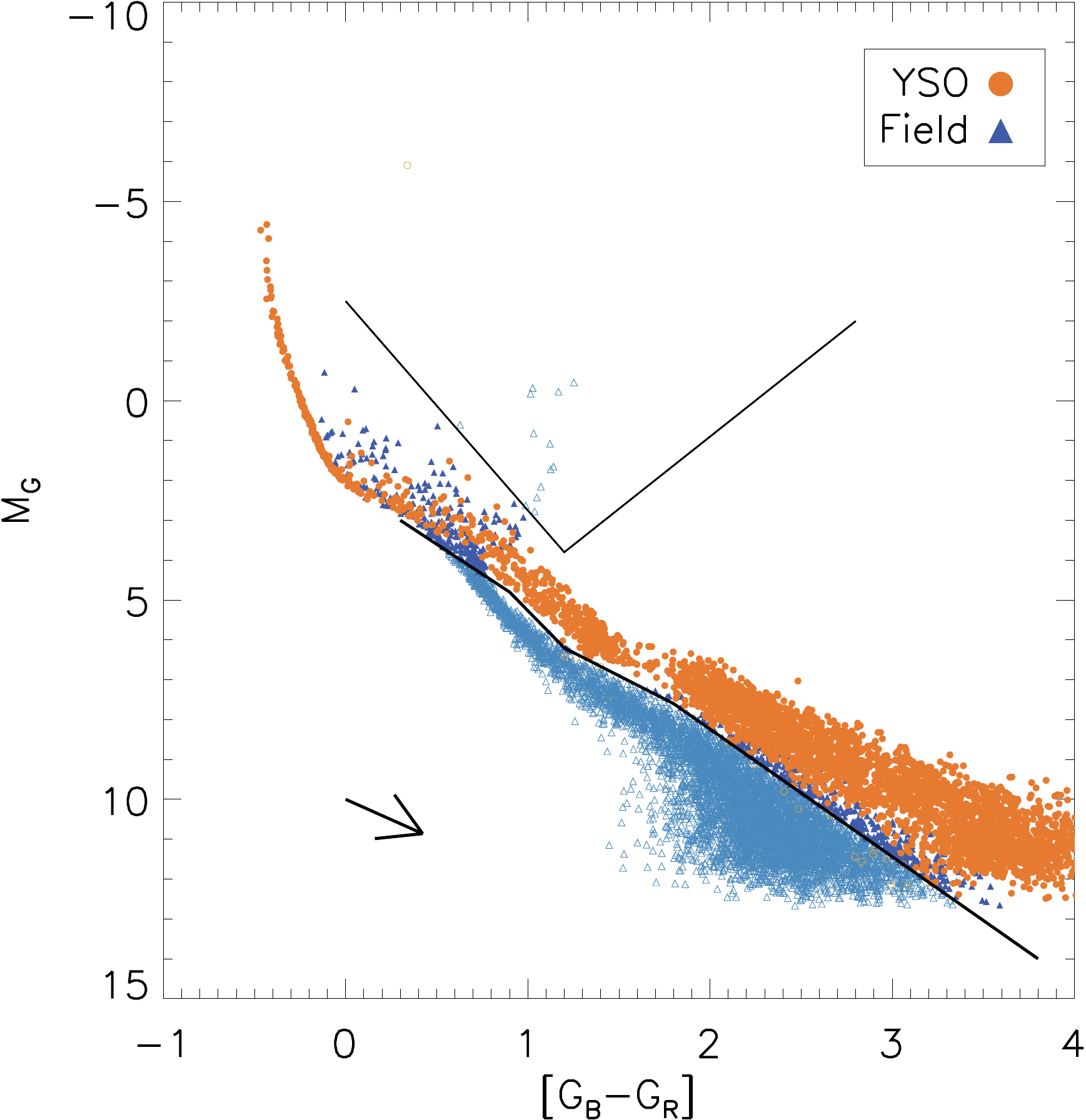}{0.45\textwidth}{}
              \fig{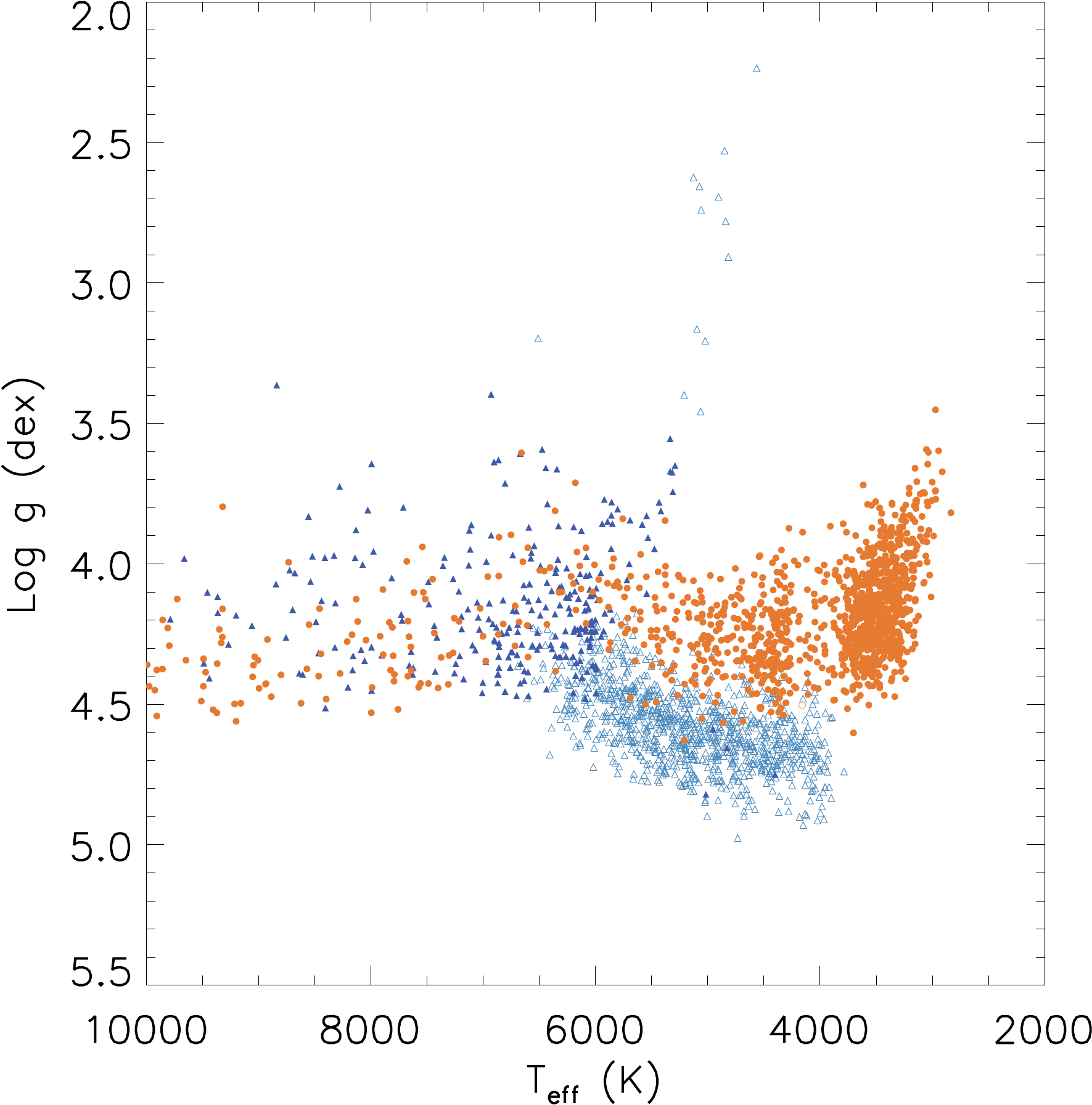}{0.46\textwidth}{}
        }
      \caption{Color-luminosity (left) and \teff-\logg (right) diagrams for the synthetic population, generated to replicate the observed distribution of parameters from APOGEE and \textit{Gaia}, details in the text. YSOs shown have ages of up to 15 Myr, field stars have ages of over 100 Myr. Dark lines show the cuts applied in the magnitude space to filter the main sequence and giant stars; open symbols correspond to the photometrically rejected sources. The arrow corresponds to a total extinction  of 1 $A_V$. \label{fig:synthetic1}}
\end{figure*}

\begin{deluxetable}{ccc}
\tabletypesize{\scriptsize}
\tablewidth{0pt}
\tablecaption{Properties of the synthetic clusters. \label{tab:synthetic2}}
\tablehead{
\colhead{Property} &\colhead{Min} & \colhead{Max}
}
\startdata
Number of stars &  15 & 75\\
Distance (pc) & 300 & 450 \\
Median velocity ($v_r,v_\alpha,v_\delta$ \kms) & -20 & 20 \\
Cluster radius (pc) & 0.4 & 5.0 \\
Velocity dispersion (\kms) & 0.3 & 2.5 \\
Age (Myr) & 3 & 15 \\
\enddata
\end{deluxetable}

We generated a synthetic population of stars with properties matched to those observed towards the Orion Complex. This population includes synthetic field stars as well as a cluster with a random age, distance, and three dimensional mean velocity. A random number cluster members are generated with peculiar position and velocities, drawn from normal distributions corresponding to the cluster characteristic radius and velocity dispersion. Both the velocity dispersion and the characteristic size are allowed to vary in 3 dimensions independently of each other, so as to allow fillamentary structures to form. All the parameters are drawn from a uniform distribution with the ranges corresponding to those listed in the Table \ref{tab:synthetic2}. This cluster is embedded into a field population, in which each field star has a random position within a 4$\times$4 deg$^2$ area on the sky, a distance of up to 1 kpc \citep[to account for the Lutz-Kelker bias,][]{lutz1973}, an age from 0.1 to 12 Gyr, and a 3-d velocity drawn from a normal distribution centered at 0 \kms\ with a 25 \kms\ velocity dispersion \citep[e.g.,]{rix2013}. The masses for both the cluster and field stars are drawn from the Initial Mass Function (IMF) from \citet{muench2002}. The various properties, such as \teff, \logg, $M_G$, $M_{G_B}$, $M_{G_R}$, and $M_H$ band luminosities, are interpolated from the PARSEC-COLIBRI isochrones in accordance with the previously assigned masses and ages \citep{marigo2017}.

We assume an average stellar density of 0.09 M$_\odot$ pc$^{-3}$ \citep{kipper2018}. The volume of space encompassed by the field stars in the simulation is $\sim1.63\times10^6$ pc$^3$. With the resulting average stellar mass of 0.3 $M_\odot$, we fixed the total number of field stars to $4.5\times10^4$.

Observational properties were then computed for the synthetic stars, with limits applied to simulate the impact of the APOGEE \& \textit{Gaia} detection limits. The positional and kinematical parameters are converted to the observable properties ($v_r,v_\alpha,v_\delta,\pi$), and the apparent magnitudes $G$ and $H$ are computed. Stars fainter than $G>20$ have their astrometric parameters discarded, and those with $H>13$ have no spectral information. If a source fails the brightness test in both parameters, it is rejected from the sample. Noise is then applied to all the measurements: 

\begin{itemize}
\item $G>15$: $\sigma_\pi$=0.04 mas, $\sigma_\mu$=0.06 \masyr
\item $G=17$: $\sigma_\pi$=0.1 mas, $\sigma_\mu$=0.2 \masyr
\item $G=20$: $\sigma_\pi$=0.7 mas, $\sigma_\pi$=1.2 \masyr \citep{katz2017}
\item Absolute magnitude $M_G$ is recomputed to incorporate the uncertainty in distance
\item $G$=13 mag: $\sigma_G$=0.001 mag, $\sigma_{G_{B}-G_{R}}$=0.004 mag
\item $G$=20 mag: $\sigma_G$=0.02 mag, $\sigma_{G_{B}-G_{R}}$=0.3 mag
\item $\sigma_v=0.25$ \kms\ for \teff$<7000 K$ ($v_r$ for sources with \teff$>7000$ are rejected as their measurements are too uncertain)
\end{itemize}
In between the magnitude bins, the value of the uncertainties is interpolated. The sample is then limited only to those sources with $2<\pi<3.5$ mas, $-25<v<25$ \kms, and $-20<v_{\alpha,\delta}<20$ \masyr (if such parameters are defined), which includes only $\sim$3500 field stars. Furthermore, photometric cuts

\[M_G<3\times[G_B-G_R]+2.1; 0.3<[G_B-G_R]<0.9\]
\[M_G<4.67\times[G_B-G_R]+0.6; 0.9<[G_B-G_R]<1.2\]
\[M_G<2.33\times[G_B-G_R]+3.4; 1.2<[G_B-G_R]>1.8\]
\[M_G<3.2\times[G_B-G_R]+8.4; [G_B-G_R]>1.8\]
\[M_G>5.25\times[G_B-G_R]-2.5; 0<[G_B-G_R]<1.2\]
\[M_G>-3.63\times[G_B-G_R]+8.15; 1.2<[G_B-G_R]<2.8\]

\noindent are applied to exclude the field main sequence stars and red giants (Figure \ref{fig:synthetic1}). After this, only $\sim$285 non-cluster sources remain in the sample. For populations of up to 15 Myr, the YSO rejection rate is $<$1\%. These photometric cuts are somewhat different from those described in section \ref{sec:data}, due to the systematic differences between the assumptions made regarding the isochrones and the real observations; however, qualitatively these cuts should yield similar results.

It should be noted that this synthetic population ignores effects of extinction (which may confuse the separation between cluster and field stars, as well as render many stars too faint to have astrometric parameters with \textit{Gaia}), or variability. However, the extinction should have a significant effect primarily on the youngest and most well-defined clusters, and it isn't expected to have a strong effect on the tests in Section \ref{sec:test} beyond the initial sample selection. The effect of multiplicity is also neglected; orbital motion does affect the kinematics measured for an individual star in a system, accelerating it to velocities where it may appear to be unrelated to a cluster. Binaries with a period of a few years are absent from the astrometric solutions in \textit{Gaia} DR2. Additionally, while close unresolved binaries with equal masses could double the apparent flux of a system, observationally only a small fraction of all sources would be affected by it. Finally, other selection biases are also ignored, such as the fiber collision within densely packed clusters or the targeting criteria that would affect the completeness of the APOGEE observations \citep{cottle2018}.

\software{IN-SYNC pipeline \citep{cottaar2014}, PHOENIX \citep{husser2013}, PARSEC-COLIBRI \citep{marigo2017}}

\acknowledgments
M.K. and K.C. acknowledge support provided by the NSF through grant AST-1449476, and from the Research Corporation via a Time Domain Astrophysics Scialog award (\#24217). J.H. acknowledges support from programs UNAM-DGAPA-PAPIIT IN103017, Mexico. ARL acknowledges partial financial support provided by the FONDECYT REGULAR project 1170476. KPR acknowledges CONICYT PAI Concurso Nacional de Inserci\'{o}n en la Academia, Convocatoria 2016 Folio PAI79160052. Support for JB is provided by the Ministry for the Economy, Development and Tourism, Programa Iniciativa Cientica Milenio grant IC120009, awarded to the Millenni. DAGH and OZ acknowledge support provided by the Spanish Ministry of
Economy and Competitiveness (MINECO) under grant AYA-2017-88254-P. Funding for the Sloan Digital Sky Survey IV has been provided by the Alfred P. Sloan Foundation, the U.S. Department of Energy Office of Science, and the Participating Institutions. SDSS-IV acknowledges
support and resources from the Center for High-Performance Computing at
the University of Utah. The SDSS web site is www.sdss.org.
SDSS-IV is managed by the Astrophysical Research Consortium for the 
Participating Institutions of the SDSS Collaboration including the 
Brazilian Participation Group, the Carnegie Institution for Science, 
Carnegie Mellon University, the Chilean Participation Group, the French Participation Group, Harvard-Smithsonian Center for Astrophysics, 
Instituto de Astrof\'isica de Canarias, The Johns Hopkins University, 
Kavli Institute for the Physics and Mathematics of the Universe (IPMU) / 
University of Tokyo, Lawrence Berkeley National Laboratory, 
Leibniz Institut f\"ur Astrophysik Potsdam (AIP),  
Max-Planck-Institut f\"ur Astronomie (MPIA Heidelberg), 
Max-Planck-Institut f\"ur Astrophysik (MPA Garching), 
Max-Planck-Institut f\"ur Extraterrestrische Physik (MPE), 
National Astronomical Observatories of China, New Mexico State University, 
New York University, University of Notre Dame, 
Observat\'ario Nacional / MCTI, The Ohio State University, 
Pennsylvania State University, Shanghai Astronomical Observatory, 
United Kingdom Participation Group,
Universidad Nacional Aut\'onoma de M\'exico, University of Arizona, 
University of Colorado Boulder, University of Oxford, University of Portsmouth, 
University of Utah, University of Virginia, University of Washington, University of Wisconsin, 
Vanderbilt University, and Yale University.
This work has made use of data from the European Space Agency (ESA)
mission {\it Gaia} (\url{https://www.cosmos.esa.int/gaia}), processed by
the {\it Gaia} Data Processing and Analysis Consortium (DPAC,
\url{https://www.cosmos.esa.int/web/gaia/dpac/consortium}). Funding
for the DPAC has been provided by national institutions, in particular
the institutions participating in the {\it Gaia} Multilateral Agreement.

\bibliographystyle{aasjournal.bst}
\bibliography{../rvstruc.bbl}

\end{document}